\documentclass[11pt,a4paper]{article}

\usepackage[usenames,dvipsnames,svgnames,x11names]{xcolor}

\usepackage[T1]{fontenc} %
\usepackage{lmodern} %
\usepackage[normalem]{ulem} %

\newcommand{\addc}[1]{{\textcolor{teal}{#1}}}
\newcommand{\addcBegin}{\color{teal}}
\newcommand{\addcEnd}{\color{black}}

\newcommand{\tempdel}[1]{}

\renewcommand{\addc}[1]{#1}
\renewcommand{\addcBegin}{}
\renewcommand{\addcEnd}{}

\usepackage[nocompress]{cite}

\usepackage{listings}

\lstdefinelanguage{XML}
{
basicstyle=\ttfamily\footnotesize,
  morestring=[b]",
  moredelim=[s][\bfseries\color{Maroon}]{<}{\ },
  moredelim=[s][\bfseries\color{Maroon}]{</}{>},
  moredelim=[l][\bfseries\color{Maroon}]{/>},
  moredelim=[l][\bfseries\color{Maroon}]{>},
  morecomment=[s]{<?}{?>},
  morecomment=[s]{<!--}{-->},
  commentstyle=\color{gray},
  stringstyle=\color{blue},
  identifierstyle=\color{red}
}
\usepackage{moreverb}

\usepackage[nounderscore]{syntax}

\usepackage[pdftex]{graphicx}
\graphicspath{{./figures/}}
\DeclareGraphicsExtensions{.pdf}

\usepackage[cmex10]{amsmath}
\usepackage{amssymb}
\usepackage{mathtools}
\usepackage{amsthm}
\usepackage{amsfonts}

\usepackage{subfig} %

\usepackage{algorithmicx}
\usepackage{algpseudocode}
\usepackage[ruled]{algorithm}
\definecolor{light-gray}{gray}{0.75}
\algrenewcommand{\algorithmiccomment}[1]{\hskip3em{{\footnotesize \textcolor{light-gray}{$\blacktriangleright$}}} #1}

\usepackage{multirow} %
\usepackage{rotating} %
\usepackage{booktabs} %
\usepackage{colortbl} %
\usepackage{tablefootnote} %

\usepackage{array}
\newcolumntype{L}[1]{>{\raggedright\let\newline\\\arraybackslash\hspace{0pt}}m{#1}}
\newcolumntype{C}[1]{>{\centering}m{#1}}

\usepackage[pdftex,colorlinks=true,urlcolor=blue,citecolor=blue]{hyperref}

\usepackage{xspace}

\usepackage{enumitem}

\newcommand{\graphdb}{$\mathcal{G}ranite$\xspace}

\usepackage{blindtext}

\newcommand{\blindfootnote}[1]{{\renewcommand\thefootnote{}\footnotetext{#1}}}

\date{}

\begin{document}

\title{A Distributed Path Query Engine for \\Temporal Property Graphs *}

\author{Shriram Ramesh, Animesh Baranawal and Yogesh Simmhan \\
\small \emph{Department of Computational and Data Sciences,} \\
\small \emph{Indian Institute of Science, Bangalore 560012, India}\\
\small \emph{Email: \{shriramr, animeshb, simmhan\}@iisc.ac.in}}
\maketitle

\blindfootnote{*An extended version of the paper that appears in IEEE/ACM International	Symposium on Cluster, Cloud and Internet Computing (CCGrid), 2020. {\tiny }doi:\href{https://dx.doi.org/10.1109/CCGrid49817.2020.00-43}{10.1109/CCGrid49817.2020.00-43}}

\begin{abstract}
Property graphs are a common form of linked data, with path queries used to traverse and explore them for enterprise transactions and mining. \emph{Temporal property graphs} are a recent variant where \emph{time} is a first-class entity to be queried over, and their properties and structure vary over time. These are seen in social, telecom, transit \addc{and epidemic} networks. However, current graph databases and query engines have limited support for temporal relations among graph entities, no support for time-varying entities and/or do not scale on distributed resources. 
We address this gap by extending a linear path query model over property graphs to include intuitive \emph{temporal predicates} \addc{and \emph{aggregation operators}} over temporal graphs. We design a \emph{distributed execution model} for these temporal path queries using the interval-centric computing model, and develop a novel \emph{cost model} to select an efficient execution plan from several.
We perform detailed experiments of our \graphdb distributed query engine using \addc{both static and dynamic} temporal property graphs as large as $52M$ vertices, $218M$ edges and \addc{$325M$} properties, and a \addc{$1600$}-query workload, derived from the LDBC benchmark. We \addc{often} offer sub-second query latencies \addc{on a commodity cluster}, which is $149\times$--$1140\times$ faster compared to industry-leading Neo4J shared-memory graph database and the JanusGraph/Spark distributed graph query engine. \graphdb also completes $100\%$ of the queries for all graphs, compared to only $32$--$92\%$ \addc{workload completion} by \addc{the} baseline systems. %
Further, our cost model selects a query plan that is within $10\%$ of the optimal execution time in $90\%$ of the cases. \addc{Despite the irregular nature of graph processing, we exhibit a weak-scaling efficiency of $\geq 60\%$ on $8$ nodes and $\geq 40\%$ on $16$ nodes, for most query workloads.}%
\end{abstract}

\section{Introduction}
\label{sec:intro}
Graphs are a natural model to represent and analyze linked data in various domains. \emph{Property graphs} allow vertices and edges to have associated \emph{key--value pair} properties, besides the graph structure. This forms a rich information schema and has been used to capture knowledge graphs (concepts, relations)~\cite{mitchell2018never}, social networks (person, forum, message)~\cite{cha2010measuring}, \addc{epidemic networks (subject, infected status, location)~\cite{liu2016epidms},} and financial and retail transactions (person, product, \addc{purchase})~\cite{haslhofer2016bitcoin}. 

\begin{figure}[!t]
	\centering
	\includegraphics[width=0.6\columnwidth]{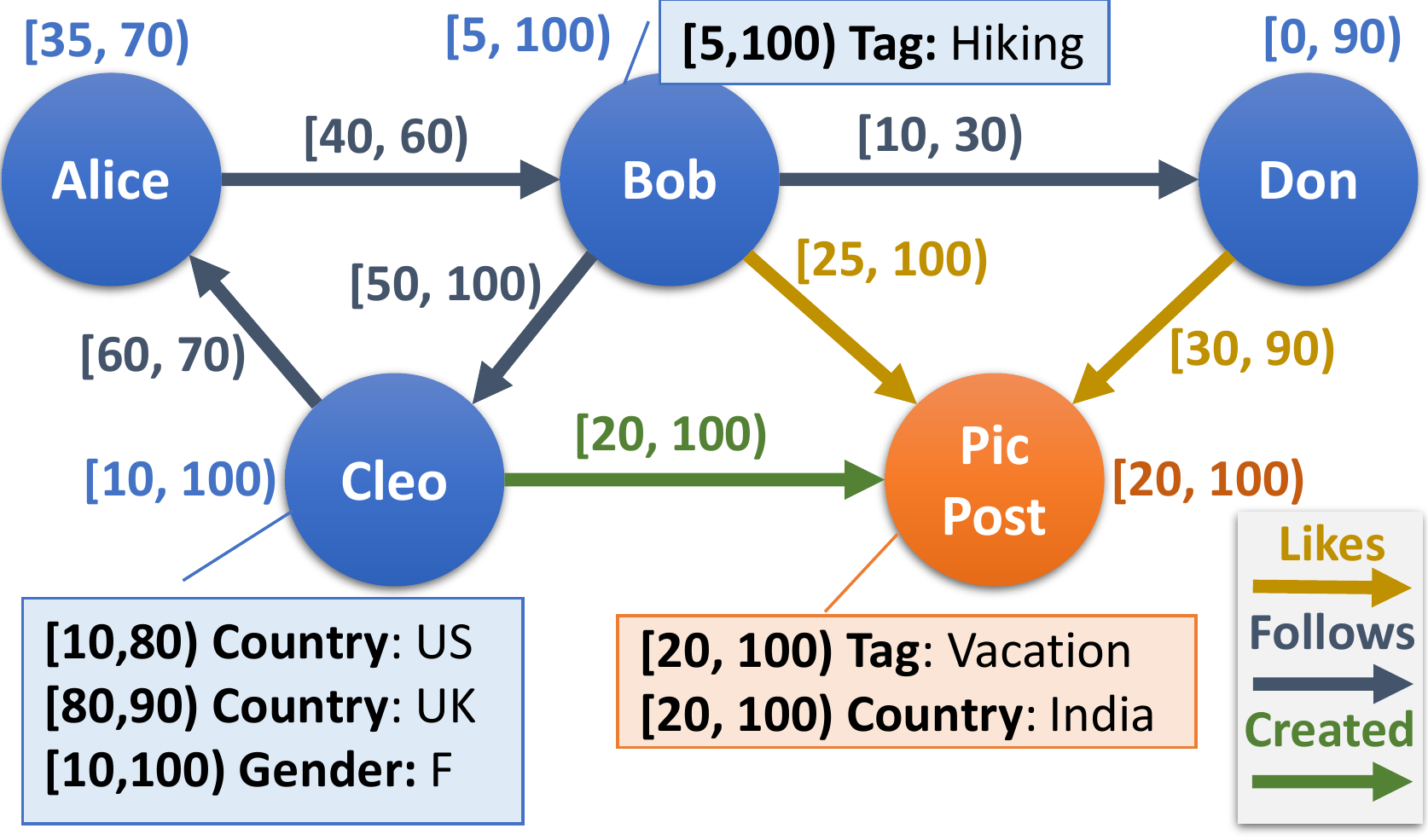}
	\caption{Sample Temporal Property Graph \addc{of a Community of Users}}
	\label{fig:tpgraph}
\end{figure}

\emph{Path queries} are a common class \addc{of queries} over property graphs~\cite{shu2017fake,Fan:2012:GPM:2274576.2274578}.~%
Here, the user defines a sequence of predicates over vertices and edges that should match along a path in the graph. E.g., in the property graph \addc{for a community of users} in Figure~\ref{fig:tpgraph}, \addc{the vertices are labeled with their \emph{ID}s, their colors indicate their \emph{type} -- blue for \emph{Person} and orange for a \emph{Post}, and they have a set of properties listed as \emph{Name:Value}. The edges are relationships, with \emph{types} such as \emph{Follows}, \emph{Likes} and \emph{Created}. We can define an example} 3-hop path query ``\textbf{[EQ1]} \emph{Find a {person} (vertex type) who lives in \addc{the {country}} `UK' (vertex property) and {follows} (edge type) a {person} who {follows} another {person} who is {tagged} with \addc{the label} `Hiking' (vertex property)}''. This \addc{query} would match \emph{Cleo$\rightarrow$Alice$\rightarrow$Bob}, if we ignore the time intervals. Path queries are used to identify concept pathways in knowledge graphs, \addc{find friends in social networks,} fake news detection, and suggest products in retail websites~\cite{shu2017fake,Fan:2012:GPM:2274576.2274578,huang2004graph}. They also need to be performed rapidly, within $\approx 1~sec$, as part of \addc{interactive} requests from websites or exploratory queries by analysts.%

While \emph{graph databases} are designed for transactional read and write workloads, we consider graphs that are updated infrequently but queried often. For these workloads, \emph{graph query engines} load and retain property graphs in-memory to service requests with low latency, without the need for locking or consistency protocols~\cite{janusgraph,castellana2015memory}. %
\addc{They may also create indexes to accelerate these searches~\cite{yan2009efficient,milo1999index}.
	Property graphs can be large, with $10^5$--$10^8$ vertices and edges, and $10$'s of properties on each vertex or edge. This can exceed the memory on a single machine, often dominated by the properties. This necessitates the use of distributed systems to scale to large graphs~\cite{junghanns2018declarative,shao2013trinity}}.%

\paragraph{Challenges} Time is an increasingly common graph feature in a variety of domains~\cite{greene2010tracking,george2008time,zhao2018stock,liu2016epidms}. However, existing property graph \emph{data models} fail to consider it as a first-class entity. Here, we distinguish between graphs with a \emph{time interval} or a \emph{lifespan} associated with their entities (properties, vertices, edges), and those where the entities themselves change over time and the history is available. We call the former \emph{static temporal graphs} and the latter \emph{dynamic temporal graphs}. \addc{Yet another class is \emph{streaming graphs}, where the topology and properties change in real-time, and queries are performed on this evolving structure~\cite{walavalkar2008streaming,ediger2011tracking}; that is outside the scope of this article.}

E.g., in the temporal graph in Figure~\ref{fig:tpgraph}, the lifespan, \emph{[start, end)}, is indicated on the vertices, edges and properties. \addc{The \emph{start} time is inclusive while the \emph{end} time is exclusive.} Other than the properties of \emph{Cleo}, the \addc{remaining entities of the graph} form a static temporal graph \addc{as they are each valid only for a single time range}. But the \addc{value of the} \emph{Country} property of \emph{Cleo} changes over time, making it a dynamic temporal graph.

This gap is reflected not just in the data model but also in the \emph{queries supported}.  %
\addc{We make a distinction between \emph{time-independent (TI)} and \emph{time-dependent (TD)} queries, both being defined on a temporal graph~\cite{then2017automatic}. TI queries are those which can be answered by examining the graph at a single point in time (\emph{a snapshot}), e.g. EQ1 executed on the temporal graph. In contrast, TD queries capture temporal relations between the entities across consecutive time intervals,} e.g., ``\textbf{[EQ2]} \emph{Find people tagged with `Hiking' who liked a post tagged as `Vacation', \underline{\smash{before}} the post was liked by a person named `Don'}~'',~ and ``\textbf{[EQ3]} \emph{Find people who started to follow another person, \underline{\smash{after}} the latter stops following `Don'~}''. Treating time as just another property fails to express temporal relations such as ensuring time-ordering among the entities on the path. While EQ2 and EQ3 should match the paths \emph{Bob$\rightarrow$PicPost$\rightarrow$Don}
and \emph{Alice$\rightarrow$Bob$\rightarrow$Don}, respectively, such queries are hard, if not impossible, to express in current graph databases.
This problem is exacerbated for \emph{path queries over dynamic temporal graphs}. E.g., the query EQ1 \addc{over the dynamic temporal graph} \emph{should not} match \emph{Cleo$\rightarrow$Alice$\rightarrow$Bob} since at the time \emph{Cleo} was living in `UK', she was not following \emph{Alice}. %

While platforms \addc{which execute snapshot at a time~\cite{malewicz2010pregel,then2017automatic}} can be adapted to support \addc{TI queries over temporal graphs, TD queries} cannot be expressed meaningfully.
\addc{Even those that support TD algorithms enforce strict temporal ordering~\cite{graphite}, requiring that the time intervals along the path should be increasing or decreasing, but not both; this limits query expressivity.}
These motivate the need to support intuitive temporal predicates to concisely express such temporal relations, \addc{and flexible platforms to execute them.}
\addc{Lastly,} the \emph{scalability} of existing graph systems is also limited, \addc{with few property graph query engines that operate on distributed memory systems with low latency~\cite{janusgraph,zeng2013distributed}, let alone on temporal property graphs.}

We make the following specific contributions in this article:
\begin{itemize}
	\item We propose a temporal property graph model, and intuitive \emph{temporal predicates} \addc{and \emph{aggregation operators}} for path queries on them (\S\ref{sec:query}).
	\item We design a \emph{distributed execution model} for these queries using the interval-centric computing model (\S\ref{sec:engine}).
	\item We develop a novel \emph{cost model} that uses graph statistics to select the best from multiple execution plans (\S\ref{sec:cost}).
	\item We \addc{conduct a detailed evaluation of} the performance and scalability of \graphdb for \addc{$8$} temporal graphs and up to \addc{$1600$} queries, %
	derived from the LDBC benchmark. We compare this against three configurations of Neo4J, and JanusGraph which uses Apache Spark (\S\ref{sec:results}). %
\end{itemize}

\noindent We discuss related work in Section~\ref{sec:related} and our conclusions in Section~\ref{sec:conclude}.

A prior version of this work appeared as a conference paper~\cite{sriram2020ccgrid}. This article substantially extends this. Specifically, it introduces the temporal aggregation operator to the query model (Section~\ref{sec:path}) and implements it within the execution model; offers details, illustrations and complexity metrics for our query model, distributed execution model and query optimizations (Sections~\ref{sec:query},~\ref{sec:engine} and~\ref{sec:cost}); and provides a rigorous empirical evaluation, including two additional large dynamic temporal graphs, aggregation query workloads, weak scaling experiments, and results on the component times of query execution, besides more detailed analysis for the cost model benefits and baseline platform comparisons (Section~\ref{sec:results}).%

\section{Related Work}%
\label{sec:related}

\subsection{Distributed and Temporal Graph Processing}
There are several distributed graph processing platforms for running graph algorithms on commodity clusters and clouds~\cite{guo2014well}. These typically offer programming abstractions like Google Pregel's \emph{vertex-centric computing model}~\cite{malewicz2010pregel} and its component-centric variants~\cite{goffish,gonzalez2014graphx} to design algorithms such as Breadth First Search, centrality scores and mining~\cite{Chen:2018:GET}. These execute using a Bulk Synchronous Parallel (BSP) model, and scale to large graphs and applications that explore the entire graph. They offer \emph{high throughput batch processing} that take $\mathcal{O}(mins)$--$\mathcal{O}(hours)$. We instead focus on exploratory and transactional path queries that are to be processed in $\mathcal{O}(secs)$. This requires careful use of distributed graph platforms and optimizations for fast responses.

There are also parallel graph platforms for HPC clusters and accelerators~\cite{boost}. These optimize the memory and communication access to scale to graphs with billions of entities on thousands of cores~\cite{dang2018lightweight}. They focus on high-throughput graph algorithms and queries over static graphs~\cite{shi2016fast}. We instead target commodity hardware and cloud VMs with 10's of nodes and 100's of total cores, and are more accessible. We also address queries over temporal property graphs.%

A few distributed abstractions and platforms support designing of temporal algorithms and their batch execution~\cite{simmhan2015distributed,han2014chronos,then2017automatic}. \addc{Most are limited to executing TI algorithms, snapshot at a time, and are unable to seamlessly model TD queries.} \addc{Our prior work \emph{Graphite} offers an interval-centric computing model (ICM) to represent TI and TD algorithms, but limits it to time-respecting algorithms~\cite{graphite}.} We use it as the base framework for our proposed distributed path query engine, \addc{while relaxing the time-ordering, including indexing and proposing different query execution plans for low-latency response.} There are also some platforms that support incremental computing over streaming graph updates~\cite{zakian2019incrementalization,cheng2012kineograph}. We rather focus on \addc{materialized} property graphs with temporal lifespans on their vertices, edges and properties that have already been collected in the past. In future, we will also consider incremental query processing over such streaming graphs. %

\subsection{Property and Temporal Graph Querying}
Query models over property graphs and associated query engines are popular for semantic graphs~\cite{chavarria2016graql, zhou2013distributed, shi2016fast}. Languages like SPARQL offer a highly flexible declarative syntax, but are costly to execute in practice for large graphs~\cite{sparql,huang2011scalable}.
Others support a narrower set of declarative query primitives, such as finding paths, reachability and patterns over property graphs, but manage to scale to large graphs using a distributed execution model~\cite{jamadagni2016godb,Sarwat:2013:HDS}. However, none of these support time as a first-class entity, during query specification or execution.

There has been limited work on querying and indexing over specific temporal features of property graph.  Semertzidis, et al.~\cite{semertzidis2018top} propose a model for finding the top-k graph patterns which exist for the longest period of time over a series of graph snapshots. They offer several indexing techniques to minimize the snapshot search space, and perform a brute-force pattern mining on the restricted set. This multi-snapshot approach limits the pattern to one that fully exists at a single time-point and recurs across time, rather than spans time-points. It is also limited to a single-machine execution, which limits scaling.

\emph{TimeReach}~\cite{semertzidis2015timereach} supports conjunctive and disjunctive reachability queries on a series of temporal graph snapshots. It builds an index from strongly connected components (SCC) for each snapshot, condenses them across time, and use this to traverse between vertices in different SCCs within a single hop. It assumes that the graph has few SCCs that do not change much over time. They also require the path to be reachable within a single snapshot rather than allow path segments to connect across time. %
Likewise, \emph{TopChain}~\cite{wu2016reachability} supports temporal reachability query using an index labeling scheme. It unrolls the temporal graph into a static graph, with time expanded as additional edges, finds the chain-cover over it, and stores the top-k reachable chains from each vertex as labels. It uses this to answer time-respecting reachability, earliest arrival path and fastest path queries. Paths can span time intervals. However, they do not support any predicates over the properties. %
Neither of these support distributed execution.

\addc{There is also literature on approximate querying over graphs. \emph{Arrow}~\cite{arrow} examines reachability queries on both non-temporal and temporal graphs using random walks. These are performed from both the source and the sink vertices, and an intersection of the two vertex sets gives the result. They use approximation by bounding the walk length based on the diameter of the graph and a tunable parameter which balances accuracy and query latency. %
	Iyer, et al.~\cite{iyer2018asap} consider approximate pattern mining on large non-temporal graphs. They use statistical techniques to sample the graph edges and estimate the number of occurrences of a specific pattern in the graph. However, their approach cannot enumerate the actual vertices and edges forming the pattern.}%

\emph{ChronoGraph}~\cite{byun2019chronograph} supports temporal \addc{traversal} queries over interval property graphs, \addc{and is the closest to our work}. They implement this \addc{by extending} the Gremlin property graph query language \addc{with temporal properties}. They propose optimizations to the Gremlin traversal operators, %
and parallelization and lazy traversals within a single machine, \addc{which are executed by the TinkerGraph engine}. %
However, they do not support novel temporal operators such as the edge-temporal relationship that we introduce. %
They also do not use indexes or query planning to make the execution plan more efficient. Their optimizations are tightly-coupled to the execution engine, which does not support distributed execution.%

\addc{Lastly, there are several open-source and proprietary graph database systems~\cite{janusgraph,neo4j,orientdb} which provide general-purpose property graph storage and querying capabilities while allowing transactional access to graph data. However, these systems do not have first class support for time-varying graphs and query models that can leverage the temporal dimension. This leads to temporal queries written in their native language which are neither intuitive in expressing temporal notion nor efficient during execution due to lack of time-aware query optimizer and execution engine.} 

In summary, these various platforms lack one or more of the following capabilities we offer: modeling time as a first-class graph and query concept; enabling temporal path queries that span time and match temporal relations across entities; and distributed execution on commodity clusters that scales to large graphs using a query optimizer \addc{that leverages the graph's structure, temporal features, and property values}.

\section{Temporal Graph and Query Models}%
\label{sec:query}

\subsection{Temporal Concepts}
The temporal property graph concepts used in this paper are drawn from our earlier work~\cite{graphite}. Time is a linearly ordered discrete domain $\Omega$ whose range is the set of non-negative whole numbers. Each instant in this domain is called a \emph{time-point} and an atomic increment in time is called a \emph{time-unit}. %
A \emph{time interval} is given by $\tau = [t_s, t_e)$ where $t_s, t_e \in \Omega$ which indicates an interval starting from and including $t_s$ and extending to but excluding $t_e$.
\emph{Interval relations}~\cite{allen1983textordfemininemaintaining} are Boolean comparators between intervals; \emph{fully before} relation is denoted by $\ll$, \emph{starts before} relation by $\prec$, \emph{fully after} relation by $\gg$, \emph{starts after} relation by $\succ$, \addc{\emph{during} relation by $\subset$, \emph{equals} relation by $=$, \emph{during or equals} relation by $\subseteq$} and \emph{overlaps} relation by $\sqcap$. %

\subsection{Temporal Property Graph Model}
\label{sec:graph}

We formally define a temporal property graph as a directed graph $G = (V, E, P_V, P_E)$.
$V$ is a set of typed vertices where each vertex $\langle vid, \sigma, \tau \rangle\in V$ is a tuple with a unique vertex ID, $vid$, a vertex \emph{type} (or schema) $\sigma$, and the \emph{lifespan} of existence of the vertex given by the interval, $\tau = [t_s, t_e)$.
$E$ is a set of directed typed edges, with $\langle eid, \sigma, vid_i, vid_j, \tau \rangle \in E$. Here, $eid$ is a unique ID of the edge, $\sigma$ its type, $vid_i$ and $vid_j$ are its source and sink vertices respectively, and $\tau = [t_s, t_e)$ is its lifespan.
We have a schema function $\mathcal{S}: \sigma \rightarrow K$, that maps a given vertex or edge type $\sigma$ to the set of \emph{property keys} (or names) it can have.
$P_V$ is a set of \emph{vertex property values}, where each $\langle vid, \kappa, val, \tau_p \rangle \in P_V$ represents a value $val$ for the key $\kappa \in K$ for the vertex $vid$, with the value valid for the interval $\tau_p \subseteq \tau$. A similar definition applies for edge property values $\langle eid, \kappa, val, \tau_p \rangle \in P_E$.%
Further, the graph $G$ must meet the \emph{uniqueness} constraint of vertices and edges, \addc{i.e., a vertex or an edge with a given ID exist at most once and for a single continuous duration;} \emph{referential integrity} constraints, where \addc{the lifespan of an edge must be contained within the lifespan of its incident vertices}; and \emph{constant edge association}, \addc{which enforces that the vertices incident on an edge remain the same during the edge's lifespan}. These are defined in~\cite{moffitt2017temporal}.%

A \emph{static temporal property graph} is a restricted version of the temporal property graph such that $\tau_p = \tau$ for the vertex and edge properties, i.e., each property key has a static value that is valid for the entire vertex or edge lifespan, \addc{formally stated as:\\
	$\forall \langle vid, \kappa, val, \tau_p \rangle \in P_V, \langle vid, \sigma, \tau \rangle \in V \implies \tau_p = \tau \text{~~~and~~~} \forall \langle eid, \kappa, val, \tau_p \rangle \in P_E, \langle eid, \sigma, vid_i, vid_j, \tau \rangle \in E \implies \tau_p = \tau$%
}
\addc{Temporal property} graphs without this restriction are called \emph{dynamic temporal property graphs}, and allow keys for a vertex or an edge to have different values for non-overlapping time intervals, \addc{i.e., $\tau_p \subseteq \tau$}. E.g., Figure~\ref{fig:tpgraph} is a dynamic temporal property graph as \emph{Cleo}'s property values change over time, but 
omitting \emph{Cleo} makes it a static temporal property graph.

\subsection{Temporal Path Query} %
\label{sec:path}

An \emph{$n$-hop} linear chain path query matches a path with $n$ vertex predicates and $n-1$ edge predicates. %
The syntax rules for this query model and its predicates  %
are given below, \addc{and illustrated for the example queries from earlier in Table~\ref{tbl:queryModelExamples}.}

\setlength{\grammarindent}{10em}
\setlength{\grammarparsep}{0.2em}
\begin{grammar}

	\addcBegin <path> ::= <ve-fragment> <ve-int-fragment>* <v-predicate> \alt <ve-fragment> <ve-int-fragment>* <v-predicate> $\oplus$ <aggregate> \addcEnd
	
	\addcBegin <ve-fragment> ::= <v-predicate> $\vdash$ <e-predicate> \addcEnd
	
	\addcBegin <ve-int-fragment> ::= <ve-fragment> $~|~$  <v-predicate> <etr-clause> $\vdash$ <e-predicate> \addcEnd

	<v-predicate> ::= <predicate>
	
	<e-predicate> ::= <predicate> <direction>
	
	<direction> ::= $\rightarrow$ $|$ $\leftarrow$ $|$ $\leftrightarrow$

	<predicate> ::= $\star$ $|$ <bool-predicate> $|$ <prop-clause> $|$ <time-clause> \addcBegin $|$ <time-clause> \texttt{AND} <bool-predicate> \addcEnd
	
	<bool-predicate> ::= <prop-clause> $|$ <prop-clause> \texttt{OR} <bool-predicate>\alt <prop-clause> \texttt{AND} <bool-predicate> 

	<prop-clause> ::= \texttt{ve-key} <prop-compare> \texttt{value}

	<time-clause> ::= \texttt{ve-lifespan} <time-compare> \texttt{interval}

	<etr-clause> ::= \texttt{el-lifespan} <time-compare> \texttt{er-lifespan}
	
	<prop-compare> ::= `==' $|$ `!=' $|$ $\ni$
	
	<time-compare> ::=  $\prec$
	$|$ 
	$\ll$ 
	$|$ 
	$\succ$ 
	$|$ 
	$\gg$ 
	$|$ 
	$\sqcap$ 
	$|$ 
	$\not\sqcap$
	
	\addcBegin
	<aggregate>  ::= <aggregate-op>[ \texttt{v-key} | $\star$ ]
	
	<aggregate-op> ::=  \texttt{count} $|$ \texttt{min} $|$ \texttt{max}
	\addcEnd
\end{grammar}	

As we can see, the property and time \emph{clauses} are the atomic elements of the \emph{predicate} and allow one to \emph{compare} in/equality and containment between a property value and the given value, and a more flexible set of comparisons between a vertex/edge/property lifespan and a given interval (\emph{time-compare}). These temporal clauses allow a wide variety of comparison within the context of a single vertex or edge, and their properties. 
These clauses can be combined using Boolean \texttt{AND} and \texttt{OR} operators. %
Edge predicates can have an optional \emph{direction}. The wildcard $\star$ matches all vertices or edges at a hop. %

A novel and powerful temporal operator we introduce is \emph{edge time relationship (ETR)}. Unlike the time clause, this \emph{etr-clause} allows comparison across \addc{edge lifespans}. Specifically, it is defined on an intermediate vertex in the path (\emph{ve-int-fragment}), and allows us to compare the lifespans of its left (\emph{el-lifespan}) and right (\emph{er-lifespan}) edges in the path.
The motivation for this operator comes from social network mining~\cite{Fan:2012:GPM:2274576.2274578} and to identify flow and frauds in transactions networks~\cite{haslhofer2016bitcoin}.
E.g., \addc{the queries} EQ2 and EQ3 from Section~\ref{sec:intro} can be \addc{concisely} captured using this.

\addc{We also support a novel \emph{temporal aggregate} operator to group the result-set from the path query. The paths are grouped on the first vertex in the resulting temporal paths, and computes a specific aggregation on a property at the last vertex of the path. The grouping is time-aware; specifically, it is based on the duration of the first vertex in the result path. E.g., if the result-set for a query contains $i=1..m$ paths of length $n$ each, $v_{1}^{i}-e_{1}^{i}-v_{2}^{i}-e_{2}^{i}-...-v_{n}^{i}$, and the first vertex $v_1^{i}$ in a result matches the query for the time period $\tau_i=[t_s^i,t_e^i)$, then we perform a ``group by'' of the result paths by the temporal vertex $\{v_{1}^{i}.id, [t_s^i,t_e^i)\}$. For all the paths $j$ in a group, we perform an aggregation operation $\oplus$ on $v_{n}^{j}.prop$, where $prop$ is a property on the last vertex that is selected by the user and may be omitted for a \emph{count} aggregation. We return the aggregated result $\{v_{1}^{i}.vid, [t_s^i,t_e^i), \operatorname*{\oplus}\limits_{j}(v_{n}^{j}.prop) \}$ for each unique temporal vertex group~\footnote{The valid duration for the first vertex can be disjoint, in which case each maximal contiguous interval for that vertex $vid$ forms a separate temporal group.}. This can help answer queries such as \emph{``[EQ4] Count the number of persons followed by a person `Bob' during his existence in the network''}. The answer to this for Figure~\ref{fig:tpgraph} varies across time, taking value $1$ during $[10, 30) \cup [50, 100)$ and $0$ during $[5, 10) \cup [30, 50)$. Our \graphdb implementation supports \emph{count}, \emph{min} and \emph{max} operations for $\oplus$, while others can also easily be added.}%
\begin{table}[t]
	\addcBegin
	\small
	\setlength{\tabcolsep}{0.35em} %
	\centering
	\caption{\addc{Query Syntax Examples}}
	\begin{tabular}{m{0.37\textwidth}|m{0.59\textwidth}}
		\hline
		\bf Example Query & \bf Query Syntax \\
		\toprule
		\textbf{EQ1~} Find a person who lives in `UK' and follows a person who follows another person who is tagged with `Hiking' & \textbf{Type} == \emph{Person} \texttt{AND} \textbf{Country} == \emph{UK} $\vdash$ \textbf{Type} == \emph{Follows} $\rightarrow$ \newline \textbf{Type} == \emph{Person} $\vdash$ \textbf{Type} == \emph{Follows} $\rightarrow$ \newline  \textbf{Type} == \emph{Person} \texttt{AND} \textbf{Tag} $\ni$ \emph{Hiking}  \\ \hline
		\textbf{EQ2~} Find people tagged with `Hiking' who liked a post tagged as `Vacation' \underline{before} the post was liked by a person named `Don' & \textbf{Type} == \emph{Person} \texttt{AND} \textbf{Tag} $\ni$ \emph{Hiking} $\vdash$ \textbf{Type} == \emph{Likes} $\rightarrow$ \newline \textbf{Type} == \emph{Post} \texttt{AND} \textbf{Tag} $\ni$ \emph{Vacation} \texttt{el-lifespan} $\prec$ \texttt{er-lifespan} \newline $~~~~~~~~\vdash$ \textbf{Type} == \emph{Likes} $\leftarrow$ \newline  \textbf{Type} == \emph{Person} \texttt{AND} \textbf{Name} == \emph{Don}  \\ \hline
		\textbf{EQ3~} Find people who started to follow another person, after they stopped following `Don' & \textbf{Type} == \emph{Person} $\vdash$ \textbf{Type} == \emph{Follows} $\rightarrow$ \newline \textbf{Type} == \emph{Person} \texttt{el-lifespan} $\gg$ \texttt{er-lifespan}  $\vdash$ \textbf{Type} == \emph{Follows} $\rightarrow$ \newline  \textbf{Type} == \emph{Person} \texttt{AND} \textbf{Name} == \emph{Don}  \\
		\hline
		\textbf{EQ4~} Count the number of persons followed by a person `Bob' during his existence in the network. & \textbf{Type} == \emph{Person} \texttt{AND} \textbf{Name} == \emph{Bob} $\vdash$ \textbf{Type} == \emph{Follows} $\rightarrow$ \newline \textbf{Type} == \emph{Person} $\oplus$ \texttt{count} $[\star]$ \\
		\hline		
	\end{tabular}
	\label{tbl:queryModelExamples}
	\addcEnd
\end{table}

\section{Distributed Query Engine}%
\label{sec:engine}

\subsection{Relaxed Interval Centric Computing}
\begin{figure}[!t]
	\centering
	\subfloat[Architecture of \graphdb]{
		\includegraphics[width=0.45\columnwidth]{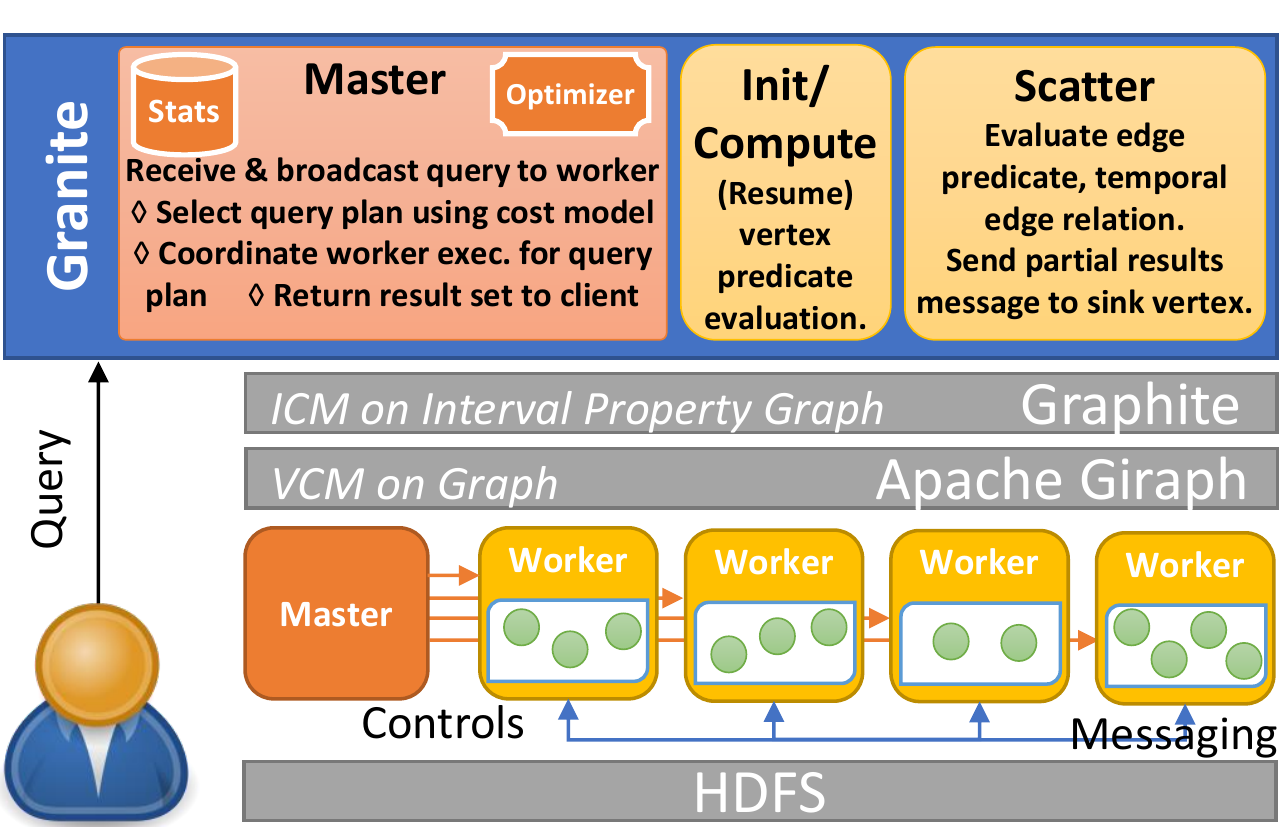}
		\label{fig:arch}
	}\qquad\qquad
	\subfloat[Iterative query execution across ICM supersteps]{
		\includegraphics[width=0.35\columnwidth]{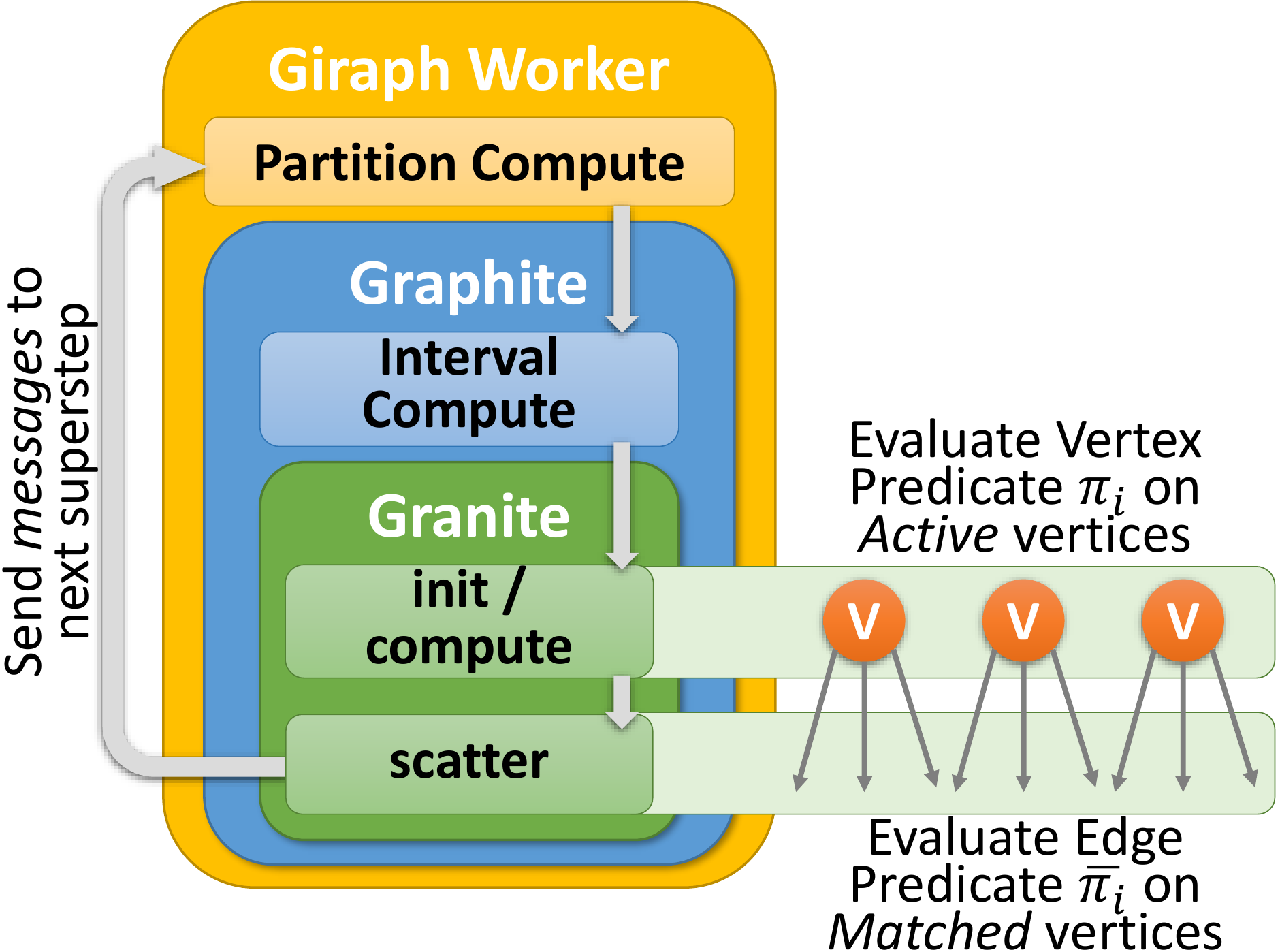}
		\label{fig:scheme}
	}
	\caption{Architecture and ICM execution model of \graphdb}
\end{figure}

\addc{The high-level architecture of our distributed query engine, \graphdb, is shown in Figure~\ref{fig:arch}.} Our query engine uses a distributed in-memory iterative execution model that extends and relaxes the \emph{Interval-centric Computing Model (ICM)}~\cite{graphite}. ICM adds a temporal dimension to Pregel's vertex-centric iterative computing model~\cite{malewicz2010pregel}, \addc{and allows} users to define their computation from the perspective of a single interval-vertex, i.e., the state and properties for a certain interval of a vertex's lifespan. In each iteration (\emph{superstep}) of an ICM application, a user-defined \texttt{compute} function is called on each active interval-vertex, which operates on its prior state and on messages it receives from its neighbors, for that interval, and updates the current state. \addc{A \emph{TimeWarp} function aligns the lifespans of the input messages to the lifespans of the partitioned interval states for an interval vertex. So each call to \texttt{compute} executes on the temporally intersecting messages and states for a vertex.}
Then, a user-defined \texttt{scatter} function is called \addc{on the out-edges of} that interval-vertex, which allows them to send temporal messages containing, say, the updated vertex state to its neighboring \addc{vertices}. The message lifespan is usually the intersection of the state and the edge lifespans.

Messages are delivered in bulk at a barrier after the \texttt{scatter} phase, and the \texttt{compute} phase for the next iteration starts after that. Vertices receiving a message whose interval overlaps with its lifespan are activated for the overlapping period. This repeats across supersteps until no messages are generated after a superstep.
\addc{The execution of \texttt{compute} and \texttt{scatter} functions are each data-parallel within a superstep, and their invocation on different interval vertices and edges can be done by concurrent threads.}

We design \graphdb using the \texttt{compute} and \texttt{scatter} primitives offered the \emph{Graphite} implementation of ICM over Apache Giraph, \addc{as illustrated in Figure~\ref{fig:scheme}}. However, ICM enforces \emph{time-respecting behavior}, i.e., the intervals between the messages and the interval-vertex state have to overlap for \texttt{compute} to be called on the messages; intervals between the states updated by the \texttt{compute} and the edge lifespans have to overlap for \texttt{scatter} to be called; and \texttt{scatter} sends messages only on edges whose lifespan overlaps with the updated states. 

But the temporal path queries do not need to meet these requirements, e.g., a query may need to navigate from a vertex to an adjacent vertex that occurs \emph{after} it. The TimeWarp operator of ICM enforces this time-respecting behavior. %
So we relax ICM to allow non-time respecting behavior between \texttt{compute}, \texttt{scatter} and messages \addc{to meet the execution requirements of our path queries}, while leveraging its other interval-centric features.

\subsection{Distributed Execution Model}

\begin{figure}[t]
	\centering        
	\subfloat[\addc{Query execution phases across different supersteps (SS) for two plans of the input query}]{
		\includegraphics[width=0.6\columnwidth]{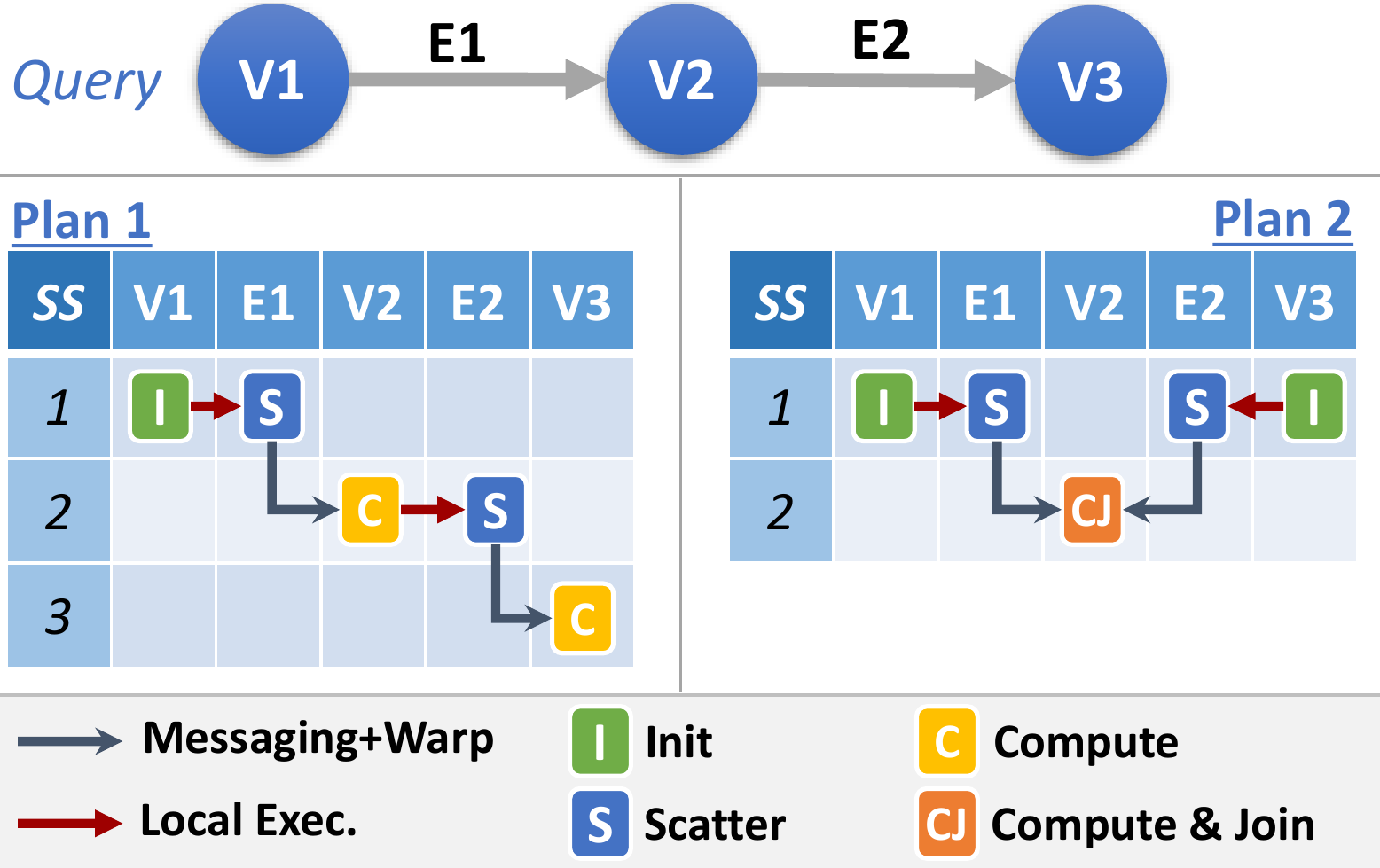}
		\label{fig:exec}
	}\qquad
	\subfloat[\addc{Splitting of query and Joining of results for \emph{EQ2}, similar to Plan 2 in Figure~\ref{fig:exec}}]{
		\includegraphics[width=0.6\columnwidth]{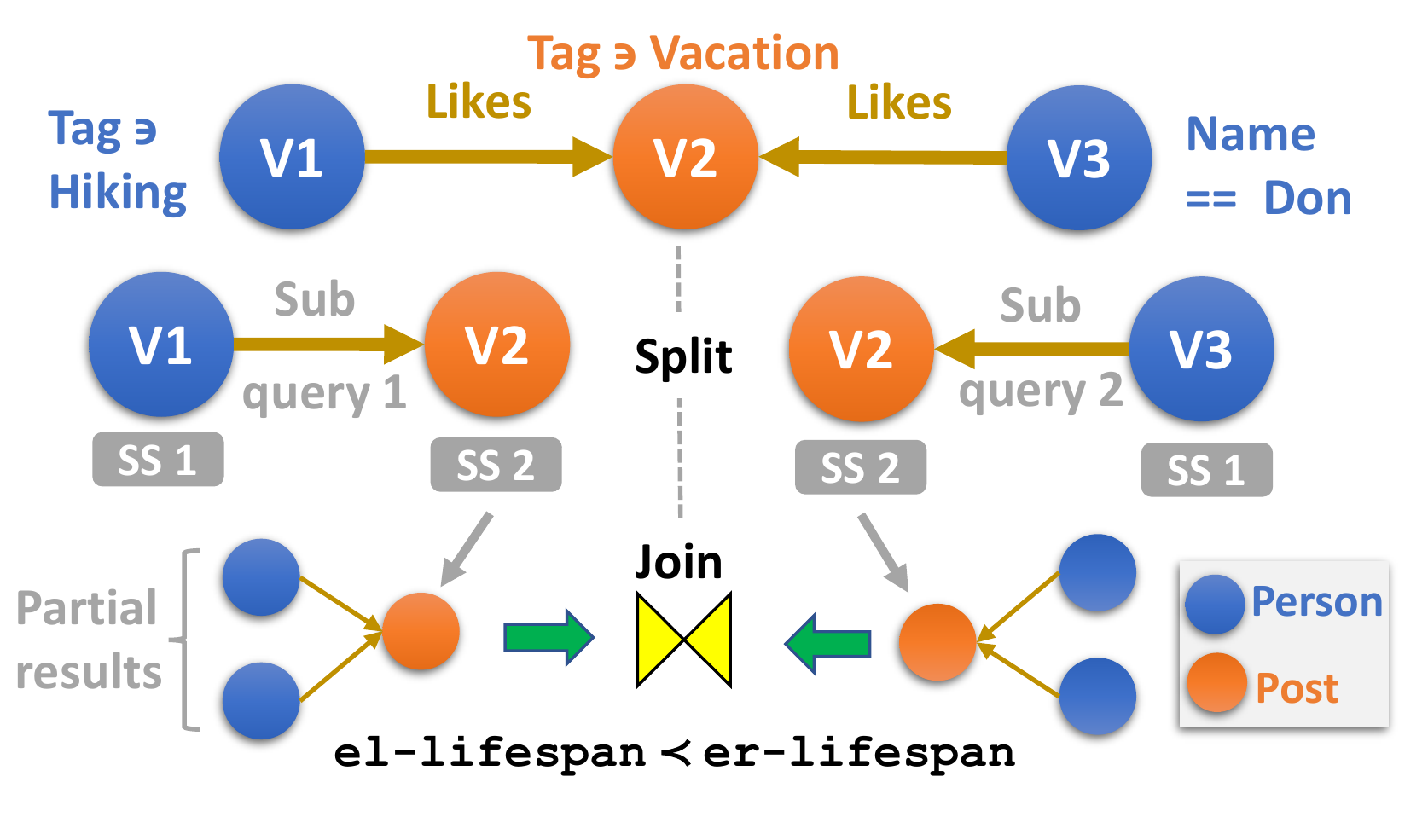}
		\label{fig:join}
	}
	\caption{Query execution plans in \graphdb}\label{fig:exec:scheme}
\end{figure}

In our execution model, each vertex predicate for a path query and the succeeding edge predicate, if any, are evaluated in a single ICM superstep. Specifically, the vertex predicates are evaluated in the \texttt{compute} function and the edge predicates in the \texttt{scatter} function. We use a specialized logic called \texttt{init} for \addc{evaluating} the first vertex predicate in a query. This is shown in Figs.~\ref{fig:scheme} and~\ref{fig:exec}.
A \emph{Master} receives the path query from the client, and broadcasts it to all \emph{Workers} to start the first superstep (Figure~\ref{fig:arch}). Each Worker operates over a set of graph partitions with one thread per partition, and each thread calls the \texttt{compute} and \texttt{scatter} functions on every active vertex in its partition. 
The \texttt{init} logic is called on all vertices in the first superstep. It resets the \emph{vertex state} for this new query and evaluates the first vertex predicate of the query. %
If the vertex matches, its state is updated with a \emph{matched} flag and \texttt{scatter} is invoked for each of its incident in or out edges, \addc{as defined in the query}. \texttt{Scatter} evaluates the next edge predicate, %
and if it matches, sends the partial path result and the evaluated path length to the destination vertex as a message. If a match fails, this path traversal is pruned.%

In the next iteration, our \texttt{compute} logic is called for vertices receiving a message. This evaluates the next vertex predicate in the path and if it matches, it puts all the partial path results from the input messages in the vertex state, and \texttt{scatter} is called on each incident edge. If the edge matches the next edge predicate, the current vertex and edge are appended to each prior partial result and sent to the destination vertex. This repeats for as many supersteps as the path length. %
In the last superstep, the vertices receiving matching paths in their messages send it to the Master to return to the client.

\addc{Figure~\ref{fig:exec} (Plan 1) illustrates this for a sample path query with vertex and edge predicates, $V1-E1-V2-E2-V3$. In superstep 1, \texttt{init} is called on all vertices to evaluate the vertex predicate $V1$, and for the ones that match, \texttt{scatter} is called to evaluate the edge predicate $E1$. Those edges that match send a message to their remote vertex in superstep 2, where all vertices that receive a message invoke their \texttt{compute} logic to evaluate the vertex predicate $V2$ of the second hop. This is (optionally) preceded by the TimeWarp operator on all the messages received by an interval vertex. Vertices that match $V2$ call \texttt{scatter} on their edges to match the predicate $E2$, and send messages if they too match. In the last superstep, vertices that receive messages evaluate the predicate $V3$, and if there is a match, return that result path to the user. Each vertex in the last superstep may return multiple matching paths based on the messages received, and different vertices may return result paths to the Master.}

\texttt{Scatter} also evaluates the \emph{edge temporal relationship}. Here, the \texttt{scatter} of the preceding edge passes its lifespan in the result message, and this is compared against the current edge's lifespan by the next \texttt{scatter} to decide on a match.  %
\addc{In the case of \emph{temporal aggregate queries}, the result set is constructed in the last superstep, similar to the non-aggregate queries. Then, the first vertex in each result path, its associated lifespan, and the count or the property value of the last vertex to be aggregated are extracted and sent to the Master. The Master temporally groups the values for each distinct temporal vertex using the \emph{TimeWarp} operator, and applies the aggregation operator on the values in each group.} %
For \emph{static temporal graphs}, we do not use any interval-centric features of ICM, \addc{such as TimeWarp}, and the entire lifespan of the vertex is treated as a single interval-vertex for execution, and likewise for edges. %
However, we do use the property graph model and state management APIs offered by the interval-vertex.

For \emph{dynamic temporal graphs} with time varying properties, %
we leverage the interval-centric features of ICM. Specifically, we enable TimeWarp of message intervals with the vertex properties' lifespans so that \texttt{compute} is called on an interval vertex with messages temporally aligned and grouped against the property intervals. %
\texttt{Scatter} is called only for edges whose lifespans overlap with the matching interval-vertex, and its scope is limited to the period of overlap. 
The \texttt{compute} or \texttt{scatter} functions only access messages and properties that are relevant to their current interval, and both can be called multiple times, for different intervals, on the same vertex and edge.

\subsection{Distributed Query Execution Plans}
Queries can be evaluated by splitting them into smaller \emph{path query segments} that are independently evaluated \emph{left-to-right}, and the results then combined. Each vertex predicate in the path query is a potential \emph{split point}. E.g., a query $V1-E1-V2-E2-V3$ can be split at $V2$ into the segments: $V1-E1-V2$ and $V3-E2-V2$; %
execution proceeds \emph{inwards}, from the outer predicates \addc{($V1$ and $V3$)} to the split vertex \addc{($V2$)} which \emph{joins} the results. \addc{This is illustrated in Figs.~\ref{fig:exec}~(Plan 2) and~\ref{fig:join}.} %
A trivial split at \addc{the last vertex predicate} $V3$ is the default execution \addc{of the query from left-to-right, shown in Figure~\ref{fig:exec}~(Plan 1)}, while an alternative split at \addc{the first vertex predicate} $V1$ evaluates this from right-to-left as $V3-E2-V2-E1-V1$. %

Each split point and plan can be beneficial based on how many vertices and edges match the predicates in the graph. Intuitively, a good plan should evaluate the most discriminating predicate first (low selectivity, few vertex/edges match) to reduce the solution space early. A \emph{cost model}, discussed in Section~\ref{sec:cost}, attempts to select the best split point. %

We modify our \graphdb logic to handle the execution of two path segments concurrently. \addc{E.g.,} for a split point $V2$, in the first superstep, we evaluate predicates $V1-E1$ and $V3-E2$ in the same \texttt{compute/init} and \texttt{scatter} logic, while in the second superstep we evaluate predicate $V2$, \addc{as shown in Figure~\ref{fig:exec}~(Plan 2)}. In the \addc{final} superstep when results from both the segments are available, we do a nested loop \emph{join} to get the cross-product of the results. This can be extended to multiple split points in the future.

\subsection{System Optimizations}

\subsubsection{Type-based Graph Partitioning}
Giraph by default does a hash-partitioning of the vertices of the graph by their vertex IDs onto workers. But we use knowledge of the entity schema types to create graph partitions hosting only a single vertex type. This helps us eliminate the evaluation of all vertices in a partition if its type does not match the vertex type specified in that hop of the query. This filtering is done before the \texttt{compute} is called, at the \texttt{partitionCompute} of Giraph. 

We first group vertices by \emph{type} to form a \emph{typed partition} each, \addc{e.g., \emph{Type A} and \emph{Type B}, as illustrated in Figure~\ref{fig:parti}}. But these can have skewed sizes, and there may be too few types (hence partitions) to fully exploit the parallelism available on the workers and their threads. So we further perform a second-level \emph{topological partitioning} of  %
each typed partition into $p$ sub-partitions  using \emph{METIS}~\cite{karypis1998fast}. This only considers the edges between vertices of the same type, i.e., within each typed partition, and uses the edge lifespan as their weight. This second-level partitioning can also reduce the network messaging cost between vertices of the same type. The sub-partitions from each typed partition are then distributed in a round-robin manner among all the workers. \addc{So if there are $w$ workers, $t$ types and $p$ sub-partition per type, each worker is expected to have $\frac{t \times p}{w}$ sub-partitions, with $\frac{p}{w}$ of each type. Since each superstep typically evaluates a query predicate for a single vertex type, this ensures load balancing of the typed sub-partitions across all workers during a superstep execution.}%

\addc{In our experiments using the 100k:A-S graph, described in Section~\ref{sec:workload}, we observe that using type-based partitioning at the first level instead of hash partitioning improves the average execution time for our query workloads by {$5.8\times$}. When we combine this with METIS partitioning in the second level, we see a further improvement of {32\%}. All our results we later report use this optimization.}%

\subsubsection{Message Optimization}
Path results can have a lot of overlaps. But each partial result path is separately maintained and sent in messages during query execution. This redundancy leads to large message sizes and more memory. Instead, we construct a \emph{result tree}, where vertices/edges that match at a previous hop are higher up in the tree and subsequent vertex/edge matches are its descendants. %
\addc{E.g., assuming a full binary tree expansion for a path query with $h$ hops and $n = 2^{h-1}$ matching paths, this reduces the result size from $\mathcal{O}(h \times n)$ to $\mathcal{O}(2n-1)$.} %
When execution completes, a traversal of this result tree gives the expanded result paths.  %

\begin{figure}[t]
	\centering
	\subfloat[\addc{Two-level load-balanced partitioning of Input Graph to Workers, by type and then by topology}]{
		\includegraphics[width=0.5\columnwidth]{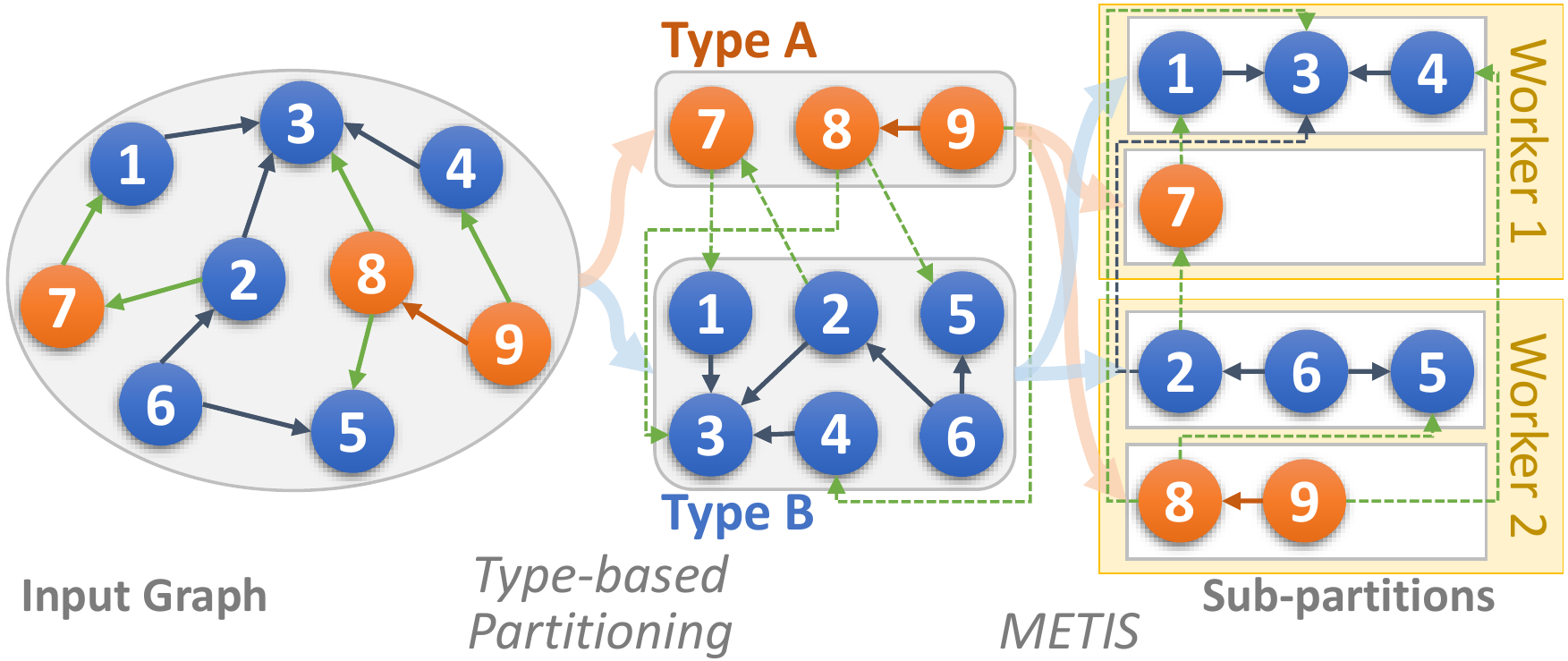}
		\label{fig:parti}
	}~~~~~~~~
	\subfloat[\addc{Message tree propagation during query evaluation}]{
		\includegraphics[width=0.41\columnwidth]{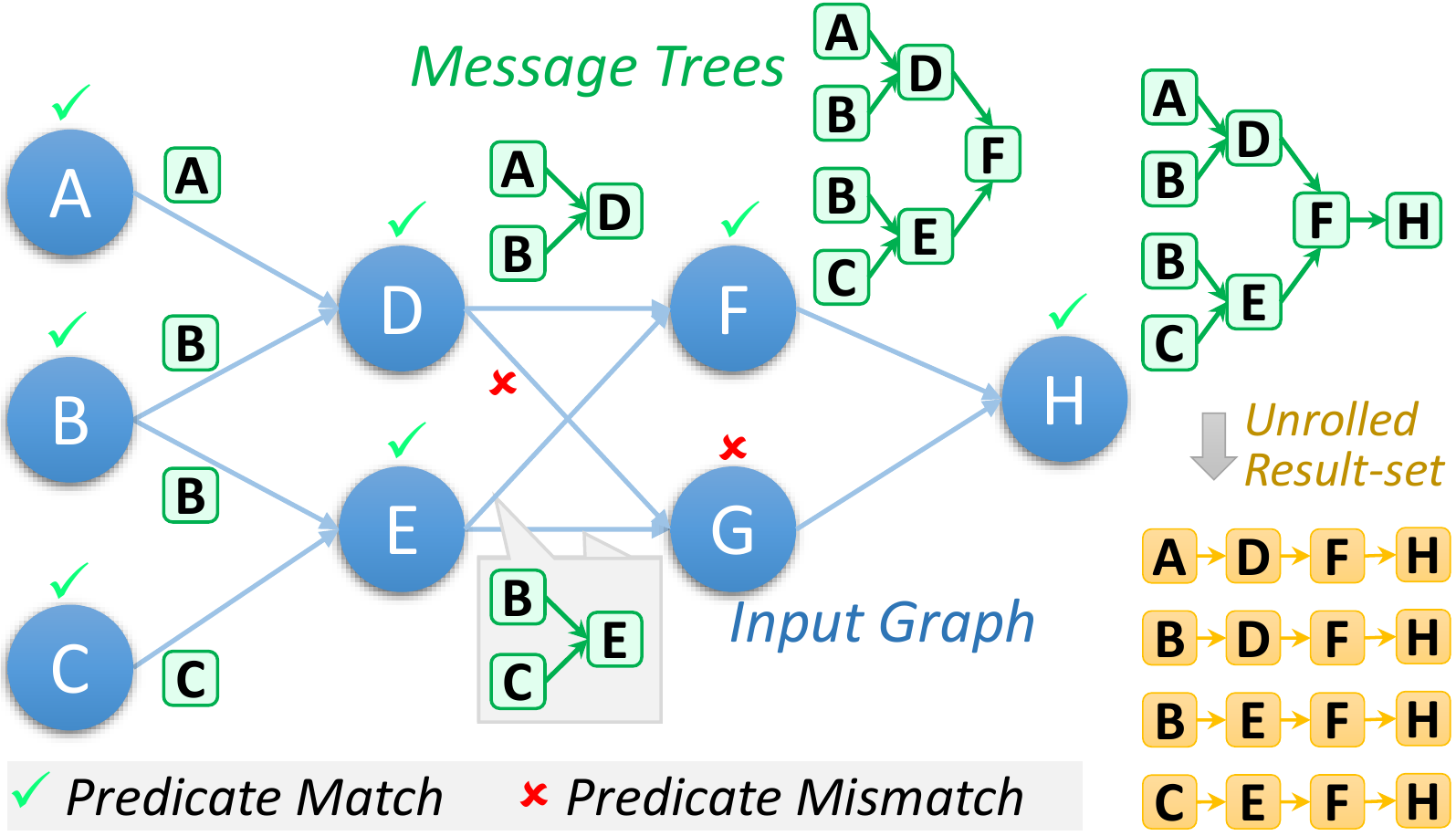}
		\label{fig:mesg}
	}
	\caption{\addc{Examples of system optimizations in \graphdb}}
\end{figure}

\addc{This is illustrated in Figure~\ref{fig:mesg}. Here, vertices $A$, $B$ and $C$ match the vertex and edge predicates in the first hop and send their partial result to their neighbors. $D$ receives the messages from $A$ and $B$ and evaluates itself for the second-hop predicate. But this execution is not unique to $A$ or $B$, but rather shared across them. If $D$ matches, rather than send a message with two sub-paths, $A-D$ and $B-D$, we instead send a sub-tree, $\{A,B\}-D$ in the message, to its neighbors. Similarly, $E$ which receives messages fro $B$ and $C$ and matches for the second predicate sends a sub-tree $\{B,C\}-E$ message. $F$ receives two sub-trees as messages, evaluates itself for the third predicate that matches, and sends a larger sub-tree, $\{\{A,B\}-D\},\{\{B,C\}-E\}-F$, to its neighbor $H$. $G$ is not a match and prunes its traversal, with no messages sent. $H$ matches the last predicate successfully, and sends the final result-tree with $H$ as the root to the Master, which unrolls the tree to return the paths from $H$ to every leaf as individual results to the client.}

\subsubsection{Memory Optimizations} 
In our graph data model, all property keys and values, excluding time intervals, are strings. In Java, string objects are memory-heavy. Since keys will often repeat for different vertices in the same JVM, we map every property key to a byte, %
and rewrite the query at the Master based on this mapping. Further, for property values that repeat, such as country, we use \emph{interning} in Java that replaces individual string objects with shared string objects. %
This works as the graph is read-only. Besides reducing \addc{the base memory usage for the graph by $\approx 5\%$, it} also allows predicate comparisons based on pointer equivalence. %

\section{Query Planning and Optimization}%
\label{sec:cost}

A given path query can be executed using different distributed execution plans, each having a different execution time. The goal of \addc{the} \emph{cost model} is to quickly estimate the expected execution time of these plans and pick the optimal plan for execution. Rather than absolute accuracy of the query execution time, what matters is its ability to \emph{distinguish poor plans} with high \addc{execution} times \emph{from good plans} with low \addc{execution} times.

We propose an analytical cost model that uses statistics about the temporal property graph, combined with estimates about the time spent in different stages of the distributed execution plan, to estimate the execution time for the different plans of a given query. We first \emph{enumerate} the possible plans, contributed by each split point in the path query. The graph statistics are then used to \emph{predict} the number of vertices and edges that will be active at each superstep of query execution, and the number of vertices that will match the predicates in this superstep and \addc{activate} the next \addc{hop of the query (superstep)}. %
Based on the number of active and matched vertices and edges, our cost model will \emph{estimate} the runtime for each superstep of the plan. Adding these up \addc{across supersteps} returns the estimated execution time for a plan. Next, we discuss the statistics that we maintain, and the models to predict the vertex and edge counts, and the execution time.

\subsection{Graph Statistics}

We maintain statistics about the temporal property graph to help estimate the vertices and edges matching a specific query predicate. Typically, relational databases maintain statistics on the frequency of tuples matching different value ranges, for a given column (property). A unique challenge \addc{for us} is that the property values can be time variant. Hence, for each property key present \addc{in the} vertex and edge types, we maintain a \emph{2D histogram}, where the \addc{Y} axis indicates \addc{the} different value ranges for the property and the \addc{X} axis \addc{the} different time ranges. Each entry in the histogram has a count of vertices or edges that fall within that value range for that time range. 

\begin{figure}[t]
		\centering
	\begin{minipage}{.62\textwidth}
		\centering%
		\subfloat[\addc{2-D Histogram of Statistics}]{
			\includegraphics[width=0.5\textwidth]{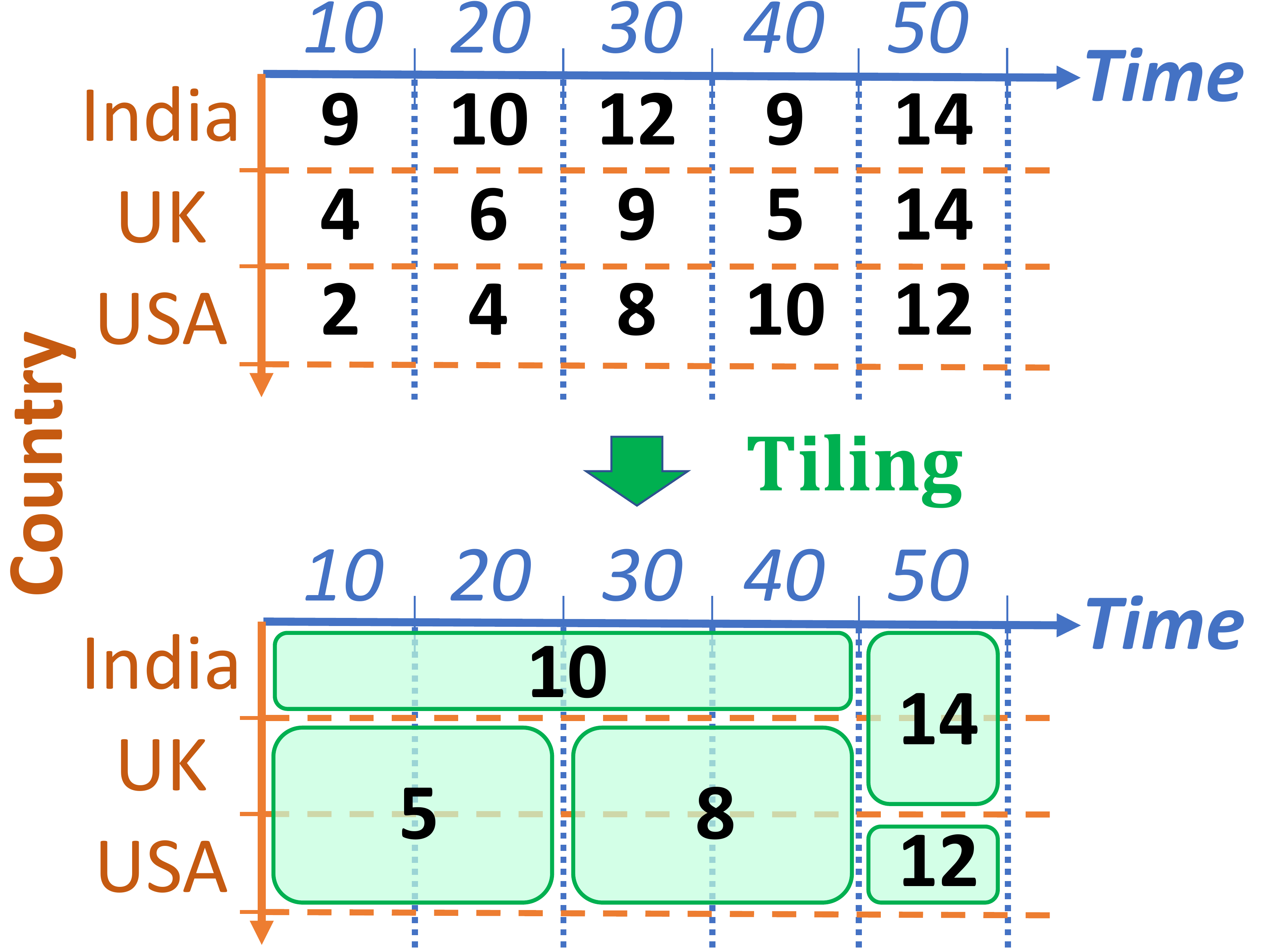}
			\label{fig:histogram}
		}~~
		\subfloat[\addc{Interval Tree for Statistics}]{
			\includegraphics[width=0.5\textwidth]{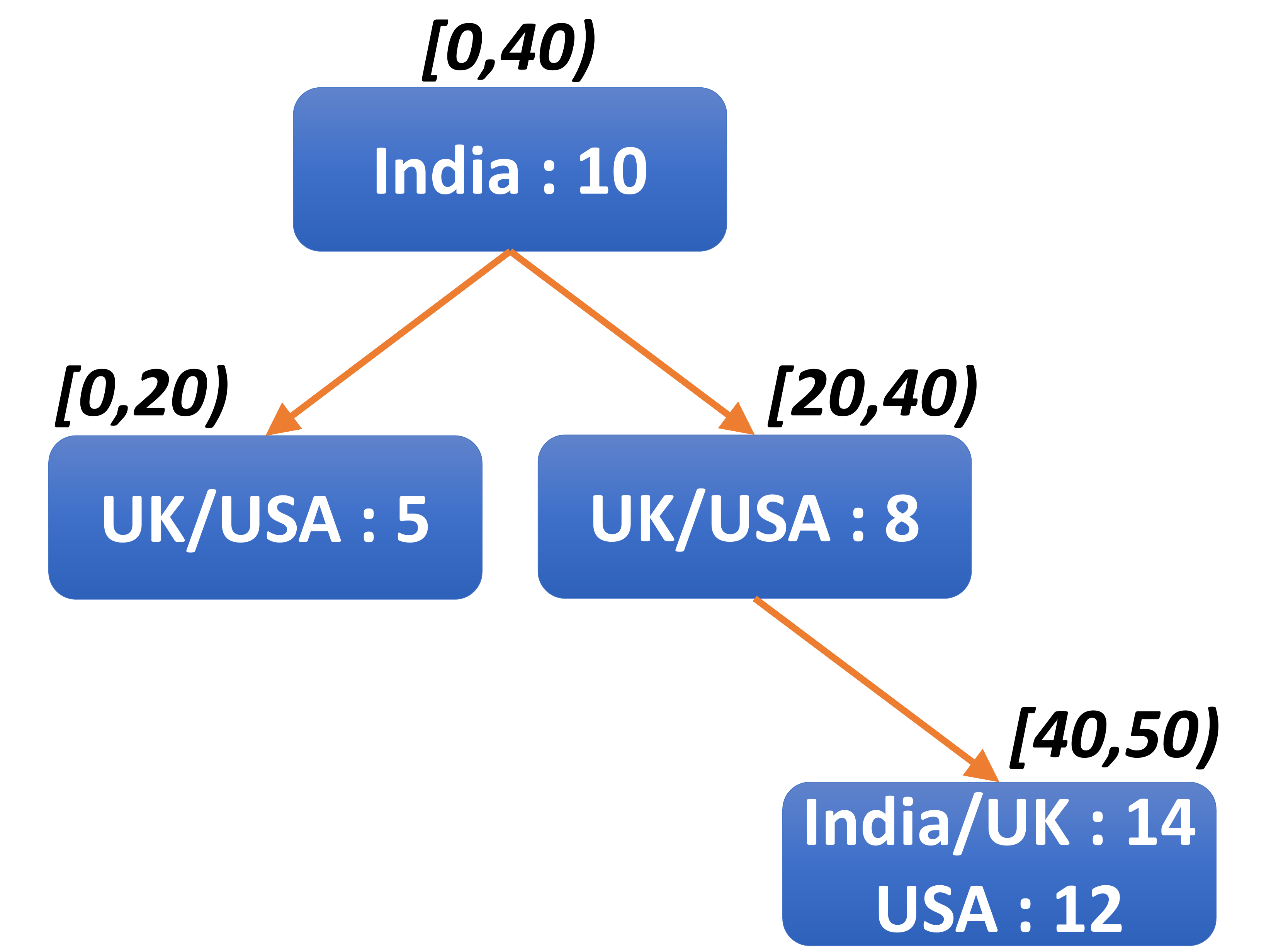}
			\label{fig:tree}
		}
		\caption{\addc{Query planning}}%
		\label{fig:statistics}
	\end{minipage}%
	~\\
	\begin{minipage}{.9\textwidth}
		\centering
		\addcBegin
		\captionof{table}{\addc{Vertex and edge \emph{count estimates} per superstep and \emph{execution time} calculated by the model for two execution plans, for query $EQ2$ on 100k:F-S graph}}
		\label{tbl:cm_example}
		\small
		\setlength\tabcolsep{3.8pt} %
		\begin{tabular}{r|r||rrr|rrr|r}
			\hline
			\textbf{Plan} & \textbf{SS} & $a_i$ & $f_i$ & $m_i$ & $\overline{a_i}$ & $\overline{f_i}$ & $\overline{m_i}$ & $T_i$ (ms) \\
			\toprule
			1 & 1 & $100k$ & $3.7\times10^{-2}$ & $3.7k$ & $6.2M$ & $35M$ & $1.3M$ & $531$ \\
			\cline{2-9}
			& 2 & $1.3M$ & $7.7\times 10^{-4}$ & $1k$ & -- & -- & -- & $132$\\
			\midrule\midrule
			2 & 1 & $51M$ & $7.7\times10^{-4}$ & $39k$ & $273k$ & $88M$ & $67k$ & $4147$\\
			\cline{2-9}
			& 2 & $67k$ & $3.7\times10^{-2}$ & $2.5k$ & -- & -- & -- & $35$\\
			\toprule
		\end{tabular}
		\addcEnd
	\end{minipage}
\end{figure}

\addc{E.g., Figure~\ref{fig:histogram}(top) shows such a histogram for the \emph{Country} property. Its Y axis lists different country values appearing in the vertices of the property graph, such as \emph{India, UK} and \emph{US}. The X axis divides the lifespan of the graph into time intervals, say, $[0,50)$ in steps of $10$. The cell values indicate the number of vertices that have these property values for those time intervals, in the entire graph. Here, $9$ vertices have the \emph{Country} property value as \emph{India} during time interval $[0,10)$ and $10$ vertices have it during $[10,20)$, and similarly for other countries and time intervals.}

Formally, for a given property key $\kappa$, we define a \emph{histogram function} $\mathcal{H}_\kappa:(val,\tau) \to \langle f, \delta_{in}, \delta_{out} \rangle$, that returns an \emph{estimate} of the frequency $f$ of vertices or edges which have the property value \emph{val} during a time interval $\tau$, and the \addc{average} in and out degrees $\delta$ of the matching vertices, which are maintained \addc{for a vertex property}.%

The \emph{granularity} of the value and time ranges has an impact on the size of the statistics maintained and the accuracy of the estimated frequencies. We make several \emph{optimizations} in this regard.
We use Dynamic Programming (DP) to coarsen the ranges of the histogram along both axes to form a \emph{hierarchical tiling}~\cite{muthukrishnan1999rectangular}. This ensures that the frequency variance among the individual \emph{value--time} pairs in each tile is no more than a threshold. \addc{For example, in Figure~\ref{fig:histogram}(bottom), the frequencies $9, 10, 12$ and $9$ for the property value \emph{India} during the interval $[10,40)$ are close to each other and hence tiled, i.e., aggregated and replaced by their average value $10$. Similarly, \emph{India} and \emph{UK} have the same frequency $14$ for the interval $[40,50)$ and are tiled. This reduces the number of entries that are maintained in the histogram, i.e., the space complexity, while bounding its impact on the accuracy of the statistics.}%

For important properties like vertex and edge types, out-degree and in-degree, we \emph{pre-coarsen the time steps} into, \addc{say,} weeks, and for other properties into, \addc{say,} months to reduce the size of the histogram -- \addc{the actual coarsening factor is} decided based on how often the properties change in the graph. %
For properties with 1000's of \emph{enumerated values} (e.g., \emph{Tag} in Figure~\ref{fig:tpgraph}), we sort them based on their frequency, cluster them into similar frequencies, and perform tiling on these clusters. %
We retain a map between property values and clusters for these, \addc{which is used to rewrite the input query to replace the property values with these cluster IDs instead.}
We use an \emph{interval tree} to maintain each histogram, with each tile inserted into this tree based on its time range. The nodes of the tree will have a set of tiles (property value ranges and their frequencies) that fall within its time interval. \addc{The invariant for all the nodes in the tree is such that the interval of a parent node will be after the left child (i.e., start time and/or end time of parent's interval is after the left child's interval), and before the right child. %
	E.g., the interval tree in Figure~\ref{fig:tree} is constructed from the 2D histogram in Figure~\ref{fig:histogram}(bottom). Every tile in the histogram becomes a node or part of a node in the interval tree. %
	We insert a tile in the right subtree if its interval is greater than the parent node's interval, in the left sub-tree if it is lesser, and in the parent if it overlaps with it. To perform a lookup, we check if the lookup interval is greater than or less than the parent interval and prune the search space accordingly, similar to a binary search tree.} %
Calling the $\mathcal{H}$ function performs a lookup in this interval tree, and matches within the set of property ranges. %

\addc{The time complexity to construct each interval tree includes the time to \emph{aggregate} the statistics from the graph, taking $\mathcal{O}(n\cdot k)$, where $n$ is the number of vertices in the graph and $k$ the average number of property names or keys per vertex type.} For each property key, the time \addc{taken} is dominated by the \emph{tiling} step that uses DP, and takes $\mathcal{O}(p^3t^3)$, where $p$ is the number of (clustered) values for the property key, and $t$ the number of (coarsened) time units they span~\cite{muthukrishnan1999rectangular}. 
\addc{The cost of building the interval tree is $\mathcal{O}(m \cdot t)$, where $m$ is the number of tiles in the coarsened histogram.} %
The lookup time is $\mathcal{O}(p\cdot t)$ in the worst case; \addc{for a balanced tree the expected lookup time is $\mathcal{O}(log\:m + k)$, where $k$ is the number of intersecting intervals in the tree.} %

The raw size of the statistics for the graphs used in our experiments ranges from $4200$--$5600~kB$ for about $13$--$15$ property keys.  %

\subsection{Estimating the Active and Matching Vertex and Edge Counts}
\label{sec:cost:counts}
A query plan contains either one or two path query segments. The query predicates on each vertex and its edges in the segment are evaluated in a single superstep. If two path segments are present, their results are \emph{joined} at the \addc{split point}. %
\addc{Aggregation operators, if any, are also evaluated in the last superstep.}
For each segment, we estimate a count of active and matching vertices and edges in each superstep, given by the recurrence relation discussed next.

Let $P = [ \pi_1, \overline{\pi_1},..., \pi_{n} ]$ denote the sequence of $n$ \emph{vertex predicates}, $\pi$, and $n-1$ \emph{edge predicates}, $\overline\pi$, for a given path query segment. Each predicate $\pi$ has a set of \emph{property clauses} $\mathcal{C}_P(\pi)=\{\langle \kappa, val \rangle \}$ and a \emph{temporal clause} $\mathcal{C}_T(\pi)=\langle lifespan, \tau \rangle$,
where $\kappa$ is a property key, $val$ is a value to compare its value against, and $\tau$ is the interval to compare that vertex/edge/property's lifespan against; \addc{and similarly for $\overline{\pi}$}. %
These clauses themselves can be combined using \texttt{AND} and \texttt{OR} Boolean operators, \addc{as described in the query syntax earlier}.

Let $\sigma_i$ ($\overline{\sigma_i}$) denote the \emph{type} of the vertex (edge) enforced by a clause of predicate $\pi_i$ ($\overline{\pi_i}$). %
Let $V_\sigma$ ($E_{\overline{\sigma}}$) denote the set of vertices (edges) of that type; if the vertex (edge) type is not specified in the predicate, these sets degenerate to all vertices (edges) in the graph.

As shown in Figure~\ref{fig:scheme}, each superstep is decomposed into 2 stages: calling \texttt{init} or \texttt{compute} on the \emph{active vertices} to find the vertices \emph{matching} the vertex predicate, and calling \texttt{scatter} on the active edges (i.e., in or out edges of the matching vertices) to identify the edges matching the edge predicates. These in turn help identify the active vertices for the next superstep of execution. Initially, all vertices of the graph are active, but if a type is specified in the starting vertex predicate, we can use the type-based partitioning to limit the active vertices to the ones having that \addc{vertex} type.

Let $a_i$ and $m_i$ denote the \emph{number of active and matched vertices}, \addc{respectively,} for vertex predicate $\pi_i$ with type $\sigma_i$, and $\overline{a_i}$ and $\overline{m_i}$ denote the \emph{number of active and matched edges}, respectively, for the edge predicate $\overline{\pi_i}$ with type $\overline{\sigma_i}$. These can be recursively defined as:

\begin{align}
a_i &~~=~~ \left\{
\begin{array}{ll}
|V_\sigma| & \text{, if }i=1 \\
\min(\overline{m_{i-1}},~|V_\sigma|) & \text{, otherwise}
\end{array}
\right. \label{eqn:cm:1}\\
\langle f_i, \delta^i_{in}, \delta^i_{out} \rangle &~~=~~ \bigotimes\limits_{\substack{\langle \kappa, val \rangle \in \mathcal{C}_P(\pi_i)\\\langle lifespan, \tau \rangle \in \mathcal{C}_T(\pi_i)}} \mathcal{H}_\kappa(val, \tau)  \nonumber \\
m_i &~~=~~ a_i \times \frac{f_i}{|V_\sigma|} \label{eqn:cm:2}\\
\overline{a_i} &~~=~~ m_i^\sigma \times (\delta^i_{in} + \delta^i_{out}) \label{eqn:cm:3}\\
\langle \overline{f_i}, -, - \rangle &~~=~~ \bigotimes\limits_{\substack{\langle \kappa, val \rangle \in \mathcal{C}_P(\overline{\pi_i})\\\langle lifespan, \tau \rangle \in \mathcal{C}_T(\overline{\pi_i})}} \mathcal{H}_\kappa(val, \tau) \nonumber\\
\overline{m_i} &~~=~~ \overline{a_i} \times \frac{\overline{f_i}} %
{|V_\sigma| \times (\bar{\delta}^{\sigma}_{in} + \bar{\delta}^{\sigma}_{out})} \label{eqn:cm:4}
\end{align}

In Equation~\ref{eqn:cm:1}, we set the active vertex count in the first superstep to be equal to the number of vertices of type $\sigma$. This reflects the localization of the search space in the \texttt{init} function to only vertices in the partitions matching that vertex type. For subsequent supersteps, the active vertex search space is upper-bounded by $|V_\sigma|$ but is usually expected to be the number of matching edges in the previous superstep~\footnote{This is in the worst case, \addc{if vertices and edges that match in the preceding hop activate mutually exclusive vertices in the next hop.}}, which would send a message to activate these vertices and call its \texttt{compute} function.

Next, in Equation~\ref{eqn:cm:2}, we use the graph statistics to find the fraction of vertices $\frac{f_i}{|V_\sigma|}$ that match the vertex predicate $\pi_i$ (also called \emph{selectivity}) and multiply this with the number of active vertices to estimate the matched vertices. This is the expected matched output count from \texttt{init} or \texttt{compute}. We use  $\mathcal{H}$ to find the selectivity by iterating through all clauses of a predicate $\pi_i$, get their frequency, average in degree and average out degree of the vertex matches for each along with any temporal clause, and then aggregate ($\otimes$) these frequencies. The aggregation between adjacent clauses can be either \texttt{AND} or \texttt{OR}, and based on this, we apply the following aggregation logic for the frequencies and degrees.
\begin{align}
f ~~~~=~~~~ \bigotimes(f1, f2) ~~=~~& \left\{
\begin{array}{ll}
\min(f1,f2) & \text{, if }\otimes = \text{\tt AND} \\
\max(f1,f2) & \text{, if }\otimes = \text{\tt OR} 
\end{array}
\right. \label{eq:freq}\\
\delta ~~~~=~~~~ \bigotimes(\langle f_i, \delta_i \rangle, ...) ~~=~~& \frac{\sum_i f_i \times \delta_i}{\sum_i f_i} \label{eq:deg}
\end{align}

\noindent Equation~\ref{eq:freq} returns the smaller of the frequencies while performing an \texttt{AND}, and the larger of the two with an \texttt{OR}; the former can be an over-estimate while the latter an under-estimate \addc{if the two properties are not statistically independent}. Equation~\ref{eq:deg} finds the weighted average of the degrees of the vertices matching the predicates.
Once the frequencies of the clauses are aggregated, we divide it by the number of vertices of this vertex type to get the selectivity for the vertex predicate.

Then, in Equation~\ref{eqn:cm:3}, we identify the number of edges for which \texttt{scatter} will be triggered by multiplying the matched vertices with the sum of the in and out degrees for the matching vertices $\delta$. %
Lastly, in Equation~\ref{eqn:cm:4} we estimate the number of edges matched by the edge predicate $\overline{\pi_i}$. Here, we get the edge selectivity using the frequency of edge matches returned by the graph statistics, and normalized by the number of preceding vertices of type $\sigma$, times the average of the in and out degrees of vertices of this type, $\delta$. The edge selectivity is multiplied by the active edge count to get the matched edges that is expected from the \texttt{scatter} call. These edges will send messages to their destination vertices, and this will feed into the active vertex count in superstep $i+1$.

\addc{E.g., Table~\ref{tbl:cm_example} shows the cost model and statistics in action for query \emph{EQ2} on graph \emph{100k:F-S} that is described later in Section~\ref{sec:workload}. It reports the counts for the active and matched vertex and edge counts ($a,m$) using Equations~\ref{eqn:cm:1}--\ref{eqn:cm:4}, and the frequency of the vertices and edges ($f$) as returned by the histogram, for each superstep of two different query plans. 
	We see that $a_1$ is higher for Plan 2 than Plan 1 since the plans start at different vertex types %
	during the \texttt{init} phase, and this will lead to different execution times for this phase ($\iota$, discussed later). The frequency $f_1$ in Plan 1 is equal to $f_2$ in Plan, 2 %
	and likewise for $f_2$ of Plan 1 and $f_1$ of Plan 2. This is expected since the predicate evaluated in superstep $1$ of Plan 1 is same as that of superstep $2$ in Plan 2. %
	The cost model also estimates the messages sent $\overline{m_1}$ to be $1.3M$ and $67k$ for the two plans.  Since we assume that the property values are independent,  %
	the selectivities remain constant. $a_{2} = \overline{m_1}$ for both plans since we assume that each message from a superstep is sent to a unique vertex in the next superstep. While the compute calls for Plan 2 is higher, and the scatter calls and messages for Plan 1 is higher. The execution time model discussed next helps decide which of these plans has a lower estimated latency.}%

The clauses for time can also have comparators like $\succ$, $\prec$, etc. and property clauses can have $!=$. These are supported by the histogram and cost model. \addc{E.g., we get the frequency for a $\prec$ operator by summing the frequencies for all values smaller than the given value, and for $!=$ by subtracting from the total frequency the frequency of values that equal the given value.}
\addc{All time-variant statistics are maintained in the histogram, while invariants such as the count of vertex and edges of each type are maintained as part of global statistics for the graph.}

\subsection{Execution Time Estimate}
Given the estimates of the active/matched vertices/edges in each superstep, we incorporate them into execution time models for the different stages within a superstep to predict the overall execution time. We use micro-benchmarks to fit a \emph{linear regression model} for the execution times, $\mathcal{I}, \mathcal{M}, \mathcal{S}, \mathcal{CC},$ and $\mathcal{IC}$, used below. These are unique to a cluster deployment of \graphdb, and can be reused across graphs and queries.

As shown in Figure~\ref{fig:scheme}, the \texttt{init} function is called on the active vertices $a_1$ in the first superstep, and generates $m_1$ outputs that affect the states of the interval vertex. Its execution time \addc{estimate} is given by \addc{the function} $\iota = \mathcal{I}(a_0, m_0)$. For subsequent supersteps $i$, the \texttt{compute} function is called similarly on the active vertices, $a_i$, to generate the matched vertices $m_i$. This has a slightly different execution logic since it has to process an estimated $\overline{m_{i-1}}$ input messages from the previous superstep and does not have to initialize data structures, unlike \texttt{init}. Its execution time \addc{estimate} is, $c_i=\mathcal{M}(a_i, m_i, \overline{m_{i-1}})$. %
In a superstep $i$, \texttt{scatter} is called on the active edges and generates matched edges, with an estimated time of $s_i = \mathcal{S}(\overline{a_i}, \overline{m_i})$.
Besides these, there are per-superstep platform overheads: for \addc{iterating over} vertices matching a given type, $cc_i = \mathcal{C}\mathcal{C}(|V_\sigma|)$ in the \emph{partitionCompute} phase, %
and a base overhead of $ic_i = \mathcal{I}\mathcal{C}(a_i)$ per active vertex for \emph{Graphite}.

Given these, the total estimated execution time of the cost model for a query path segment with $n$ hops is:
\begin{equation*}
T = (\iota + s_1 + cc_1 + ic_1) + \sum_{i=2}^{n} c_i + s_i + cc_i + ic_i
\end{equation*}

\addc{In practice, these functions are determined by fitting simple linear regression models over query micro-benchmarks performed on the cluster on which the platform will be deployed. This is done once, and the functions are common for different graphs and query workloads on that cluster.} 
\addc{E.g., Table~\ref{tbl:cm_parameters} shows the coefficients for the linear equations that we fit for these functions, for the experiment setup in Section~\ref{sec:setup}. Also, Table~\ref{tbl:cm_example} shows the estimated execution time $T_i$ in each superstep $i$ for the two execution plans, using these coefficients. Plan 1 takes lesser time than Plan 2 due to the latter taking $7.8\times$ longer in superstep 1. This is caused by a high \texttt{init} execution time, $\iota$, since it has to evaluate $51M$ vertices ($a_1$) compared to only $100k$ in Plan 1. Since the total time is dominated by the \texttt{init} time, our cost model will choose Plan 1 for executing of this query.}

\begin{table*}[t]
	\small
	\centering
	\caption{\addc{Cost model coefficients for linear regression fit for each execution phase, as used in our experiments}}
	\label{tbl:cm_parameters}
	\begin{tabular}{rrr|rrrr|rrr}
		\toprule
		\multicolumn{3}{c|}{\bf Init ($\mathcal{I}$)} & \multicolumn{4}{c|}{\bf Compute ($\mathcal{C}$)} & \multicolumn{3}{c|}{\bf Scatter ($\mathcal{S}$)} \\
		\hline
		$a_0$ & $m_0$ & cons. &
		$a_i$ & $m_i$ & $\overline{m_{i-1}}$ & cons. &
		$\overline{a_i}$ & $\overline{m_i}$ & cons. \\
		\toprule
		9.4e-5 & -3.1e-5 & 3.83 &
		7.2e-5 & 3.3e-5 & 1.8e-5 & 1.63 &
		7.9e-5 & 0 & -3.81 \\
		\bottomrule
	\end{tabular}
	\begin{tabular}{rr|rr}
		\toprule
		\multicolumn{2}{p{2cm}|}{\bf Interval \mbox{Compute} ($\mathcal{IC}$)} & \multicolumn{2}{p{2cm}}{\bf Partition Compute ($\mathcal{CC}$)} \\
		\hline
		$a_i$ & cons. &
		$V_\sigma$ & cons.\\
		\toprule
		-5.1e-6 & 8.6e-2 &
		-8.0e-6 & 28.7
		\\
		\bottomrule
	\end{tabular}
\end{table*}

\addc{We exclude the time to perform join and aggregation (for aggregate queries) from the cost model equation. This is based on our observation that this time is negligible (e.g., $20$--$30~ms$ in our experiments) compared to the overall execution time of a query ($1000~ms$) in most cases. In contrast, the execution time for \texttt{init} and the three \texttt{compute} functions together take about $900~ms$. Further, the join and aggregate costs are proportional to the result set size. Even with a large result set size, there would inevitably be a large number of intermediate compute calls, and so the relative time taken by join and aggregate will remain low. Avoiding their inclusion helps keep the model concise, with only the most significant costs included.}
\addc{The time taken to find the optimal split point for a query using the approach described in this section is $2$--$9$ms.}

\section{Results}%
\label{sec:results}
\subsection{Workload}%
\label{sec:workload}
We use the \emph{social network benchmark} from the Linked Data Benchmark Council (LDBC)~\cite{LdbcTechSpecification} for our evaluation of \graphdb. It is a community-standard workload with realistic transactional path queries over a social network property graph. There are two parts to this benchmark, a social network graph generator and a suite of benchmark queries. 

\begin{figure}[t]
	\centering
	\includegraphics[width=0.8\columnwidth]{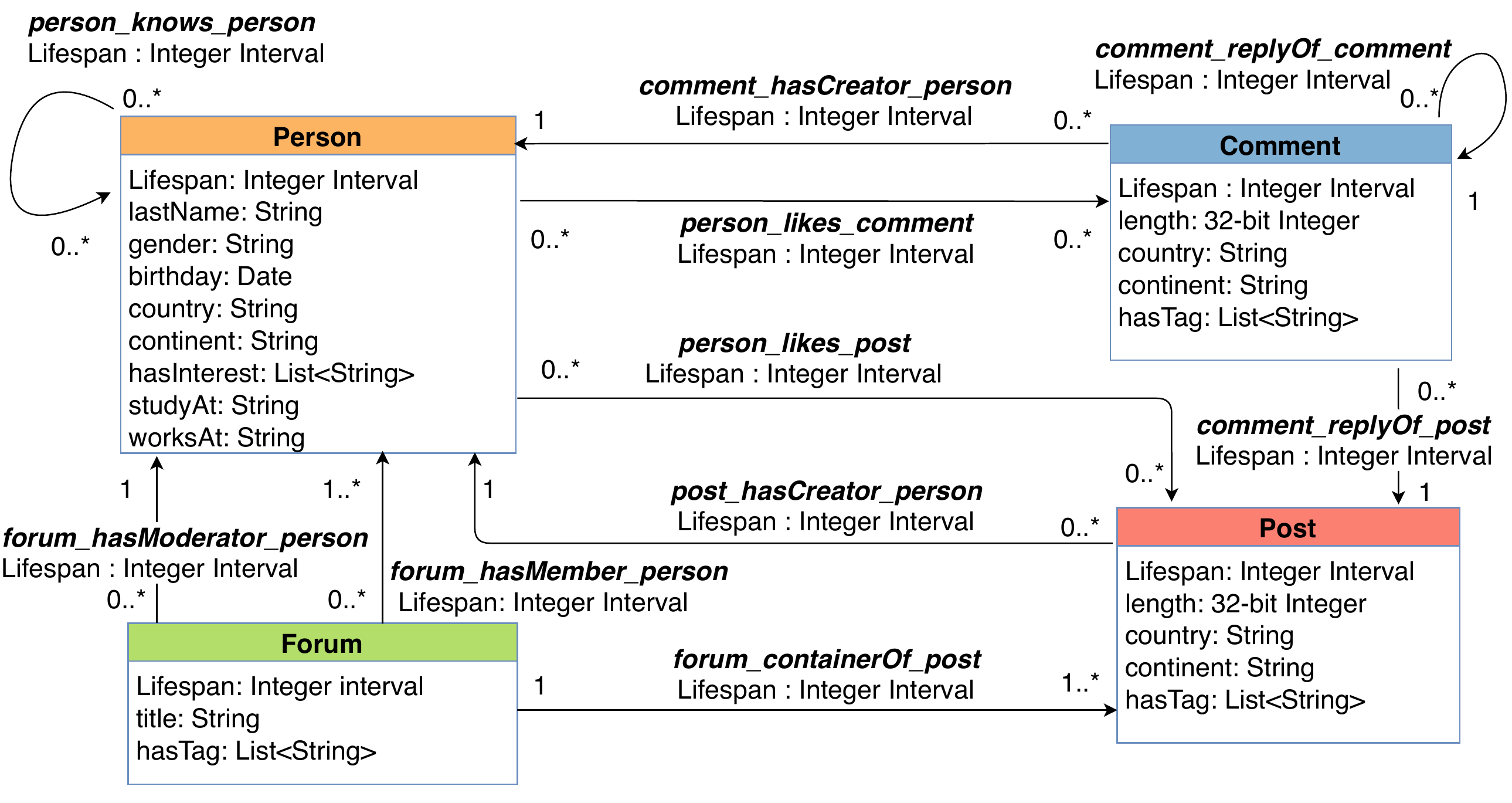}
	\caption{Modified LDBC Temporal Property Graph schema used in the evaluation}%
	\label{fig:schema}
\end{figure}

\paragraph{Property Graph Datasets} The graph generator S3G2~\cite{s3g2}
models a social network as a large correlated directed property graph with diverse distributions. Vertices and edges have a schema type and a set of properties for each type. Vertex types include \emph{person, message, comment, university, country,} etc., \addc{while edge types are \emph{follows}, \emph{likes}, \emph{isLocatedIn}, etc.} The graph is generated for a given number of persons in the network, and a given degree distribution of the \emph{person--follows--person} edge: \emph{Altmann (A), Discrete Weibull (DW), Facebook (F)} or \emph{Zipf (Z)}.

\begin{table}[t]
	\small
	\setlength{\tabcolsep}{0.35em} %
	\centering
	\caption{Characteristics of graphs used in the experiments} \label{tbl:graphs}
	\begin{tabular}{c||r|r||r|r|r|r}
		\hline
		&&&\multicolumn{4}{c||}{\em Frequent Vertex Types} \\
		\bf Graph & $\mathbf{|V|}$~~ & $\mathbf{|E|}$~~ & \bf Persons & \bf Posts & \bf Comments & \bf Forums \\
		\toprule
		\multicolumn{6}{c}{\emph{Static Temporal Graphs}}\\
		\hline
		\bf 10k:DW-S & 5.5M  & 20.8M  & 8.9k  & 1.1M  & 4.3M & 82k \\ %
		\bf 100k:Z-S & 12.1M  & 23.9M  & 89.9k  & 7.4M  & 2.3M  & 815k \\ %
		\bf 100k:A-S & 25.4M  & 78.2M  & 89.9k & 8.7M & 15.7M & 816k  \\ %
		\bf 100k:F-S & 52.1M & 217.6M & 100k & 12.6M  & 38.3M & 996k \\ %
		\toprule
		\multicolumn{6}{c}{\emph{Dynamic Temporal Graphs}}\\
		\hline
		\bf 10k:DW-D  & 6.6M  & 29.3M  & 10k  & 1.4M  & 5.1M  & 100k   \\ %
		\bf 100k:Z-D  & 15.2M  & 37.1M  & 100k  & 9.3M & 4.8M  & 995k  \\ %
		\bf \addc{100k:A-D}  & \addc{32.0M}  & \addc{112.2M}  & \addc{100k}  & \addc{10.8M} & \addc{20.1M}  & \addc{995k}   \\ %
		\bf \addc{100k:F-D}  & \addc{52.0M}  & \addc{216.5M}  & \addc{100k}  & \addc{12.6M} & \addc{38.2M}  & \addc{995k}   \\ %
		\hline
	\end{tabular}
	\begin{tabular}{c||r|r||r}
		\hline
		&\multicolumn{2}{c||}{\em Frequent Edge Types} & \bf Unrolled \\
		\bf Graph & \bf hasMember$^*$ & \bf hasCreator$^\dagger$ & \bf Properties$^\#$  \\
		\toprule
		\multicolumn{4}{c}{\emph{Static Temporal Graphs}}\\
		\hline
		\bf 10k:DW-S & 3.3M & 4.3M  & 35M \\ %
		\bf 100k:Z-S & 1.5M & 2.3M & 60M \\ %
		\bf 100k:A-S & 12.7M & 15.8M & 157M \\ %
		\bf 100k:F-S & 52.2M & 38.4 M & 325M \\ %
		\toprule
		\multicolumn{4}{c}{\emph{Dynamic Temporal Graphs}}\\
		\hline
		\bf 10k:DW-D  & 7.2M & 5.1M & 30M  \\ %
		\bf 100k:Z-D  & 3.2M & 4.8M & 57M \\ %
		\bf \addc{100k:A-D} & \addc{25.6M} & \addc{20.1M} & \addc{132M} \\ %
		\bf \addc{100k:F-D}  & \addc{51.8M} & \addc{38.3M} & \addc{222M} \\ %
		\hline
		\multicolumn{2}{l}{$^*$~\emph{forum\_hasMember\_person}} & 
		\multicolumn{2}{l}{$^\dagger$~\emph{comment\_hasCreator\_person}} \\
		\multicolumn{4}{l}{$^\#$~\emph{Unrolls multi-valued properties into individual ones}}
	\end{tabular}
\end{table}

We make two changes to the LDBC property graph generator. One, we \emph{denormalize} the schema to embed some vertex types such as \emph{country, company, university} and \emph{tag} directly as properties inside \emph{person, forum, post} and \emph{comment} vertices. This simplifies the data model. Two, while LDBC vertices are assigned a creation timestamp that can fall within a 3-year period, we include an end time of $\infty$ to form a time interval. We also add \emph{lifespans} to the edges incident on vertices based on their referential integrity constraints, and \addc{replace time-related} properties like \emph{join date} and \emph{post date} \addc{with the built-in lifespan property instead}. The vertex and edge lifespans are also inherited by their properties. Figure~\ref{fig:schema} shows this modified graph schema.%

However, this is still only a static temporal property graph. To address this, we introduce temporal variability into the properties, \emph{worksAt}, \emph{country} and \emph{hasInterest} of the \emph{person} vertex. For \emph{worksAt}, we generate a new property every year using the LDBC distribution; the \emph{country} is correlated with \emph{worksAt}, and hence updated as well. %
We update the \emph{hasInterest} property %
based on the list of tags for a forum that a person joins, at different time points. %

Table~\ref{tbl:graphs} shows the vertex and edge counts, the number of vertices of each type and the total number of property \addc{values}, for graphs we generate with $10^4$ (10k) or $10^5$ (100k) persons, %
with different distributions (DW, Z, A, F), and with static (S, top) and dynamic (D, bottom) properties.
\addc{As we see, the \emph{Comments} type dominates the number of vertices, with up to 400 comments per person over a 3~year period, followed by about 100 \emph{Posts} per person. The most frequent edge types are \emph{forum\_hasMember\_person} and \emph{comment\_hasCreator\_person}, while each person \emph{Follows} $10.2$ other friends on average. Properties such as \emph{hasInterest} for person and \emph{hasTag} for comment take up the most space since they are multi-valued, with an average of $23$ interests per person and $1.22$ tags per comment.}%

\begin{figure*}[!t]
	\centering%
	\subfloat[10k:DW]{
		\includegraphics[width=0.43\textwidth]{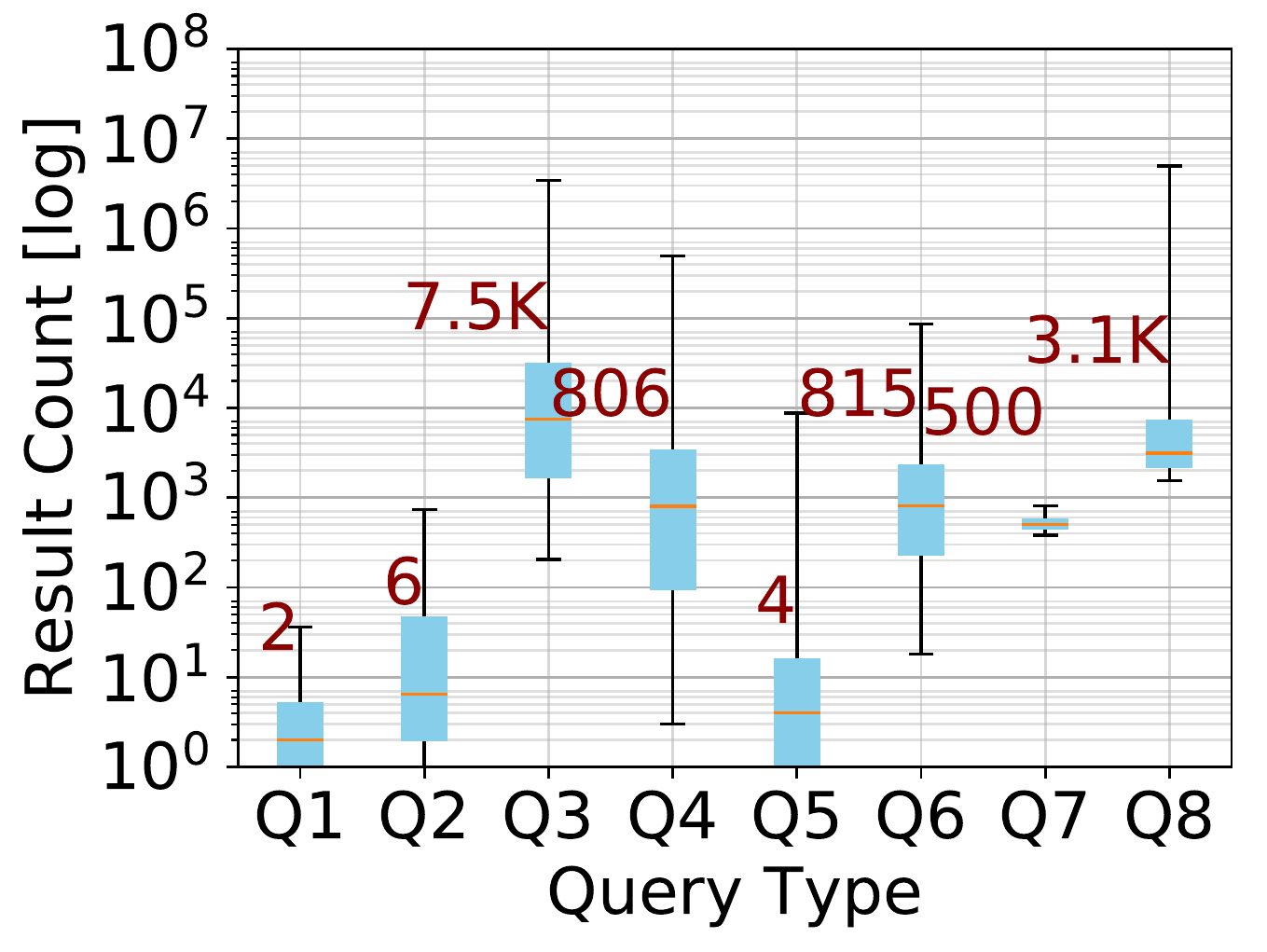}
		\label{fig:10k_dw_result_dist}
	}~~
	\subfloat[100k:Z]{
		\includegraphics[width=0.43\textwidth]{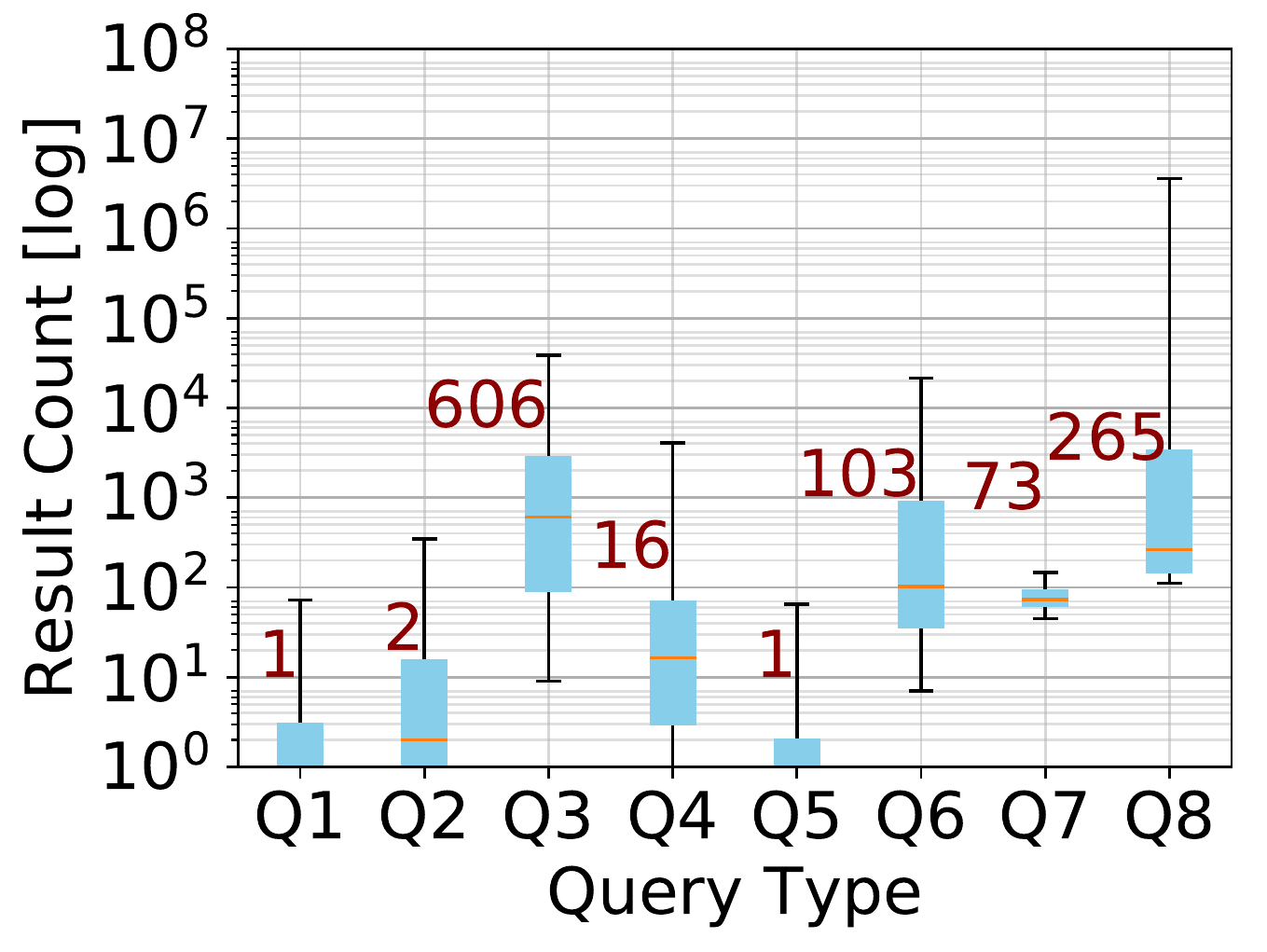}
		\label{fig:100k_zf_result_dist}
	}~~\\
	\subfloat[100k:A]{
		\includegraphics[width=0.43\textwidth]{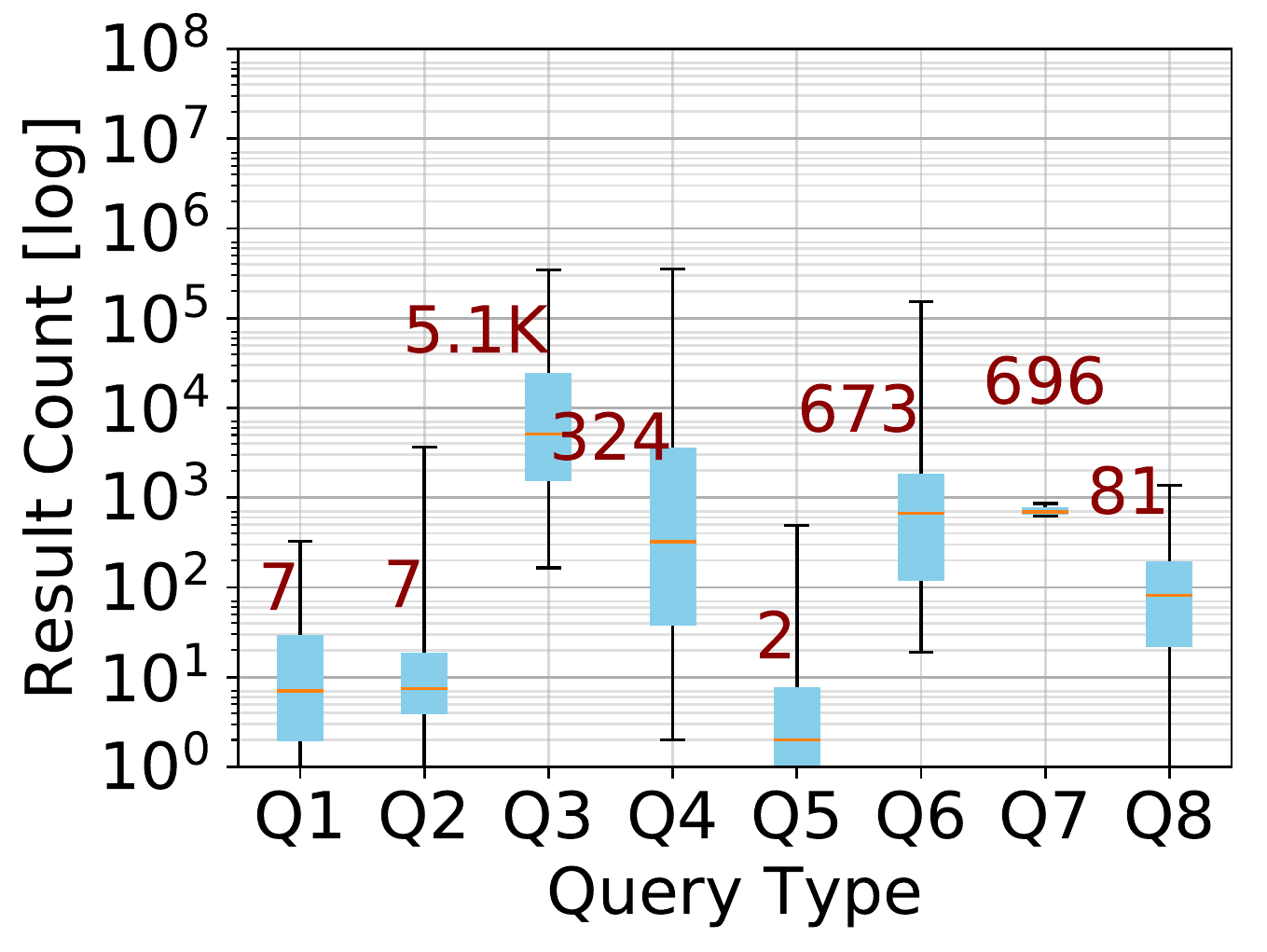}
		\label{fig:100k_a_result_dist}
	}~~
	\subfloat[100k:F]{
		\includegraphics[width=0.43\textwidth]{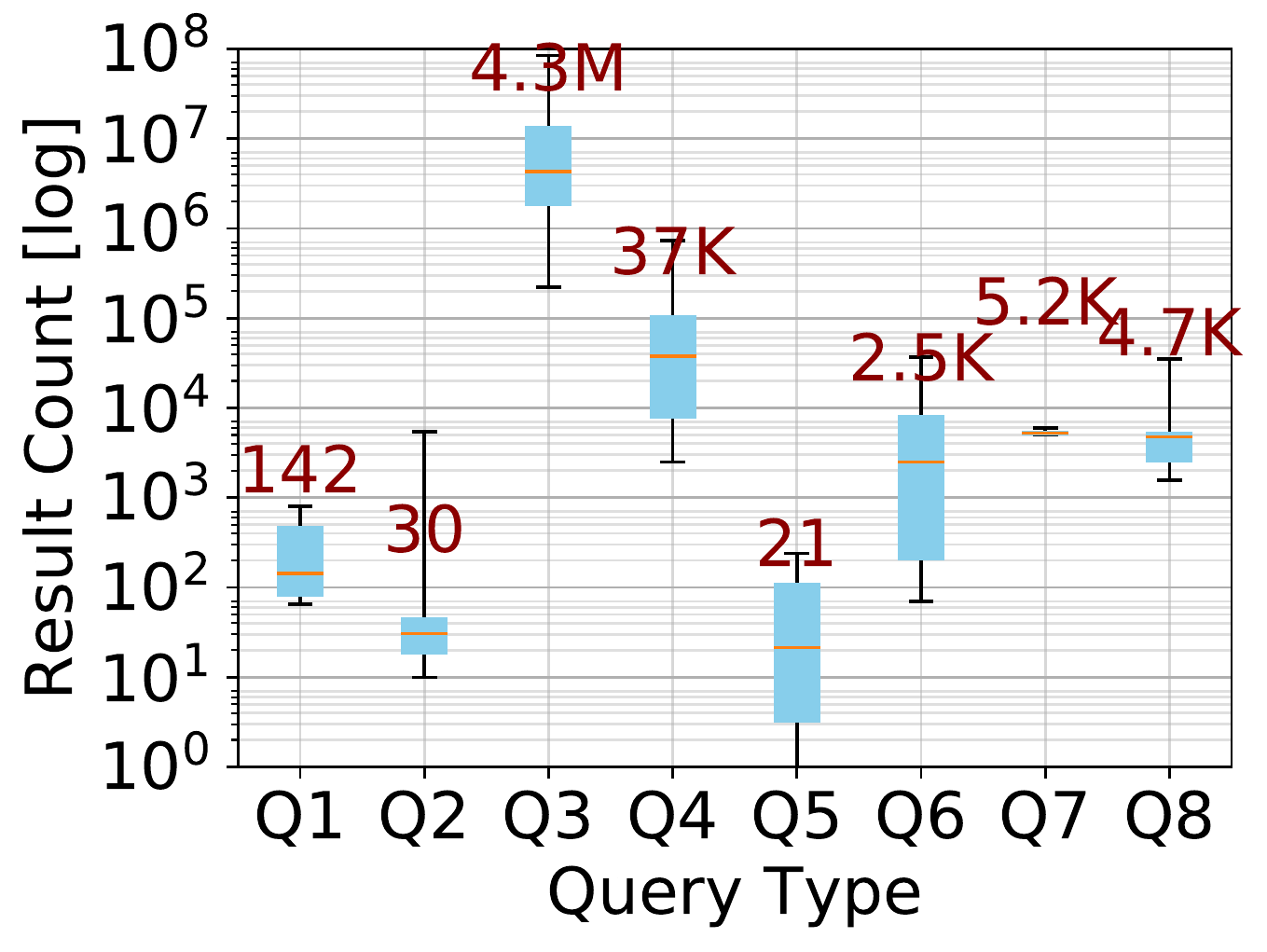}
		\label{fig:100k_f_result_dist}
	}
	\caption{\addc{Box and whiskers distribution plot of the result set count for the 100 instances of each non-aggregate query type. Q1--Q7 are reported on the static graphs while Q8 is on the dynamic graphs. The median result set count is labeled.}}
	\label{fig:result_dist}
\end{figure*}

\begin{table*}[t!]
	\small
	\setlength{\tabcolsep}{0.2em} %
    \centering
	\caption{Description of query workload used in the experiments} 
	\label{tbl:queries}
	\begin{tabular}{c|c|c|C{1.4cm}|C{1.4cm}|C{1.5cm}|L{5.5cm}}
		\hline
		\bf Query & \bf LDBC ID & \bf Hops & \bf Property Predicates & \bf Time Predicates & \bf Has ER Predicate? & \textbf{Description of path to find}~~\emph{(Parameterized property values are underlined)} \\%Result-size \\
		\toprule
		\bf Q1 & BI/Q9 & 3 & 4 & 1 & Yes &  Two \emph{messages} with different \underline{\emph{tags}} belong to the same \emph{forum}, with a time ordering between the messages \\ \hline
		\bf Q2 & BI/Q10 & 2 & 6 & 1 & No & A \emph{person} with a given \underline{\emph{tag}} creates a \emph{message} with the same \underline{\emph{tag}} after a given \underline{date}.\\ \hline
		\bf Q3 & BI/Q16 & 3 & 6 & 1 & Yes & A \emph{person} from a given \underline{\emph{country}} has commented or liked a \emph{post} before a \emph{person} from another given \underline{\emph{country}}.\\ \hline
		\bf Q4 & BI/Q17 & 4 & 3 & 2 & Yes & Mutual friendships between three \emph{persons}, but with a time-respecting order in which they befriend each other.\\ \hline
		\bf Q5 & -- & 5 & 7 & 3 & Yes & A \emph{person} posts a \emph{message} with a given \underline{\emph{tag}} to a \emph{forum} and, after a \underline{time offset}, they post another \emph{message} to the same \emph{forum} with a different \underline{\emph{tag}}.\\ \hline
		\bf Q6 & -- & 5 & 7 & 1 & Yes & A \emph{person} with a specific \underline{\emph{gender}} replies to a \emph{post} after another \underline{\emph{person}} replies to it.\\ \hline
		\bf Q7 & BI/Q23 & 4 & 5 & 3 & Yes & A \emph{person} posts a \emph{message} from outside their home \underline{\emph{country}}, then befriends another \emph{person}, and that \emph{person} then posts another \emph{message} from outside their home \underline{\emph{country}}. \\ \hline
		\bf Q8 & IW/Q11 & 3 & 3 & 1 & Yes & Two \emph{persons} working in different \underline{\emph{companies}} have a common \emph{friend} at a time-point.\\
		\hline
	\end{tabular}
\end{table*}

\paragraph{Query Workload}
We select a subset of query templates provided in the LDBC query workload~\cite{LdbcTechSpecification} that conform to a linear path query, and adapt them for our temporal graphs. Table~\ref{tbl:queries} describes the query templates. These are either from the \emph{Business Intelligence (BI)} or the \emph{Interactive Workload (IW)}. We also include two additional query templates $Q5$ and $Q6$ to fully exercise our query model. Also, query template $Q8$ depends on \emph{worksAt} which is a dynamic property and so it is only evaluated for dynamic temporal graphs.

Each template has some parameterized property or time value. We generate $100$ query instances for each template by randomly selecting a value for the parameters, evaluating the query on the temporal graph, and ensuring that there is at least 1 valid result set in most cases.
\addc{Query instances are generated for both the static and dynamic graphs. In addition to these \emph{non-aggregate queries}, we also create another workload that includes a \texttt{count} %
	temporal aggregate operator to these query templates, i.e., it will group the results of the original query by the first vertex and its time intervals, and return the count for each vertex-interval. This helps evaluate the performance of \emph{aggregate queries}. For brevity, we limit these aggregate queries to the two largest graphs, \emph{100k:A} and \emph{100k:F}.}
\addc{Figures~\ref{fig:10k_dw_result_dist},~\ref{fig:100k_zf_result_dist},~\ref{fig:100k_a_result_dist} and~\ref{fig:100k_f_result_dist} show the distribution of the result set count for the (non-aggregate) queries on the different graphs in our workload.}

These illustrate the expressivity of our query model, and ability to intuitively extend it to the time domain. \addc{The query length varies between $2$ and $5$ hops, allowing us to evaluate the cost model and \graphdb performance for different lengths. All the vertex types appear as predicates in our workload. The queries filter on both single-valued properties like \emph{country} and \emph{lastName}, and multi-valued properties like \emph{hasInterest} and \emph{hasTag}. All edge types except \emph{forum\_hasModerator\_person} are used in the workload. $7$ out of the $8$ query types have ETR predicate and all the queries have at least $1$ time predicate. They are diverse with respect to result sizes too, as shown in Figure~\ref{fig:10k_dw_result_dist},~\ref{fig:100k_zf_result_dist},~\ref{fig:100k_a_result_dist}~and~\ref{fig:100k_f_result_dist}, and the result counts span several orders of magnitude, from $10^0$--$10^4$.}

In our experiments, each query is given an execution budget of $600~secs$, after which it is terminated and marked as \emph{failed}. \addc{The average execution times are only reported on the successful queries.}
We verify the correctness of all queries on \graphdb and baseline platforms. For the performance evaluations, the queries only return the count of the result sets for timeliness.

\subsection{Experiment Setup}
\label{sec:setup}
Our commodity cluster has 18 \emph{compute nodes}, \addc{each} with one Intel Xeon E5-2620 v4 CPU with 8 cores (16 HT) @ 2.10GHz, 64~GB RAM and 1~Gbps Ethernet, running CentOS v7. For some shared-memory experiments on other baseline graph platforms, we also use a ``big memory'' \emph{head node} with 2 similar CPUs and 512~GB RAM. \graphdb is implemented over our in-house Graphite v1.0 \addc{ICM platform}~\cite{graphite}, Apache Giraph v1.3.0, Hadoop v3.1.1 and Java v8. By default, our distributed experiments use 8~compute nodes in this cluster, run one \graphdb Worker JVM per machine with 8~threads per Worker, and \addc{have} 50~GB RAM available to the JVM. The graphs are initially loaded into \graphdb from JSON files stored in HDFS, with their pre-computed cost model statistics, \addc{and the query workloads run on this distributed in-memory copy of the graph}.%

\subsection{Baseline Graph Platforms}
We use the widely-used \emph{Neo4J Community Edition v3.2.3}~\cite{neo4j} as a baseline graph database to compare against. This is a single-machine, single-threaded platform. We use three variants of this. One specifies the workload queries using the community standard \emph{Gremlin} query language (\emph{N4J-Gr}, in our plots), and the other uses Neo4J's native \emph{Cypher} language (\emph{N4J-Cy}). Both these variants run on a single compute node with 50~GB heap size. A third variant uses Cypher as well, but is allocated $8\times50=400~GB$ of \addc{heap space} on the head node (\emph{N4J-Cy-M}). As graph platforms are \addc{often} memory bound, this \addc{configuration} matches the total distributed memory available to our \graphdb setup by default.
We build indexes on all properties in Neo4J.

There are few open source \emph{distributed} graph engines available. \emph{JanusGraph}~\cite{janusgraph}, a fork from Titan, is popular, and uses \emph{Apache Spark v2.4.0} as a distributed backend engine to run Gremlin queries (\emph{Spark}, in our plots). It uses \emph{Apache Cassandra v2.2.10} to store and access the input graph. %
Spark runs on 8 compute nodes with 1 Worker each and 50~GB heap memory per Worker. Cassandra is deployed on 8 additional compute nodes. \addc{This is based on the recommended configuration for JanusGraph on Cassandra~\footnote{https://docs.janusgraph.org/storage-backend/cassandra/}. Spark initially loads the graph from Cassandra into its distributed memory present on its 8 compute nodes. This load time is not considered as part of the query execution time. So effectively, only the 8 Spark nodes are used during query execution.}
For all baselines, we follow the standard performance tuning guidelines provided in their documentation~\footnote{https://neo4j.com/docs/operations-manual/3.2/performance/}~\footnote{https://docs.janusgraph.org/advanced-topics/hadoop/}.%

Since these platforms do not natively support temporal queries over \emph{dynamic} temporal graphs, we transform the graphs into a static temporal graph using techniques described by Wu, et al. and used earlier by Graphite~\cite{wu2016reachability,graphite}. \addc{This static property graph converts the time-intervals on vertices and edges of the original interval graph into an expanded set of vertices and edges that are valid for just a single discrete time point.} This lets us adapt the query to operate on the static graph, albeit a bloated one. \addc{Also, temporal aggregation is not feasible internally on these platforms. So we perform the final aggregation at the client side for queries with an aggregate operator.} JanusGraph/Spark is unable to load these two large graphs in-memory, and hence was not evaluated for the aggregate queries. %
The results from all platforms for all queries are verified to be identical.

\begin{figure}[!t]
	\centering%
	\subfloat[Distribution of queries that exceed the optimal plan's time by a \% (Y axis), for each fixed plan and for the cost model, for \emph{100k:A-S} graph on $Q4$ type queries]{%
		\includegraphics[width=0.43\textwidth]{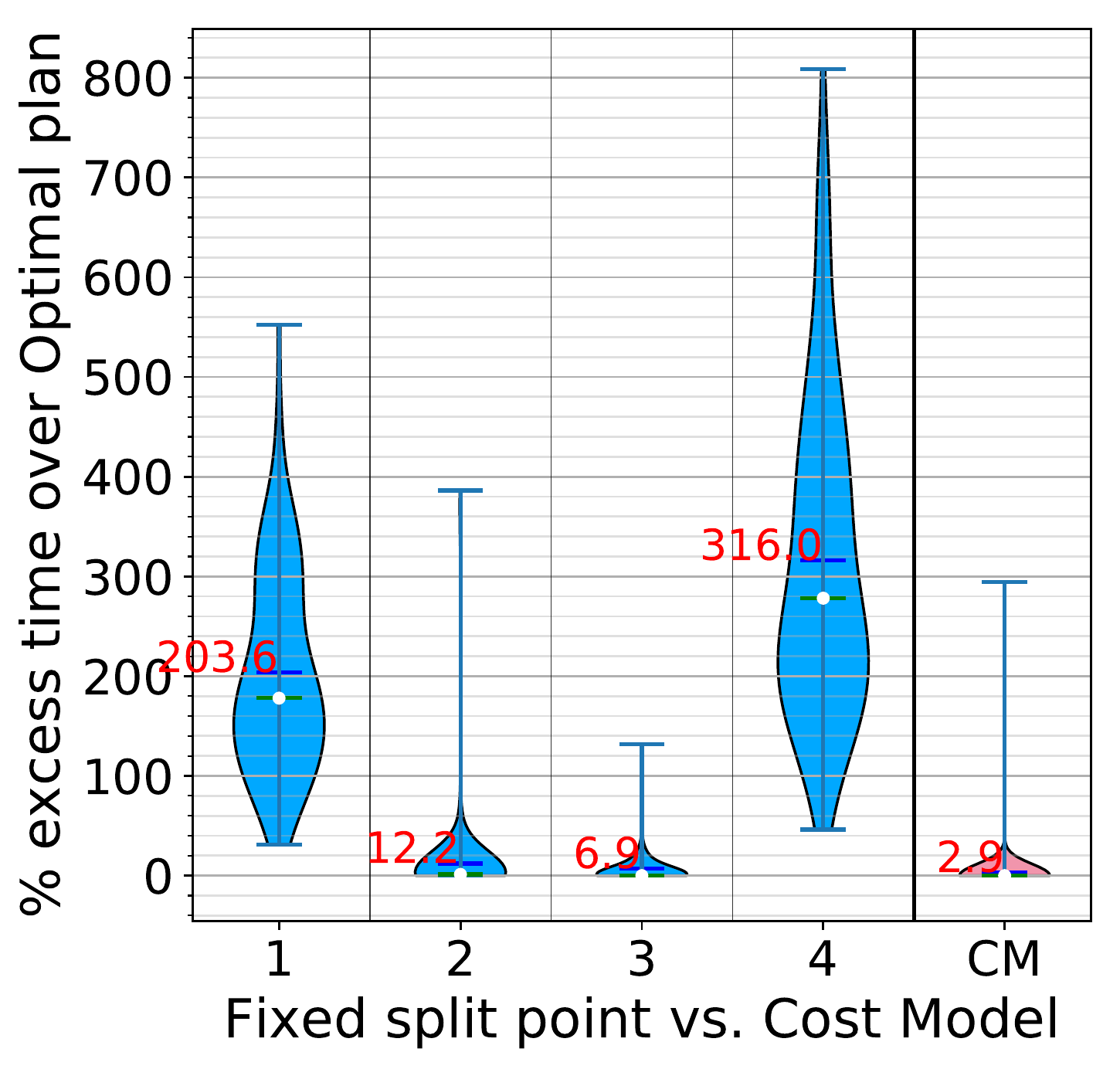} 
		\label{fig:cm_100k_a_4}
	}~~
	\subfloat[\addc{Actual vs. Cost model estimated execution time for all query instances of \emph{100k:A-S} graph, with correlation coefficient of $\rho=0.87$}%
	]{
		\includegraphics[width=0.43\textwidth]{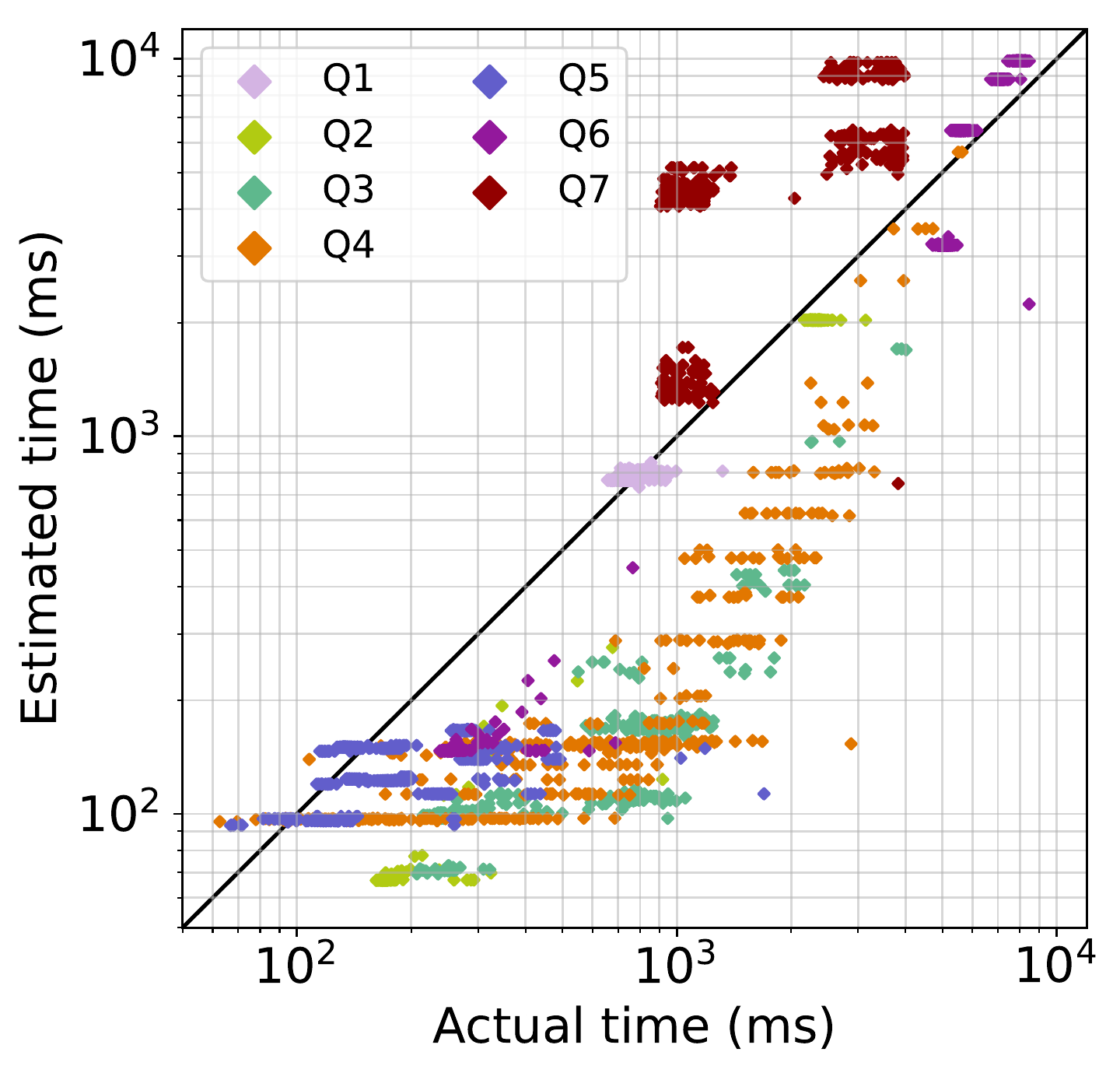} 
		\label{fig:scatter_100k_a}
	}~~\\
	\subfloat[\addc{Ratio of estimated average execution cost of the other plans relative to the optimal plan, for all query types of \emph{100k:A-S} graph}]{
		\includegraphics[width=0.43\textwidth]{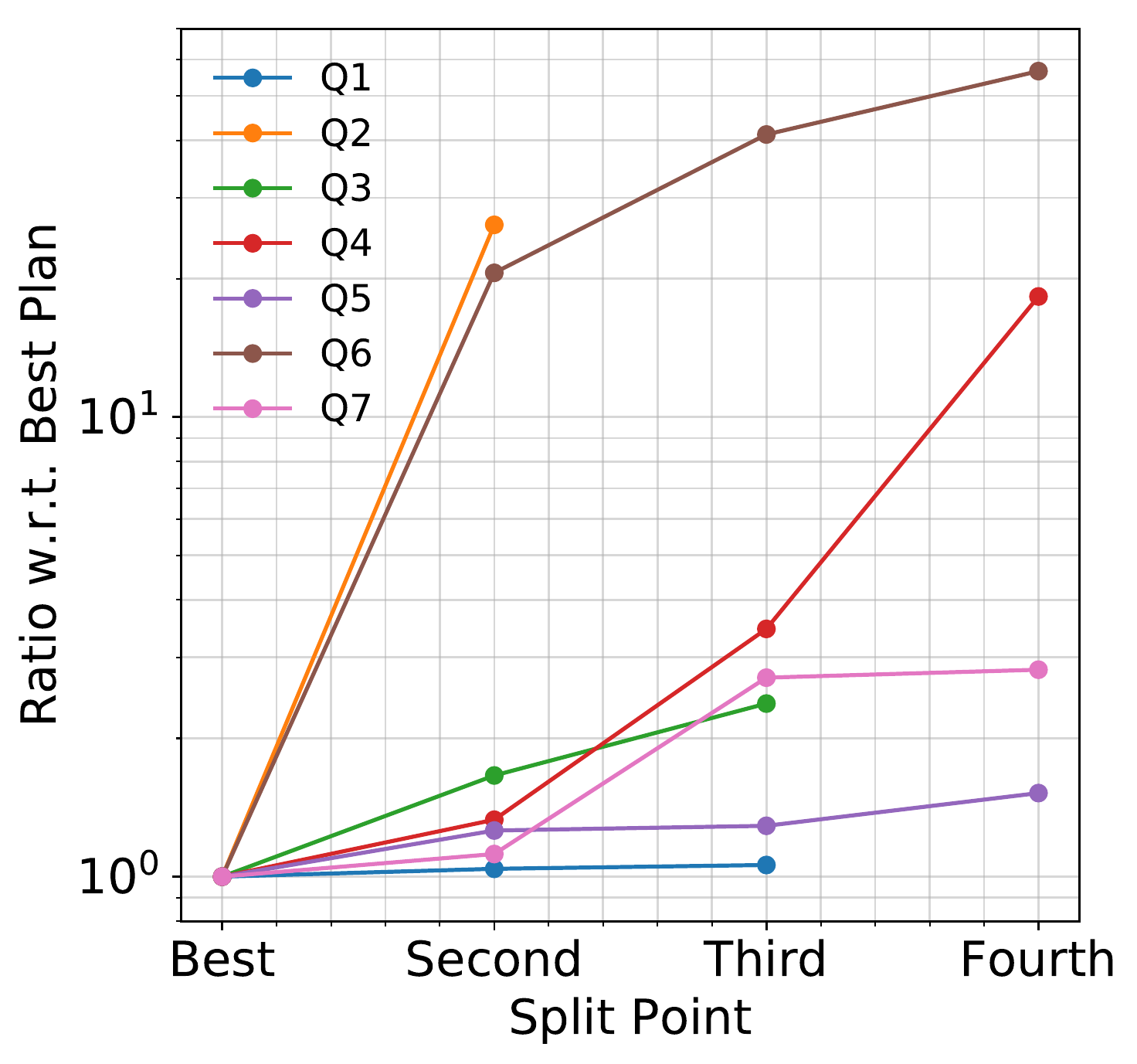} 
		\label{fig:cost_diff_100k_a}
	}~~
	\subfloat[Cost Model Accuracy. \% of times the optimal plan, $2^{nd}$ best plan and other plans were selected by our model for all graphs]{%
		\includegraphics[width=0.43\textwidth]{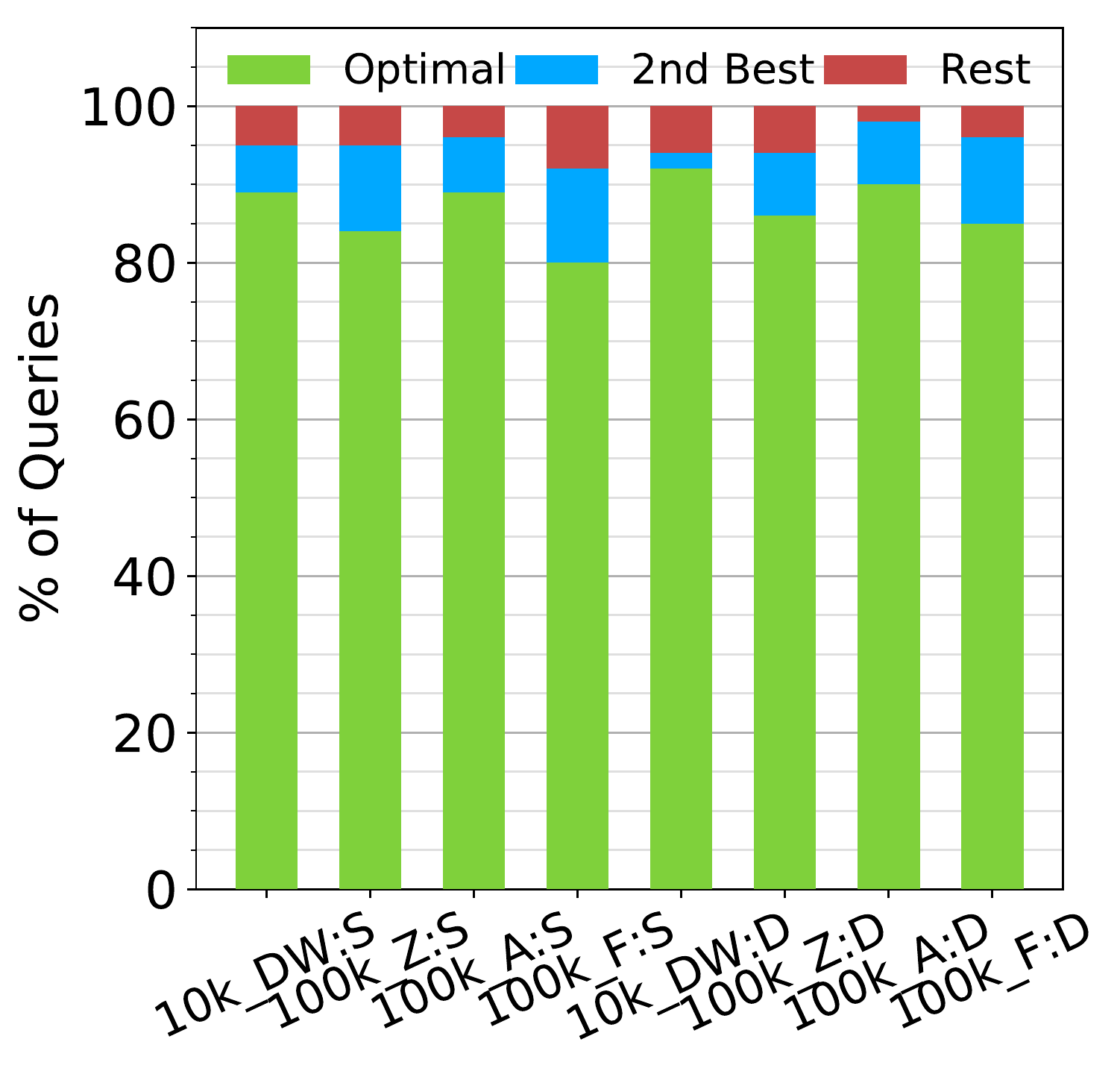}
		\label{fig:cm_acc}
	}
	\caption{Effectiveness of cost model in picking the best plan for non-aggregate queries}%
\end{figure}

\subsection{Effectiveness of Cost Model}%

\begin{figure}[!thp]
	\resizebox{!}{0.307\textwidth}
	{
		\begin{minipage}{1.1\textwidth}
			\setlength\tabcolsep{1pt} %
			\centering%
			\captionof{table}{\small \% excess time spent over the Optimal plan by the Cost Model selected plan, for different query percentiles of each query type}\label{tbl:100k_a}
			\small
			\subfloat[100k:A-S]{%
				\begin{tabular}{r||r|r|r|r|r|r|r}
					\toprule
					\emph{\%le} & \emph{Q1} & \emph{Q2} & \emph{Q3} & \emph{Q4} & \emph{Q5} & \emph{Q6} & \emph{Q7}  \\
					\hline
					\hline
					75 & 1.8 & 0 & 2.2 & 0 & 0 & 0 & 0\\ \hline
					90 & 6.8 & 0 & 12.6 & 0 & 0 & 0 & 0\\  \hline
					95 & 8.5 & 0 & 24.6 & 0 & 56 & 0 & 0\\  \hline
					99 & 17.6 & 0 & 47.1 & 0 & 123 & 0 & 195\\  
					\bottomrule
				\end{tabular}
				\label{tbl:100k_a_per_opt}
			}~
			\subfloat[100k:A-D]{
				\begin{tabular}{r||r|r|r|r|r|r|r|r }
					\toprule
					\emph{\%le} & \emph{Q1} & \emph{Q2} & \emph{Q3} & \emph{Q4} & \emph{Q5} & \emph{Q6} & \emph{Q7} & \emph{Q8} \\
					\hline
					\hline
					75 & 0 & 0 & 0 & 0 & 0 & 0 & 0 & 0\\ \hline
					90 & 0 & 0 & 42 & 0 & 7.1 & 0 & 0 & 59\\ \hline
					95 & 2.4 & 0 & 124 & 66 & 8.8 & 0 & 0 & 112\\ \hline
					99 & 3.6 & 0 & 198 & 191 & 12 & 0 & 0 & 277\\ 
					\bottomrule
				\end{tabular}
				\label{tbl:100k_a_dyn_per_opt}
			}~\\
			\subfloat[\addc{100k:A-S (Temporal Aggregate)}]{
				\begin{tabular}{r||r|r|r|r|r|r|r}
					\toprule
					\emph{\%le} & \emph{Q1} & \emph{Q2} & \emph{Q3} & \emph{Q4} & \emph{Q5} & \emph{Q6} & \emph{Q7}  \\
					\hline
					\hline
					75 & 0 & 0 & 0 & 0 & 0 & 0 & 0 \\ \hline
					90 & 5.7 & 0 & 20 & 0 & 19 & 0 & 0\\  \hline
					95 & 6.3 & 0 & 24 & 0 & 24 & 0 & 0\\  \hline
					99 & 8.3 & 0 & 30 & 0 & 52 & 0 & 0\\  
					\bottomrule
				\end{tabular}
				\label{tbl:100k_a_per_opt_agg}
			}~
			\subfloat[\addc{100k:A-D (Temporal Aggregate)}]{
				\begin{tabular}{r||r|r|r|r|r|r|r|r}
					\toprule
					\emph{\%le} & \emph{Q1} & \emph{Q2} & \emph{Q3} & \emph{Q4} & \emph{Q5} & \emph{Q6} & \emph{Q7}& \emph{Q8}  \\
					\hline
					\hline
					75 & 0 & 0 & 0 & 0 & 0 & 0 & 0 & 0\\ \hline
					90 & 4 & 0 & 6 & 0 & 28 & 0 & 0 & 57\\  \hline
					95 & 12 & 28 & 21 & 132 & 39 & 0 & 0 & 175\\  \hline
					99 & 18 & 145 & 84 & 166 & 57 & 0 & 0 & 643\\  
					\bottomrule
				\end{tabular}
				\label{tbl:100k_a_dyn_per_opt_agg}
			}
		\end{minipage}
	} ~\\
	\begin{minipage}{1\textwidth}
		\addc{
			\centering%
			\subfloat[100k:A-S]{
				\includegraphics[width=0.43\textwidth]{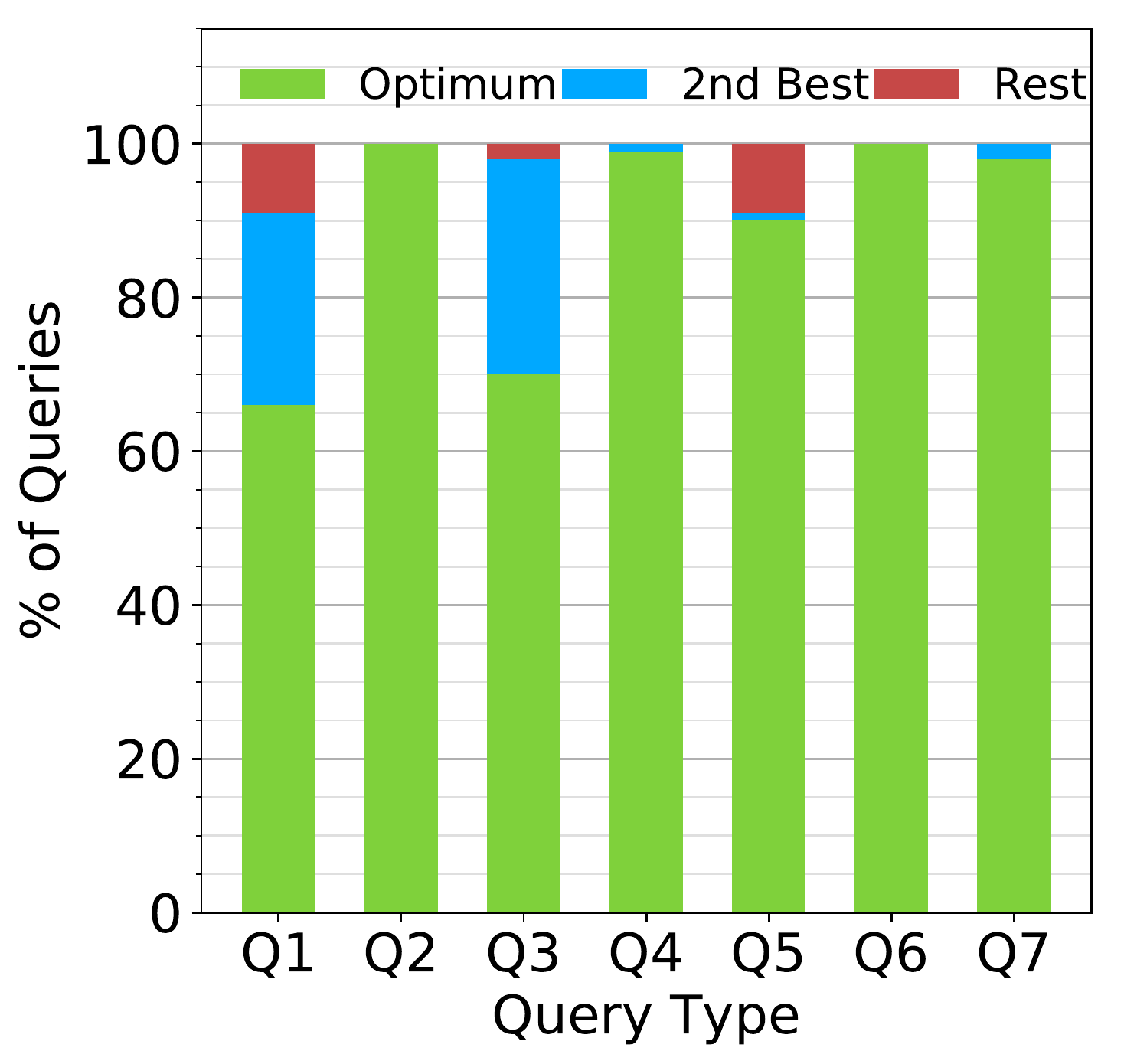}
				\label{fig:cm_100k_a}
			}~
			\subfloat[100k:A-D]{
				\includegraphics[width=0.43\textwidth]{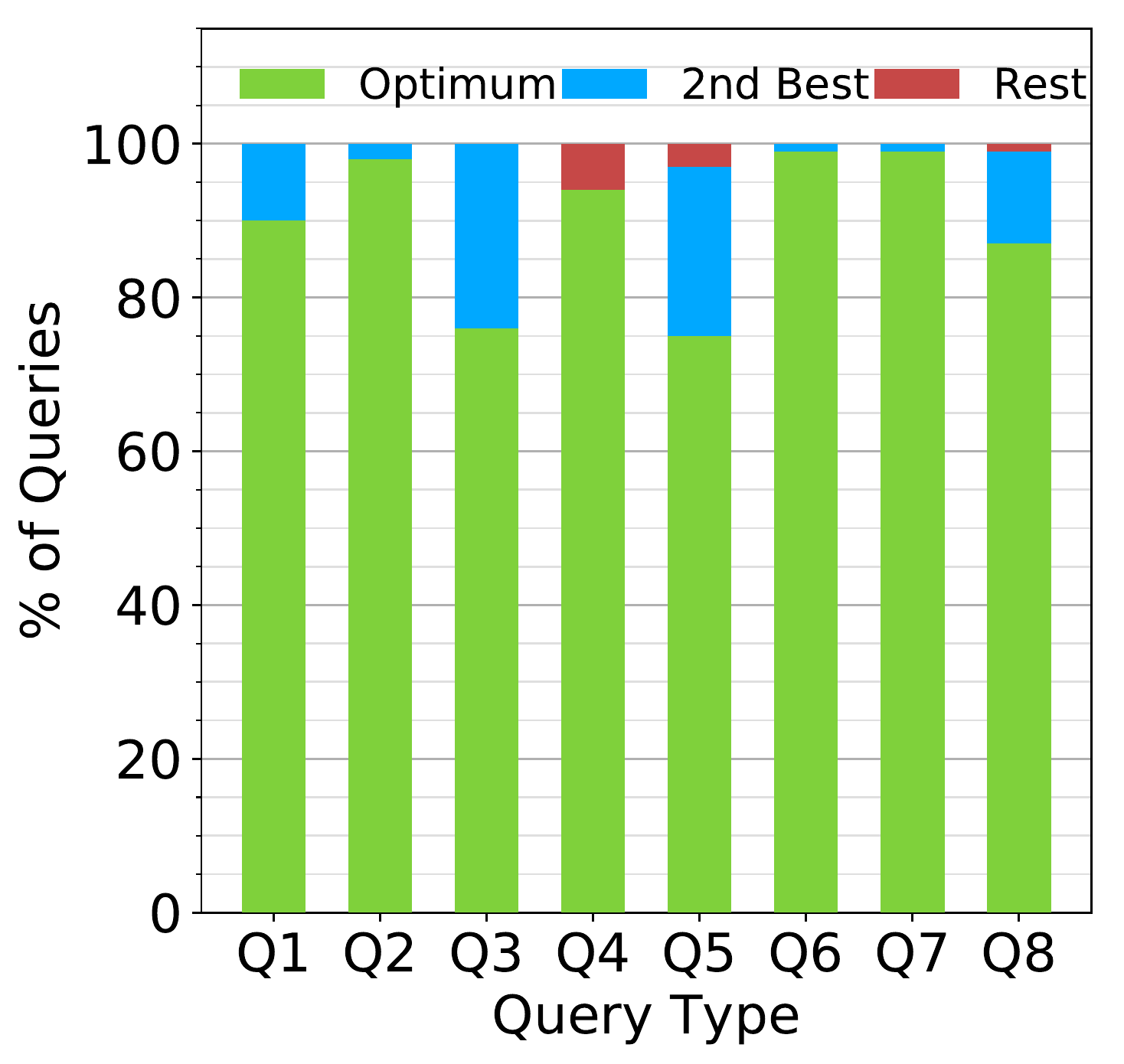}
				\label{fig:cm_100k_a_dyn}
			}~\\
			\subfloat[100k:A-S, Temporal Agg.]{
				\includegraphics[width=0.43\textwidth]{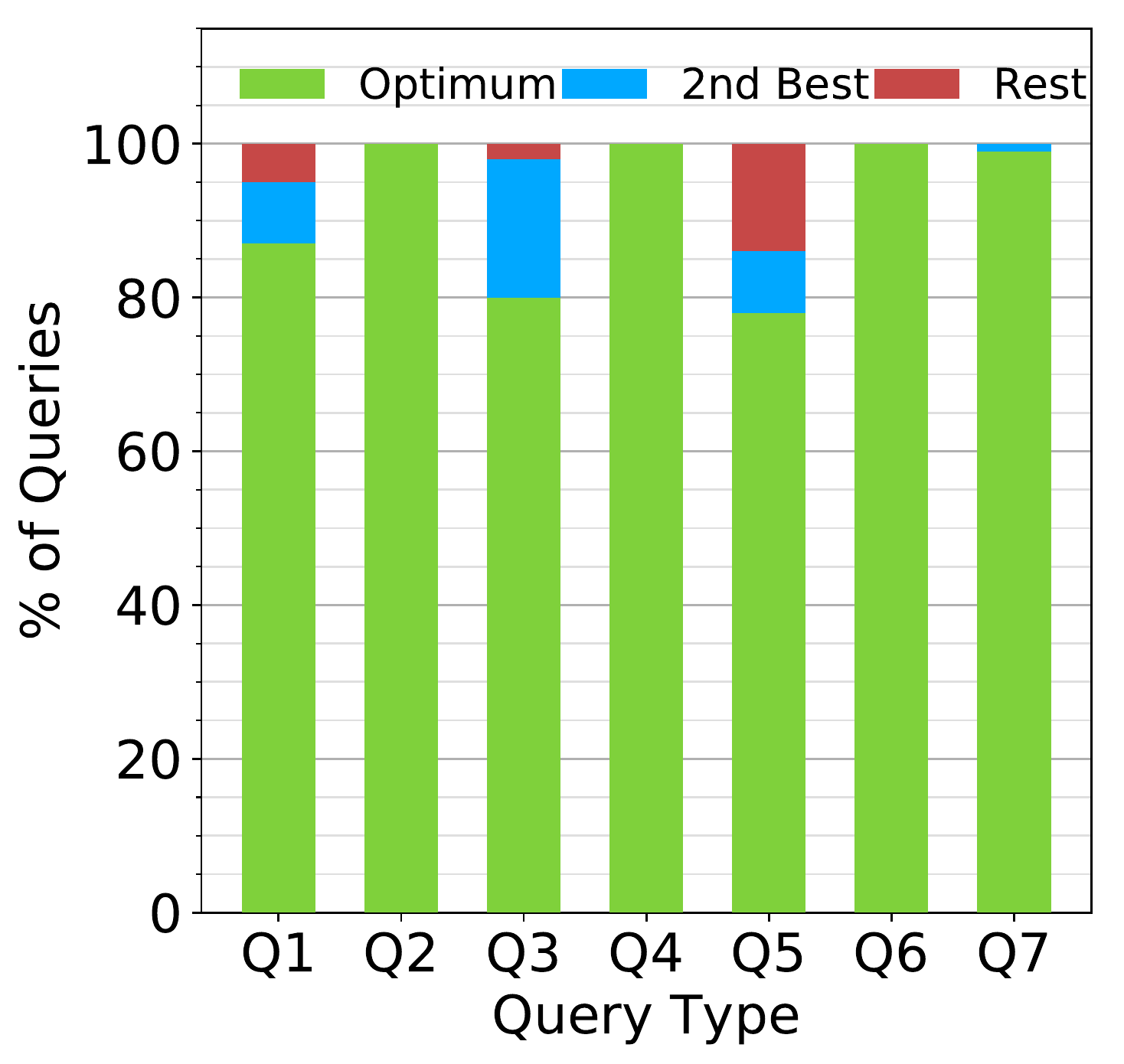}
				\label{fig:cm_100k_a_agg}
			}~
			\subfloat[100k:A-D, Temporal Agg.]{
				\includegraphics[width=0.43\textwidth]{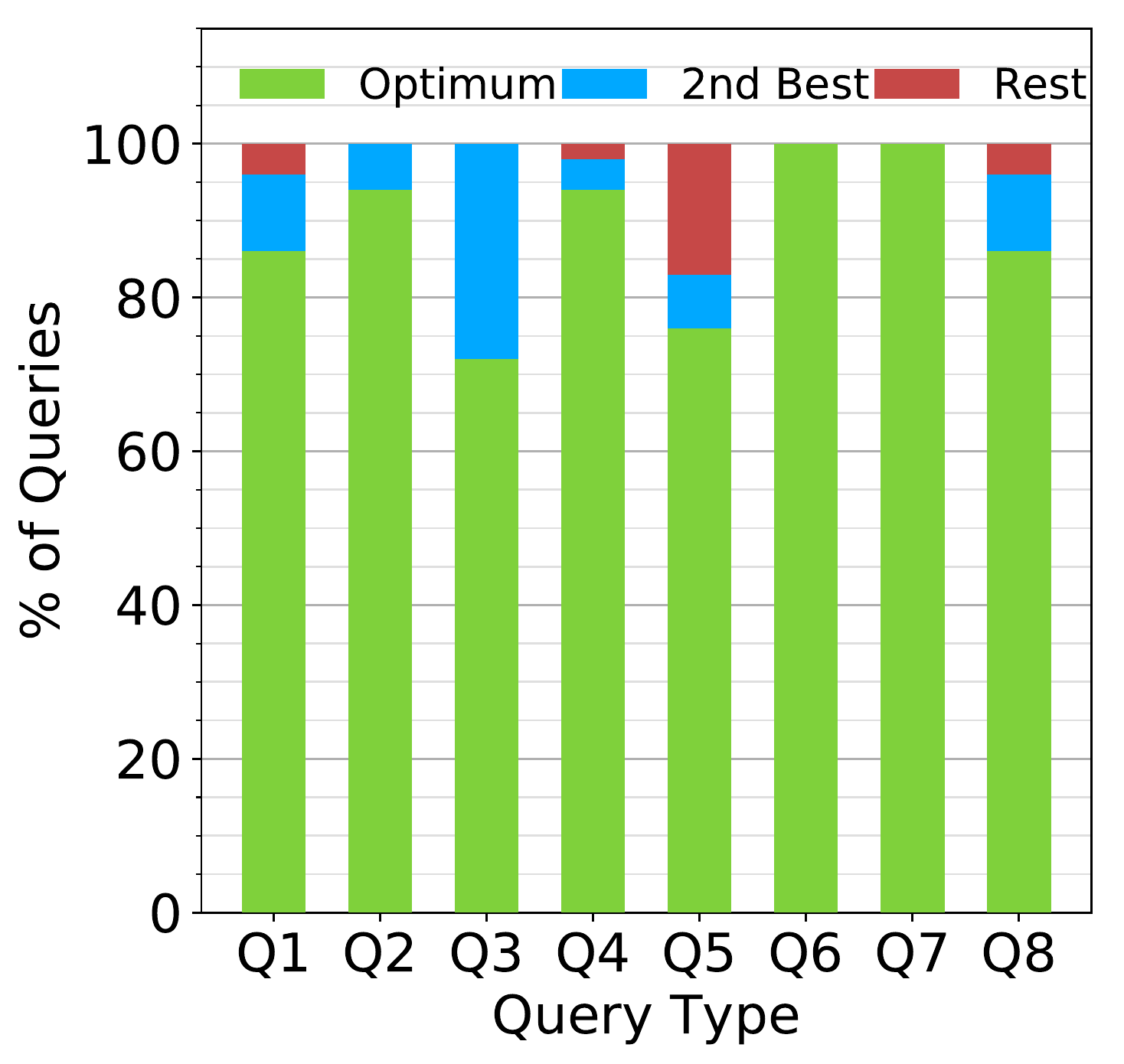}
				\label{fig:cm_100k_a_dyn_agg}
			}
			\caption{\addc{Cost Model Accuracy. \% of times the optimal plan, the second best plan and the other plans were selected by our model}}%
			\label{fig:cm_100k}
		}
	\end{minipage}
\end{figure}

We first evaluate the effectiveness of \graphdb's cost model in identifying the optimal split point for the distributed query execution. For each query type \addc{(template)}, we execute its $100$ query \addc{instances} using \emph{all their possible query plans}, \addc{i.e., every possible split point is considered for each query}. From the execution time of all plans for a query, we pick the smallest as its \emph{optimal plan}. We compare this against the plan selected by our \emph{cost model}, and report the \emph{\% of excess execution time} that our model-selected plan takes above the optimal plan. This is the effective time penalty when we select a sub-optimal plan.

Figure~\ref{fig:cm_100k_a_4} shows a violin plot of the the distribution of this \% excess time over optimal, for the different \emph{fixed} split points $1$--$4$ executed for the 100 queries of type $Q4$ (non-aggregate) on graph \emph{100k:A-S}, compared to %
the plan selected by our cost model (CM) -- \addc{lower this value, closer to optimal the performance.} We see that the execution time varies widely across the plans, with some taking $8\times$ longer than optimal. Also, some split points like 2 and 3 are in general better than the others, but among them, neither is consistently better. In contrast, our cost model plan has a low mean excess time of 2.9\%, relative to 12.2\% and 6.9\% excess time taken by these other split points. %
Also, it is not possible to \emph{a priori} find a single fixed split point which is generally better than the rest, without running the queries using all split points. These motivate the need for an automated analytical cost model for query plan selection.

\addc{We analyze the accuracy of the cost model for \emph{100k:A}, the second largest graph, in more detail, and its impact on the execution cost.} %
\addc{First, Figure~\ref{fig:scatter_100k_a} shows a scatter plot between the actual and the model-estimated execution time for the \emph{100k:A-S} static graph; the plot has $\approx 2500$ points. %
	Overall, we see a high correlation coefficient of $\rho=0.87$. %
	There is an over-estimation for $Q7$ (maroon) due to an inaccurate estimation of the number of matching edges in the second hop, %
	and under-estimates for $Q1$ to $Q5$. %
	But $Q6$ (purple) shows a high correlation of $\rho=0.94$. %
}

\addc{Given these execution time inaccuracies of the model, we examine its effect on: (1) picking the optimal execution plan, and (2) on the latency penalty when it does not pick the optimal plan. Figures~\ref{fig:cm_100k} show the fraction of times the cost model selects the optimal plan, the second best plan, and the rest of the plans, for the static and dynamic variants of \emph{100k:A}, and for non-aggregate and aggregate queries. We also have corresponding data in Tables~\ref{tbl:100k_a} which report for different query types (columns), and for different percentiles of their queries (rows), what is the \% excess execution time over the optimal spent by the plan chosen by the cost model.}

\addc{For the \emph{non-aggregate queries}, the best or the second best plan were selected over $97\%$ of the time across all queries, as seen in Figures~\ref{fig:cm_100k_a} and~\ref{fig:cm_100k_a_dyn}. For queries $Q2$, $Q4$, $Q6$ and $Q7$, the optimal plan was chosen $99\%$ of the time.
	In $Q2$, this is due to a short query length of 2 that reduces the cumulative errors in the model, as well as a high difference in cost between the best and the second best plans. This is seen in Figure~\ref{fig:cost_diff_100k_a}, which gives the ratio of the $2^{nd}$, $3^{rd}$ and $4^{th}$ best plan relative to the optimal. For $Q2$, the best plan evaluates the \emph{person} vertices first, which are $500\times$  fewer than the \emph{message} vertices evaluated first by the other plan. As a result, the optimal execution time is $10\times$ smaller than the other and the model easily selects the former plan. %
	Similarly, $Q6$ also exhibits a high difference in cost between the optimal plan and the remaining three. %
	But the top two plans for queries $Q4$ and $Q7$ have a similar cost. For $Q4$, starting at either ends causes a high fan-out and hence the plans that start at the two intermediate hops have a lower, but similar, cost. %
	In such cases, as Figures~\ref{fig:cm_100k_a} and~\ref{fig:cm_100k_a_dyn} show, we may occasionally select the second best plan.}

\addc{However, the consequence of choosing the second best plan on the actual execution latency is low when the top-2 plans have a similar model cost.} In fact, for \emph{100k:A-S}, we see from Table~\ref{tbl:100k_a_per_opt} that the execution time of the model-selected plan is within 2\% of the optimal execution time for the $75^{th}$ percentile query, within a query type, and within 13\% for the $90^{th}$ percentile query. Its only at the $95^{th}$ percentile query that we see higher penalties of $8$--$56\%$ for 3 of the 7 query types. %
\addc{Even for the dynamic graph \emph{100k:A-D}, 6 of the 8 query types have negligible time penalties at the $90^{th}$ percentile query in Table~\ref{tbl:100k_a_dyn_per_opt}, while two, $Q3$ and $Q8$, have higher penalties of $42$--$59\%$.
	The sub-optimal behavior happens when the execution model predicts a similar cost for the top-2 plans but selects the actual second-best, \emph{and} the observed runtime for the second-best is much worse than the best. E.g., 
	for the \emph{100k:A-S} graph, the difference in actual execution cost between the optimum and second best plans %
	for query $Q3$ is $18\%$. This causes the model to select the second best plan $\approx 28\%$ of the time, and causes $\approx 5\%$ of the queries to take $25\%$ or longer to execute than the optimal plan.} %

\addc{
	We see similar trends for the \emph{temporal aggregate queries} as well, in Figures~\ref{fig:cm_100k_a_agg} and~\ref{fig:cm_100k_a_dyn_agg}, and Tables~\ref{tbl:100k_a_per_opt_agg} and~\ref{tbl:100k_a_dyn_per_opt_agg}.  The models predict the same costs for these aggregate queries since it ignores the aggregate operation and join costs due to their negligible overheads. Despite that, these queries perform on par or better than the equivalent queries without the aggregation step.%
}
In fact, this is broadly applicable to all the graphs, as observed in Figure~\ref{fig:cm_acc}. It reports that across all queries and graphs evaluated, our cost model picks the best (optimal) or the second best plan over $95\%$ of the time. %

In summary, the cost model is accurate when the query is of shorter length, and accurate enough to distinguish between the similar good plans and the rest when certain predicate have \addc{high} cardinalities. So we pre-dominantly pick a plan that is optimal, or has an execution time that is close to the optimal plan. \addc{Thus,} while our cost model is not perfect, it is accurate enough to discriminate between the better and the worse plans, and consequently reduce the actual query execution time.

\subsection{Comparison with Baselines}%
\begin{figure*}[!t]
	\centering%
	\subfloat[10k:DW-S]{
		\includegraphics[width=0.435\textwidth]{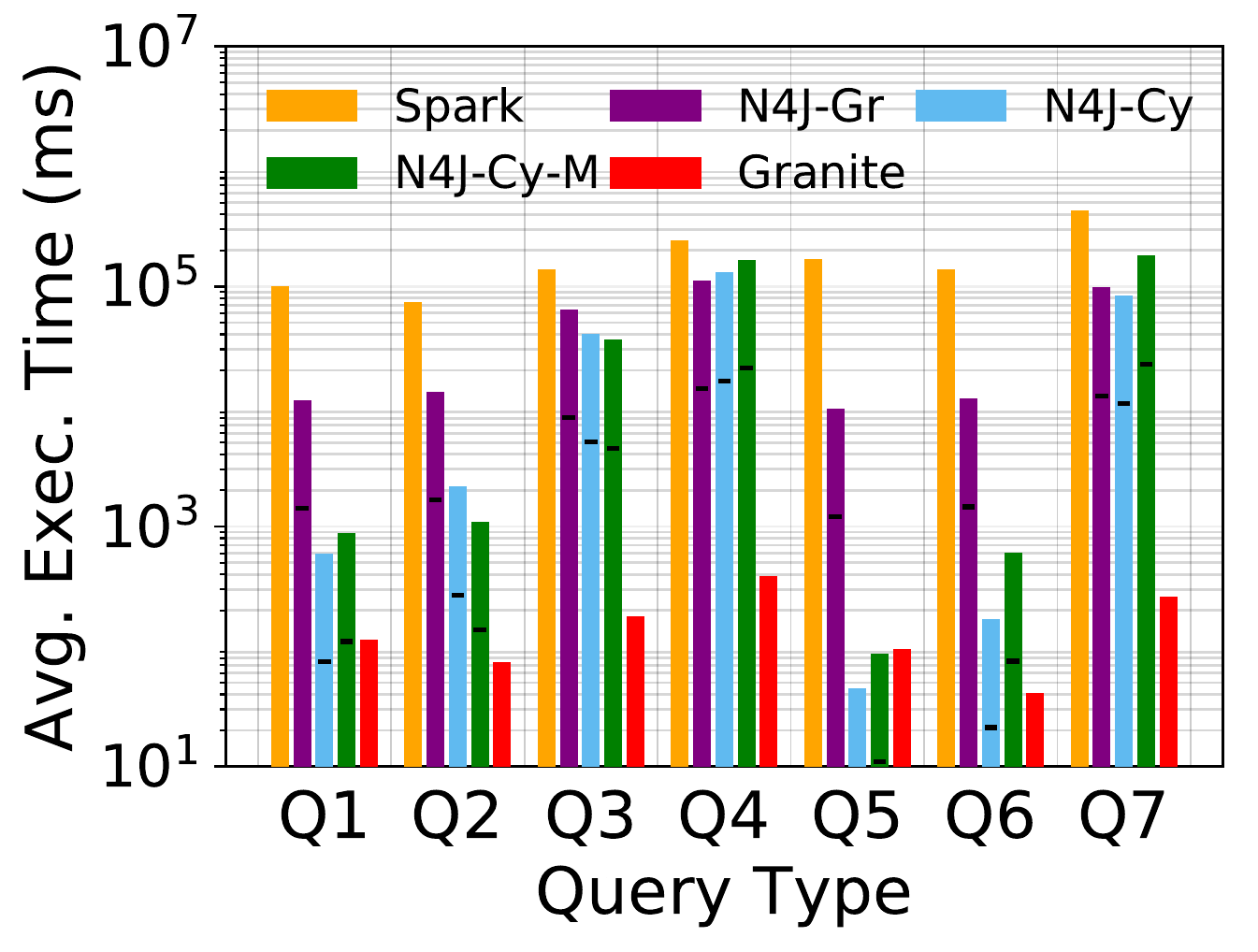}
		\label{fig:bc_10k_dw}
	}~
	\subfloat[100k:Z-S]{
		\includegraphics[width=0.435\textwidth]{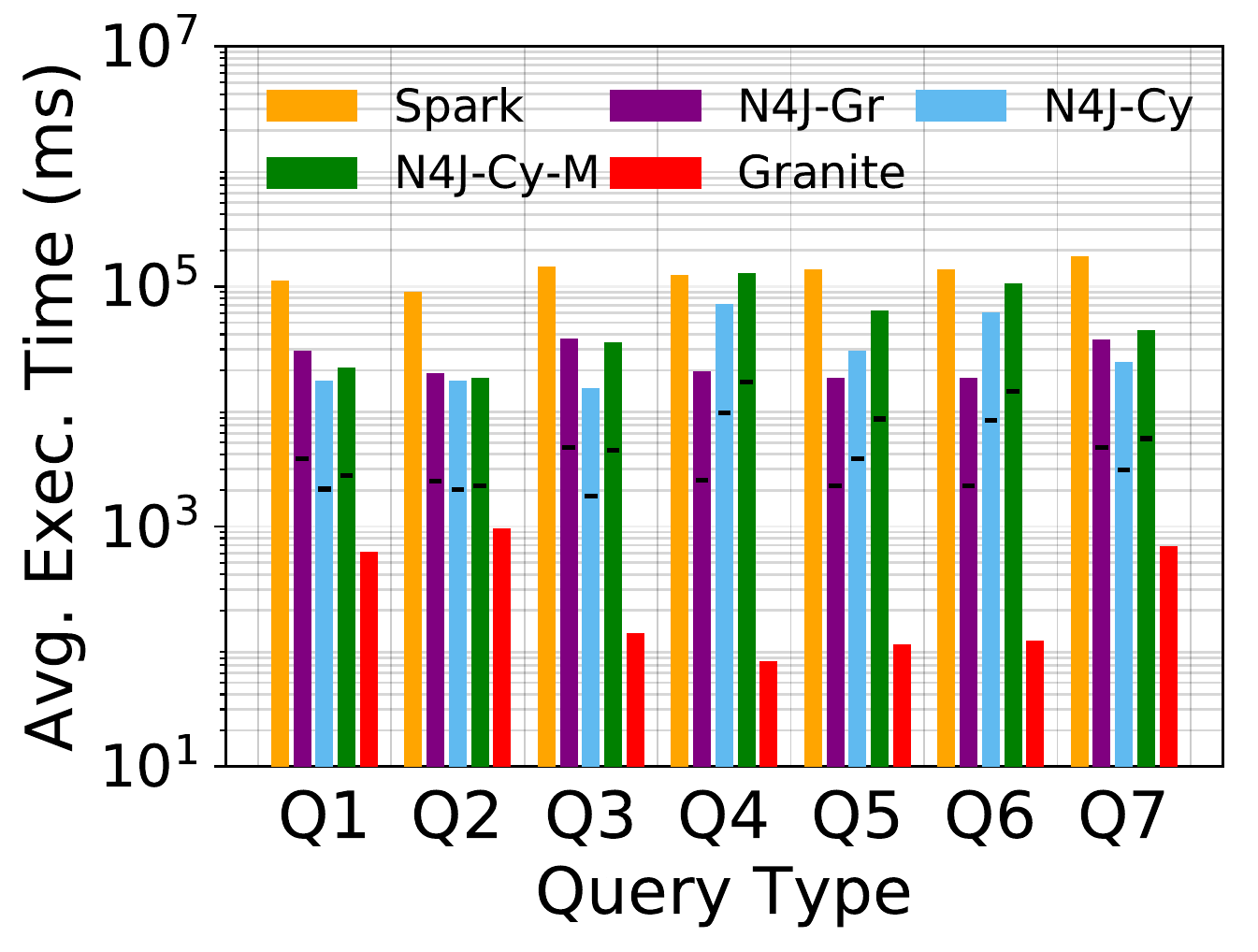}
		\label{fig:bc_100k_zf}
	}~\\
	\subfloat[100k:A-S]{%
		\includegraphics[width=0.435\textwidth]{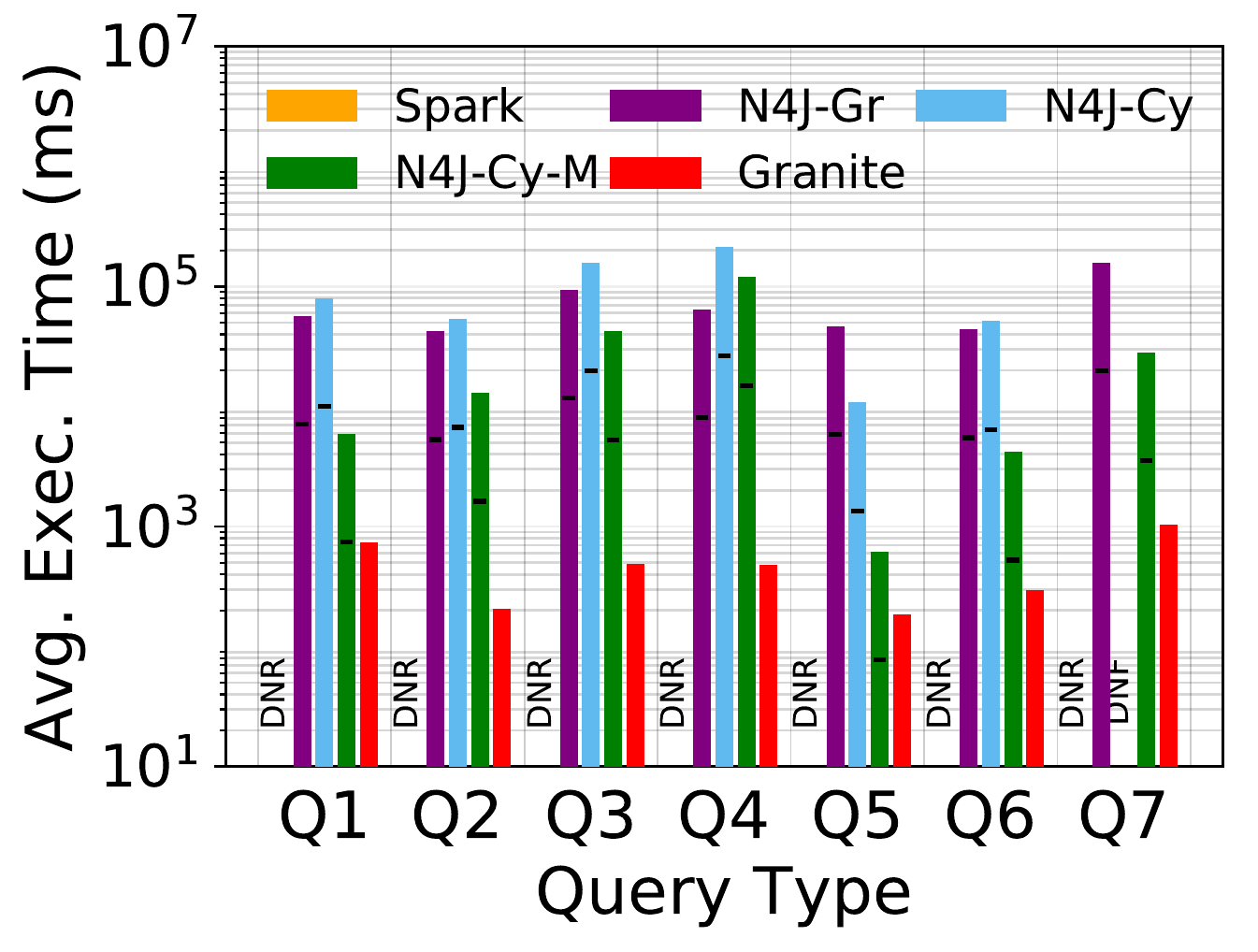}
		\label{fig:bc_100k_a}
	}~
	\subfloat[100k:F-S]{%
		\includegraphics[width=0.435\textwidth]{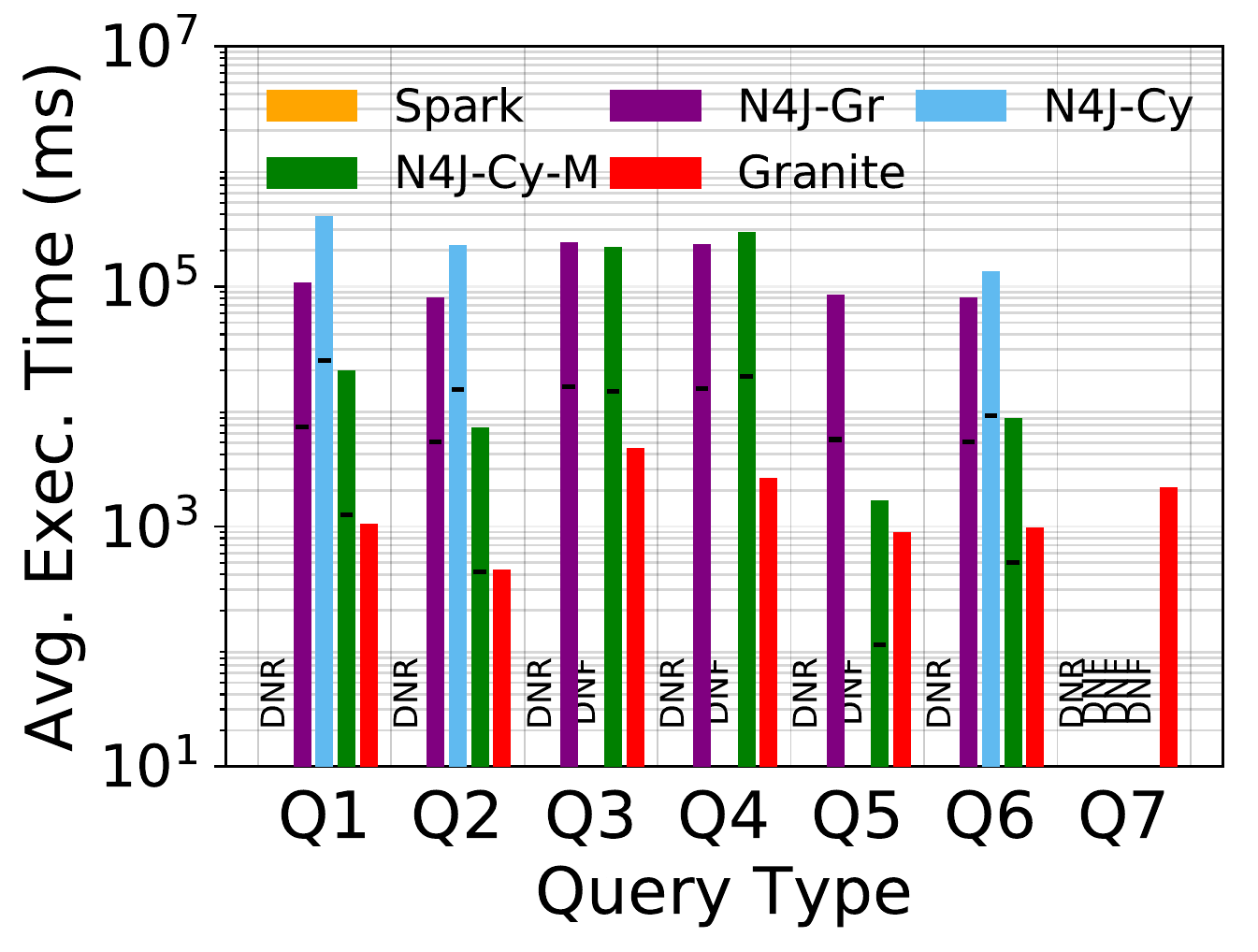}
		\label{fig:bc_100k_f}
	}
	\caption{Comparison of average execution time of \graphdb with baseline systems for non-aggregate query types, on \emph{Static Temporal Graphs}}
	\label{fig:bc_stat}
\end{figure*}

\begin{figure*}[!t]
	\subfloat[10k:DW-D]{
		\includegraphics[width=0.43\textwidth]{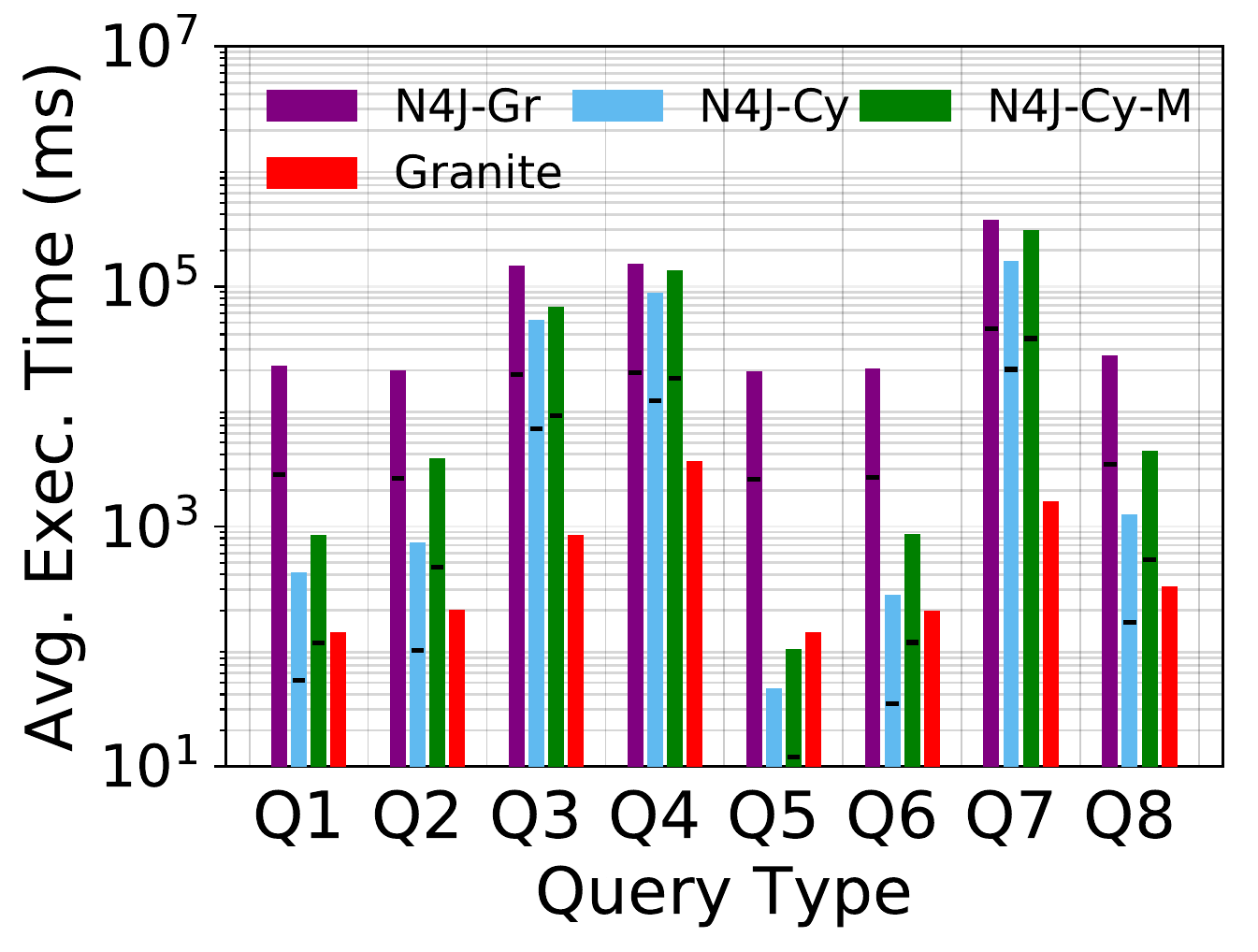}
		\label{fig:bc_10k_dw_dyn}
	}~
	\subfloat[100k:Z-D]{
		\includegraphics[width=0.43\textwidth]{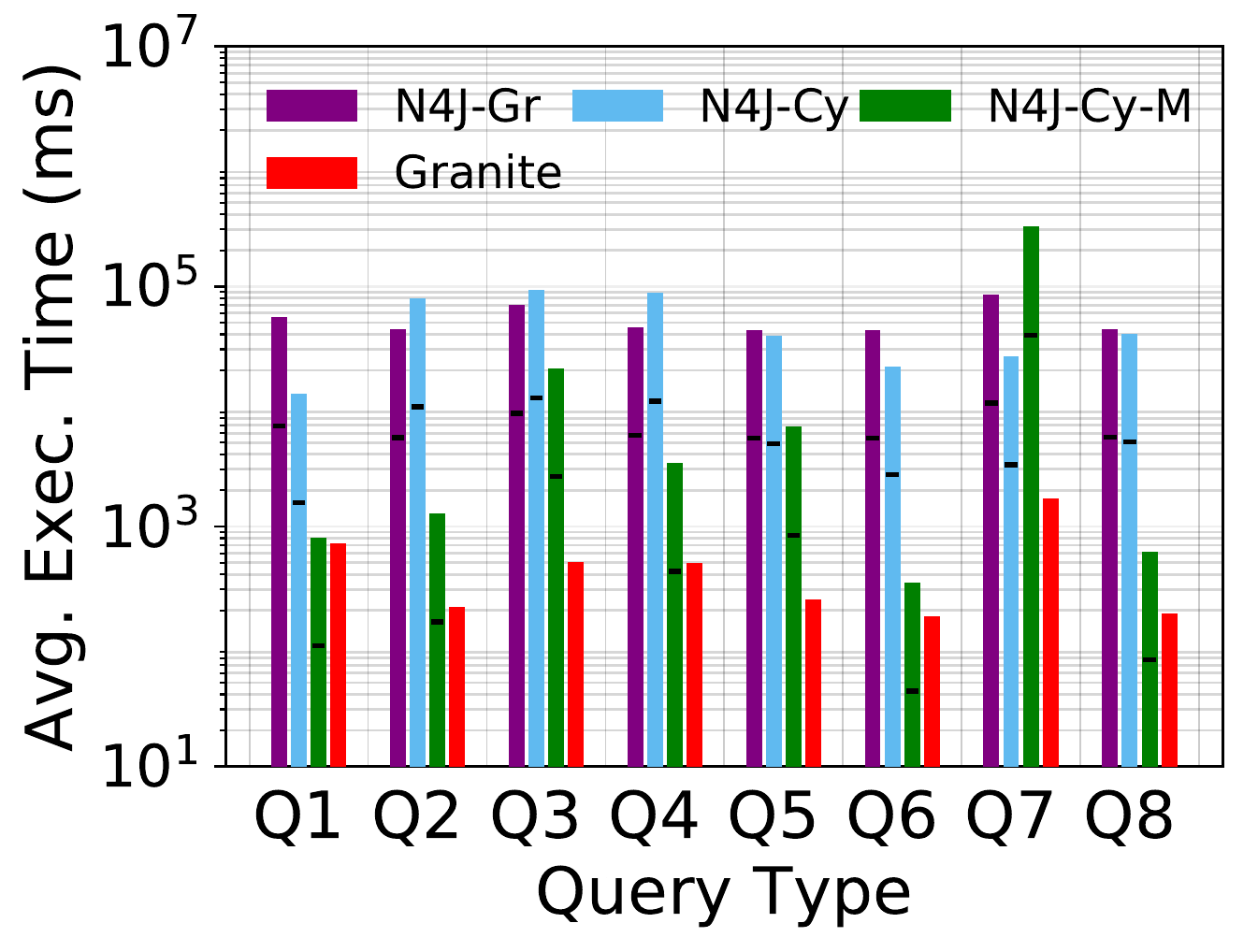}
		\label{fig:bc_100k_zf_dyn}
	}~\\
	\subfloat[\addc{100k:A-D}]{
		\includegraphics[width=0.43\textwidth]{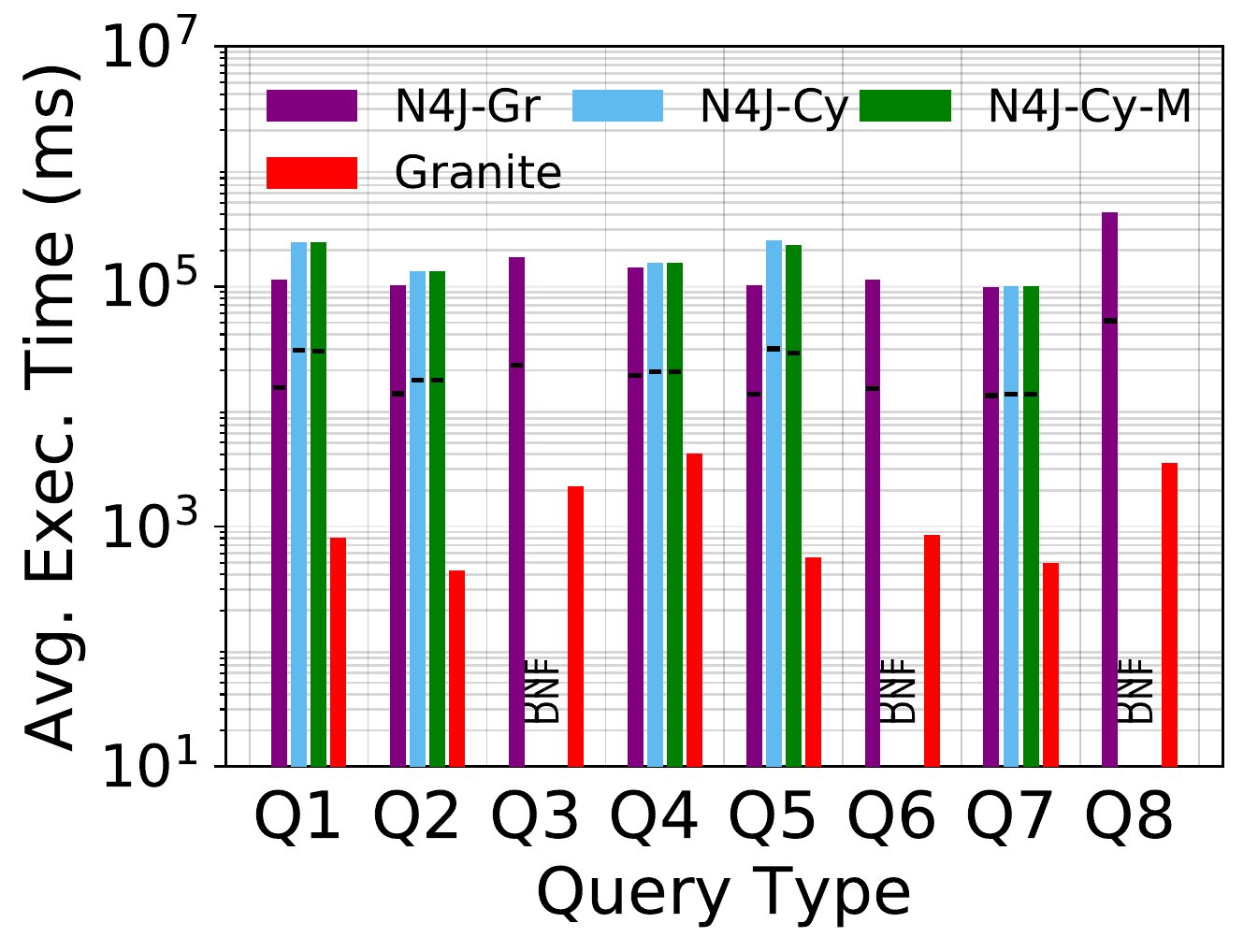}
		\label{fig:bc_100k_a_dyn}
	}~
	\subfloat[\addc{100k:F-D}]{
		\includegraphics[width=0.43\textwidth]{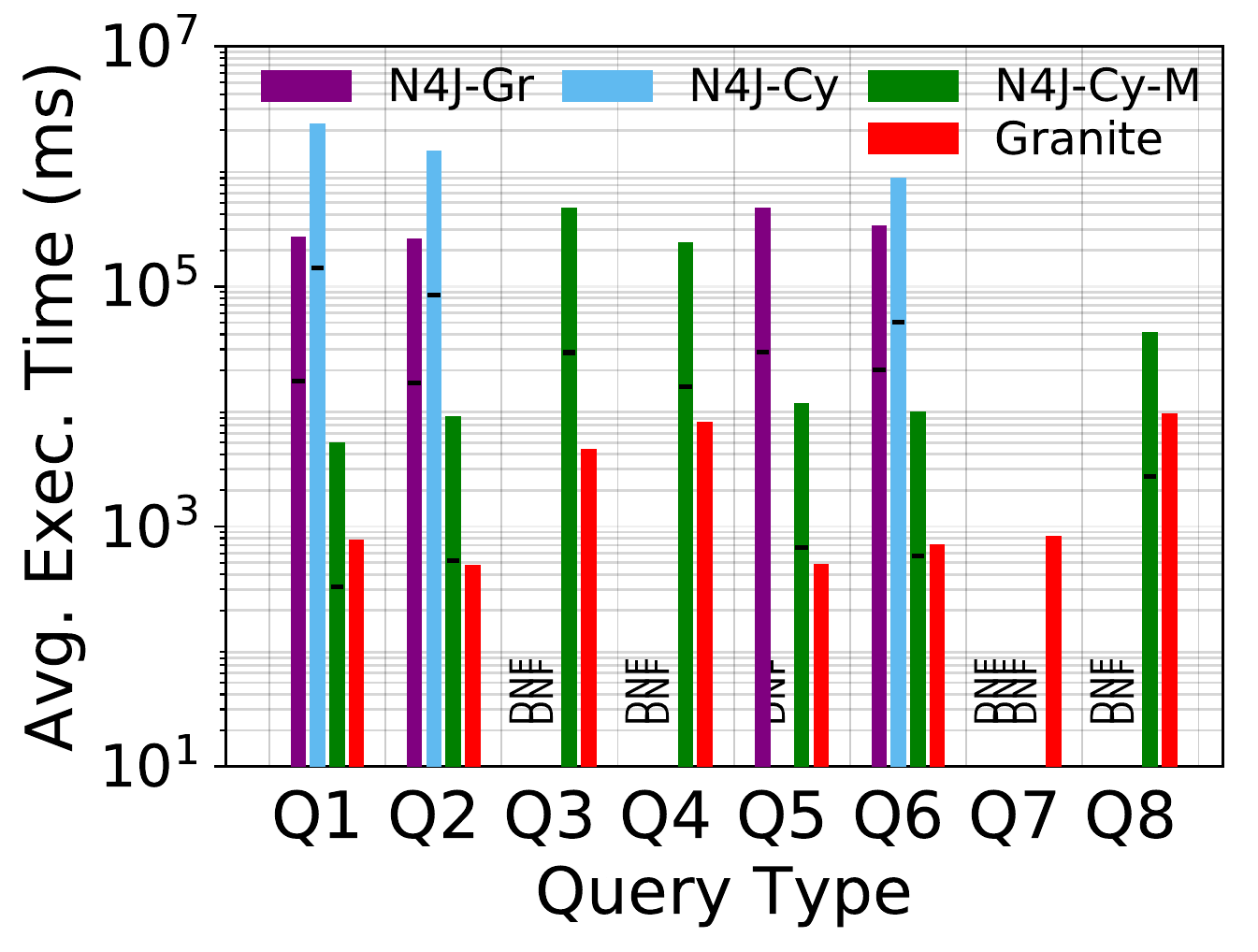}
		\label{fig:bc_100k_f_dyn}
	}
	\caption{Comparison of average execution time of \graphdb with baseline systems for non-aggregate query types, on \emph{Dynamic Temporal Graphs}}
	\label{fig:bc_dyn}
\end{figure*}

Figures~\ref{fig:bc_stat} show the average execution time on \graphdb and the baseline platforms (Y axis, log scale) for the different \emph{non-aggregate} query types (X axis) for the \emph{static temporal graphs}, and Figures~\ref{fig:bc_dyn} for the \emph{dynamic temporal graphs}. Only queries that complete in the $600~sec$ time budget are plotted. As Table~\ref{tbl:perc_completion} shows, Janus/Spark did not run (DNR) for several larger graphs due to resource limits when loading the graph in-memory from Cassandra. $32$--$79\%$ of queries did not finish (DNF) \addc{within the time budget} on Neo4J for \emph{100k:F-S}, the largest graph. \graphdb completes all queries on all graphs, often within $1~sec$. For the largest graph \emph{100k:F-S}, 
\graphdb uses 16 nodes to ensure that the graph fits in distributed memory.

\addc{
	\begin{figure*}[!t]
		\centering%
		\subfloat[100k:A-S]{
			\includegraphics[width=0.43\textwidth]{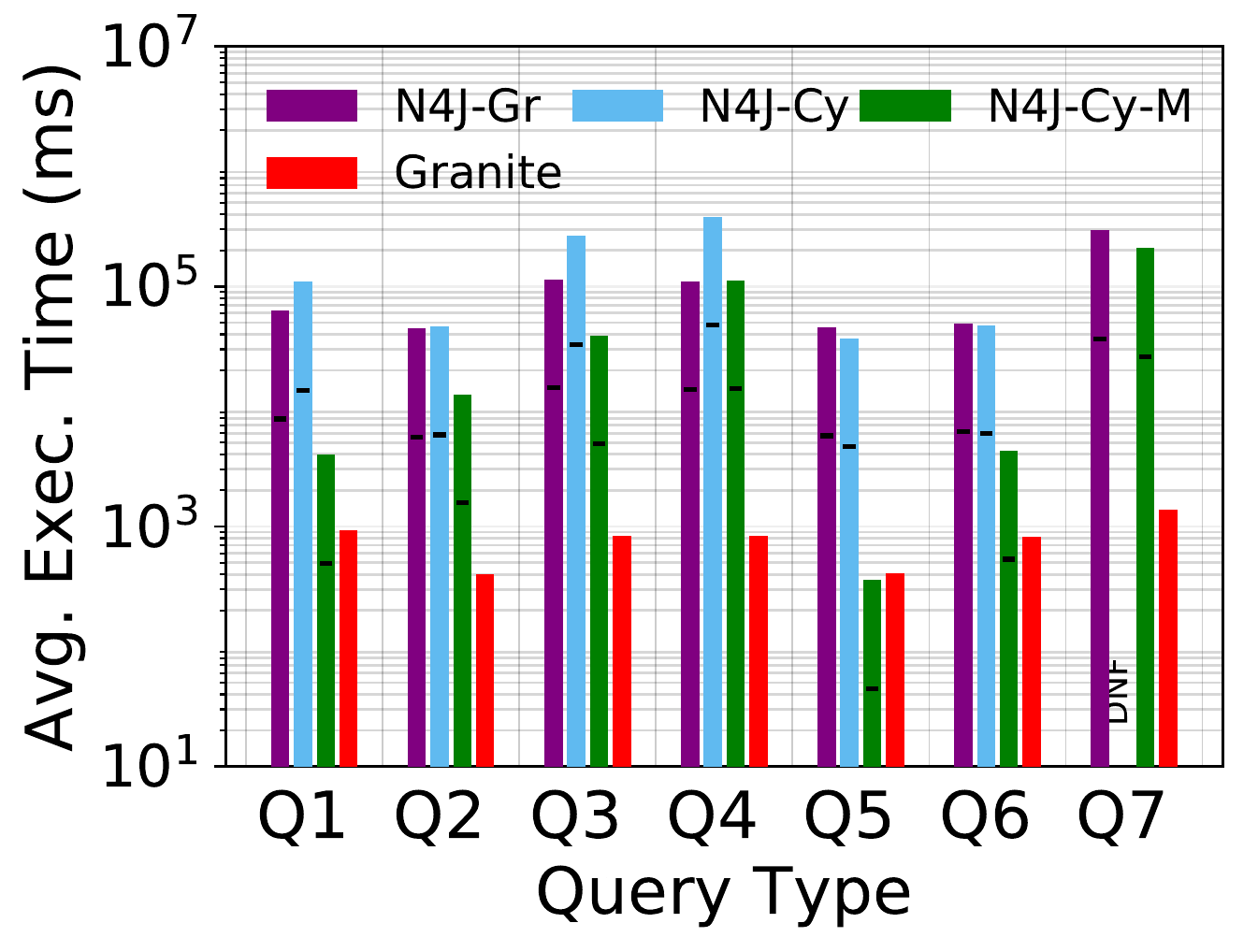}
			\label{fig:bc_100k_a_agg}
		}~
		\subfloat[100k:F-S]{
			\includegraphics[width=0.43\textwidth]{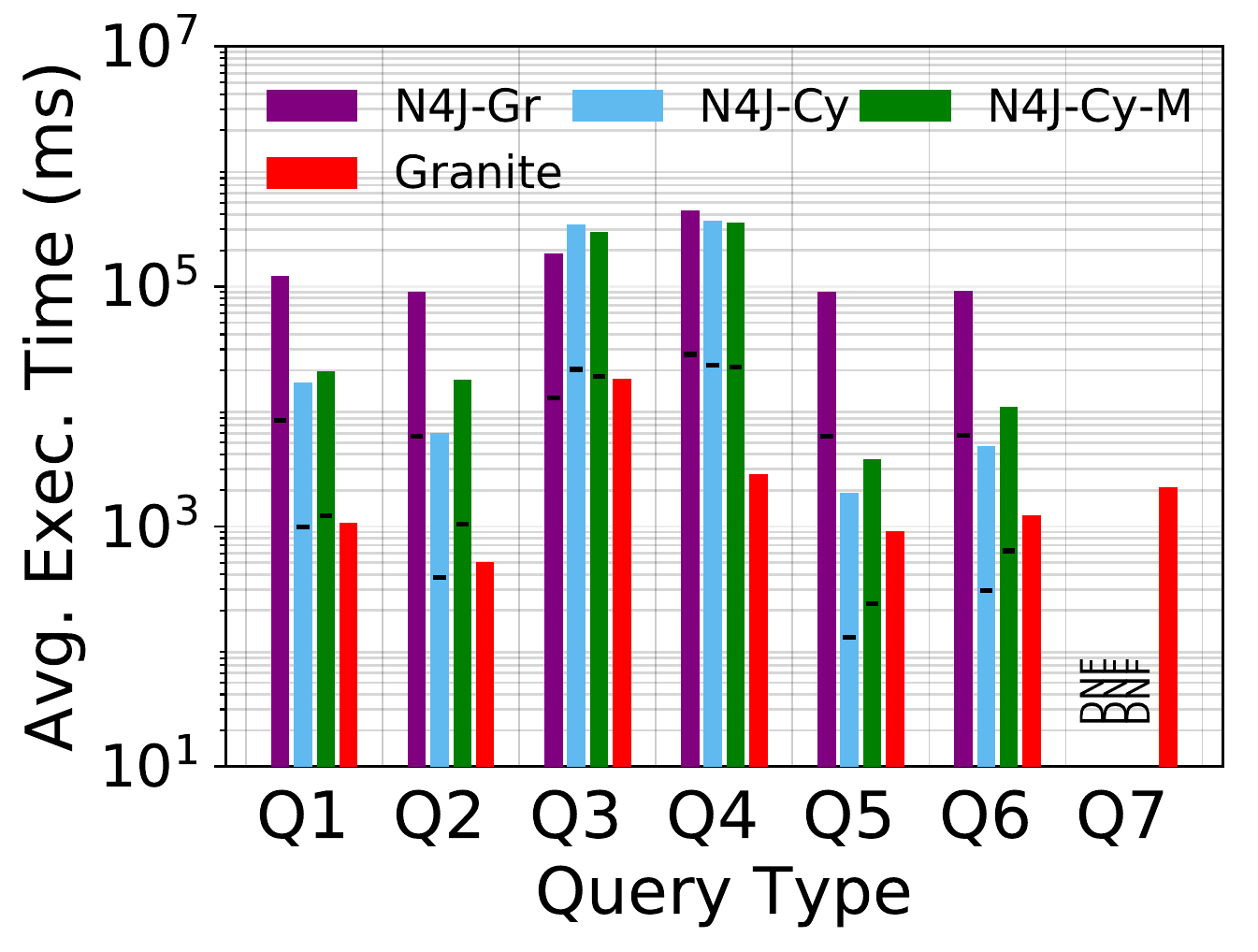}
			\label{fig:bc_100k_f_agg}
		}~\\
		\subfloat[100k:A-D]{
			\includegraphics[width=0.43\textwidth]{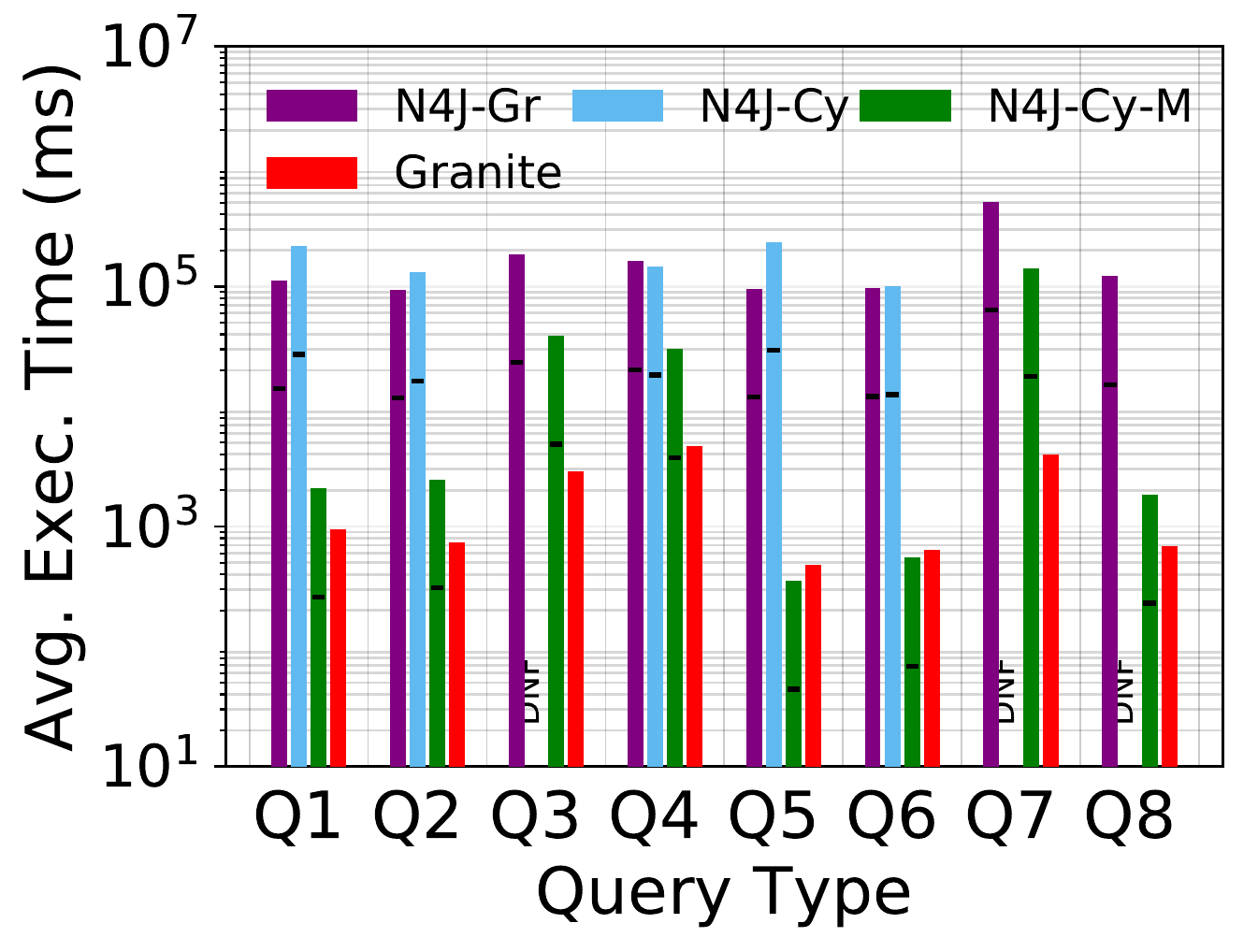}
			\label{fig:bc_100k_a_dyn_agg}
		}~
		\subfloat[100k:F-D]{
			\includegraphics[width=0.43\textwidth]{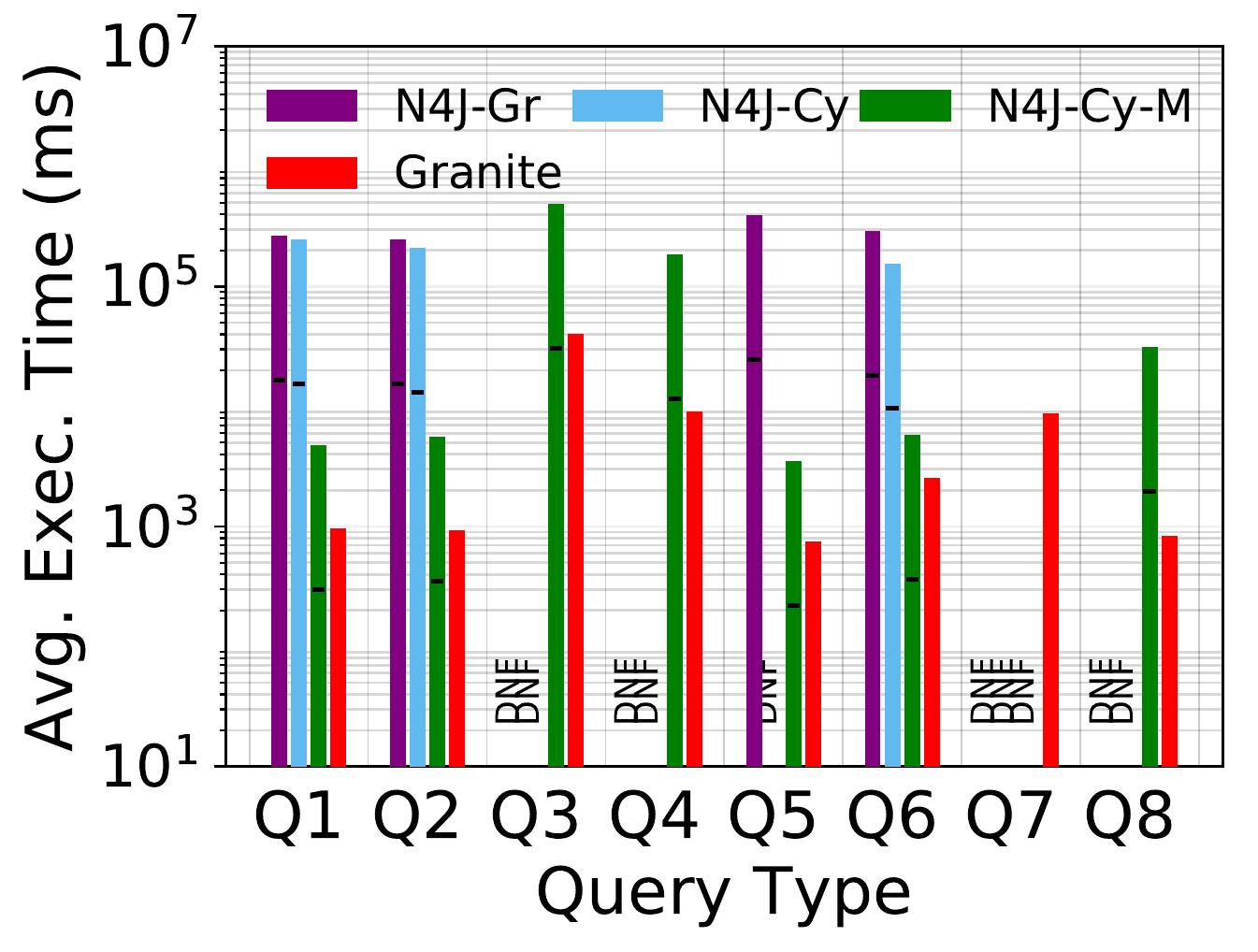}
			\label{fig:bc_100k_f_dyn_agg}
		}~
		\caption{\addc{Comparison of average execution time of \graphdb with baseline systems for \emph{Temporal Aggregate} query types}}
		\label{fig:bc_agg}
	\end{figure*}
} 

\begin{table}[!t]
	\small
	\setlength\tabcolsep{5pt} %
	\centering
	\caption{\% of queries that complete within $600~seconds$ for different platforms on the temporal graphs}
	\begin{tabular}{l||rrrrr}
		\hline
		\textbf{Graph} & \textbf{Spark} & \textbf{N4J-Gr} & \textbf{N4J-Cy} & \textbf{N4J-Cy-M} & \textbf{\graphdb} \\
		\toprule
		&		\multicolumn{5}{c||}{\em Static Graphs, Non-aggregate queries} \\ \hline
		10k:DW & 100 & 99 & 99 & 80 & 100
		\\
		\hline
		100k:Z & 93 & 90 & 100 & 100 & 100
		\\
		\hline
		100k:A & DNR & 100 & 90 & 98 & 100
		\\
		\hline
		100k:F & DNR & 66 & 21 & 68 & 100 \\
		\toprule		
		& \multicolumn{5}{c}{\em \addc{Dynamic Graphs, Temporal aggregate queries}} \\ \hline
		\addc{100k:A} & DNR & 98 & 65 & 99 & 100
		\\
		\hline
		\addc{100k:F} & DNR & 60 & 68 & 65 & 100
		\\	
		\bottomrule
	\end{tabular}
	\begin{tabular}{l||rrrrr}
		\hline
		\textbf{Graph} & \textbf{Spark} & \textbf{N4J-Gr} & \textbf{N4J-Cy} & \textbf{N4J-Cy-M} & \textbf{\graphdb} \\
		\toprule
		& \multicolumn{5}{c}{\em Dynamic Graphs, Non-aggregate queries} \\ \hline
		10k:DW 
		& DNR & 96 & 98 & 96 & 100 \\
		\hline
		100k:Z
		& DNR & 100 & 90 & 98 & 100 \\
		\hline
		100k:A 
		& DNR & 97 & 46 & 46 & 100 \\
		\hline
		100k:F 
		& DNR & 30  & 20 & 75 & 100 \\
		\toprule		
		& \multicolumn{5}{c}{\em \addc{Dynamic Graphs, Temporal aggregate queries}} \\ \hline
		\addc{100k:A}
		& DNR & 95 & 46 & 99 & 100 \\
		\hline
		\addc{100k:F}
		& DNR & 36 & 19 & 78 & 100 \\	
		\bottomrule
	\end{tabular}
	\label{tbl:perc_completion}
\end{table}

The bar plots show that \graphdb is much faster than the baselines, across \emph{all} graphs and \emph{all} query types, except for $Q5$ on the smallest graph, \emph{10k:DW-S}. %
On average, we are $149\times$ faster than N4J-Cy-M, $192\times$ faster than N4J-Cy, $154\times$ faster than N4J-Gr and $1140\times$ faster than Spark. 
Other than the largest graph, \graphdb completes on an average within $500~ms$ for all static graphs and most query types, and on an average within $1000~ms$ for \emph{100k:F-S} and all the dynamic graphs.

\addc{Focusing on specific query types for the largest static temporal graph, \emph{100k:F-S}, $Q2$ takes the least time for \graphdb due to its short path length of 2. The left-to-right execution by the baseline platforms is the optimal query plan, but we are still able to out-perform them due to the parallelism provided by partitioning.}
\graphdb takes $\approx 5~secs$ for $Q3$ due to the huge number of results, $\approx 5.9M$ on average. %
But this \addc{query} does not even complete for N4J-Cy and Spark. \graphdb's tree-based result structure is more compact, reducing memory and communication costs.
$Q4$ for this graph is also $89$--$112\times$ better in \graphdb than the baselines, \addc{with large} result sizes of $\approx 72k$ on average. Here, there is a rapid fan-out of matching vertices followed by a fan-in as they fail to match downstream predicates, leading to high costs.  %
\addc{
	$Q7$ queries are able to complete only in \graphdb and not on the baseline platforms. This query has an optimal split point of $1$ or $2$ which is not adopted by the baselines. In fact, baselines use the worst possible left-to-right plan, which we see is $4\times$ slower than the optimal for \graphdb.
}

\graphdb is also consistently better for the \emph{dynamic graphs}. %
\addc{Similar to the static graphs,} the only time that our average query time is slower than a baseline is for $Q5$ on \emph{10k:DW-D}. 
Here, the default left-to-right execution is near-optimal, \addc{and the query has} a low traversal fan-out and $<10$ results. So the baselines are in an ideal configuration while \graphdb has overheads for distributed execution.
\emph{Neo4J using Cypher}, on the single compute node \addc{(N4J-Cy)} and the big memory node \addc{(N4J-Cy-M)}, are the next best to \graphdb. The large memory variant gives similar performance as the regular memory one for the smaller graphs, %
but for larger graphs like \emph{100k:A} and \emph{100k:F}, it out-performs. For the latter graph, N4J-Cy could not finish several query types. Though Neo4J uses indexes to  help filter the vertices for the first hop, query processing for later hops involves a breadth first traversal and pruning of paths based on the predicates. There are also complex joins between consecutive edges along the path to apply the temporal edge relation. These affect their execution times.
Gremlin and Cypher variants of Neo4J are comparable \addc{in performance}, with no strong performance skew either way. Interestingly, the Gremlin variant of Neo4J is able to run most query workloads for all graph, albeit with slower performance.

The \emph{Janus/Spark} distributed baseline takes the most time for all these queries. %
This is despite omitting its initial graph RDD creation time ($\approx 80~secs$).
\graphdb persists the graph in-memory across queries. Despite using distributed machines, Spark is unable to load large graphs in memory and often fails to complete execution within the time budget. A similar challenge was seen even for alternative engines like, Hadoop, used by JanusGraph and Spark was the best of the lot.

In the bar plots, we also show a black bar for the single-machine baselines, which is \addc{marked at} the $1/8^{th}$ execution time-point -- this shows the theoretical time that would be taken by these platforms \addc{if they had} perfect \addc{parallel} scaling on 8 machines, though they do not support \addc{parallel execution}. %
As we see, %
\graphdb is often able to complete its execution %
within that mark, showing that our distributed engine shows scaling performance comparable or better than highly optimized single-machine platforms, \addc{even if they had ideal scaling}. %

\addc{
	Lastly, we compare the performance of \emph{temporal aggregate queries} for the two largest static and dynamic graphs, \emph{100k:A} and \emph{100k:F}. Their execution times on the different platforms are shown in Figure~\ref{fig:bc_agg}. %
	For the static graphs, %
	we observe from Figures~\ref{fig:bc_100k_a_agg} and~\ref{fig:bc_100k_f_agg} that \graphdb is much faster than all the baselines for most query types. On average, we are $165\times$ faster than N4J-Cy-M, $175\times$ faster than N4J-Cy and $95\times$ faster than N4J-Gr. This is $10\times$ faster even when compared with %
	the perfect scaling extrapolation for the baselines.
}

\addc{These temporal aggregate queries are slower compared to their non-aggregate equivalents. Specifically, for \emph{100k:A-S}, \graphdb takes {$64\%$} ($\approx 315~ms$) more on average while the baseline platforms on average are {$56\%$} (N4J-Cy), {$42\%$} (N4J-Gr) and {$78\%$} (N4J-Cy-M) slower, %
	which translates to $\approx 24$--$53~secs$ longer, per query. %
	The baselines' time increase considerably due to the additional overhead of sending the entire result set back to client to perform the temporal aggregation, as opposed to just sending the total number of results for the non-aggregate queries. Since \graphdb does this natively in a distributed manner, we mitigate this cost.}

\addc{
	\graphdb completes all these queries when executed using the plan selected by the cost model (Table~\ref{tbl:perc_completion}). %
	The baseline platforms are only able to complete, on average, $79\%$ (N4J-Gr), $67\%$ (N4J-Cy) and $82\%$ (N4J-Cy-M) of the queries on the static graphs, and this is worse for the dynamic graphs, ranging from $33\%$--$89\%$.}%

\addc{Also, as Figures~\ref{fig:bc_100k_a_agg} and~\ref{fig:bc_100k_f_agg} show, we take under $1~sec$ to run all queries on \emph{100k:A-S} except Q7, and within $2.1~secs$ for all queries on the largest graph, \emph{100k:F-S}, except Q3 -- query Q3 takes longer due to the large result count of $\approx 4.3M$ (Figure~\ref{fig:100k_f_result_dist}). For dynamic graphs, we take under $3~secs$ for all queries on \emph{100k:A-D} except Q4 and Q7, and within $9.2~secs$ for all queries on the largest graph, \emph{100k:F-D}, except Q3 (Figures~\ref{fig:bc_100k_a_dyn_agg} and~\ref{fig:bc_100k_f_dyn_agg}).
	None of the baseline platforms could finish query type $Q7$ for \emph{100k:F-S} or \emph{100k:F-D}. This query starts and ends with the \emph{Post} vertex type, which has a high cardinality. Also, these queries on the baseline platforms need to accumulate all the results for client-side aggregation. Both of these lead to memory-pressure for the larger graphs. %
}

\subsection{Components of Execution Time}

\addc{Next, we briefly examine where the time is spent in distributed execution. As an exemplar, Figure~\ref{fig:stacked_7} shows a stacked bar plot of the time taken by $Q7$ in different supersteps, and within different workers in a superstep, for the 100k:A-S graph. The stacks represent the time taken by the \emph{init/compute}, \emph{scatter}, and \emph{join} phases of \graphdb, the \emph{interval compute} parent phase of Graphite (ICM), the \emph{partition compute} grand-parent phase of Giraph (VCM), and \emph{other} residual time such as barrier synchronization and JVM garbage collection (GC), in each superstep. These times are averaged across all $100$ instances of the query type. For deterministic execution, we select a fixed split point for the execution plan that is optimal for a majority of the queries, which, for $Q7$ is at the third vertex in the path.}

\addc{For Q7, the first superstep time is dominated by the \emph{init} logic as the predicate operates on the \emph{Post} vertex type, which has $8.7M$ vertices. Its \emph{scatter} time is minimal as only $71k$ out edges match out of $250k$ and are used to send messages. The overheads of \emph{interval compute} are small, but \emph{partition compute} takes longer at $140~ms$. In the latter, the Giraph logic which we extend selects the active partitions based on the vertex type of the query predicate (\emph{Post}, in the case of $Q7$), iterates through its active vertices, invokes interval compute on each with the incoming messages, and clears the message queue.  %
	The \emph{other} time is non-trivial at $145~ms$. This is caused by GC triggering due to memory pressure, and taking $110~ms$, with the rest going to the superstep barrier.}

\addc{In superstep 2, the \emph{compute} time is negligible at $1.5~ms$ as only $3.4k$ \emph{Person} vertices are active across both branches of the query plan, but \emph{scatter} takes $247~ms$ since $2.83M$ edges are processed along one branch of the plan -- the \emph{Person} vertex has a high out-edge degree -- out of which $31k$ satisfy the predicate. About $100~ms$ is taken by \emph{partition and interval computes}, for selecting and iterating over the relevant active vertices, and for performing TimeWarp and state initialization, %
	while there is a GC overhead of $64~ms$ in \emph{other}.
	In the last superstep, there is a small time taken for \emph{compute} and to \emph{join} the results.}

\addc{Interestingly, the time taken by each phase is similar across the different workers in a superstep for this query. This indicates that the partitioning manages to balance the load for this query type. However, for other queries like $Q4$ (not shown for brevity), we observe that in some supersteps, scatter takes 79\% longer for the slowest worker compared to the fastest due to a skew in the number of edges activated per worker. %
	Also, queries like $Q4$ take less time for the first superstep but a larger time in superstep 2 due to a high fanout, going from $36k$ edges processed in the first step to $1.48M$ edges in the second step. In others like $Q3$, the first superstep is dominated by scatter since the initial vertex type \emph{Person} has only $89k$ vertices with $770$ of them matching, but these cause $950k$ edges to be processed of which $122k$ match and trigger messaging.%
}

\addc{In summary, the different supersteps have high variability in execution times and there is also variability in the time taken by each phase. Despite that, the cost model is able to discriminate and select near-optimal plans. The load is mostly balanced across workers in a superstep, though this depends on the query type. Much of the time is spent directly in processing the query using \emph{compute} and \emph{scatter}, with some additional overheads for the other phases.}

\subsection{Weak Scaling}

\begin{figure}[!t]
	\addcBegin
	\centering
	\begin{minipage}{.75\textwidth}
		\centering%
		\includegraphics[width=1\columnwidth]{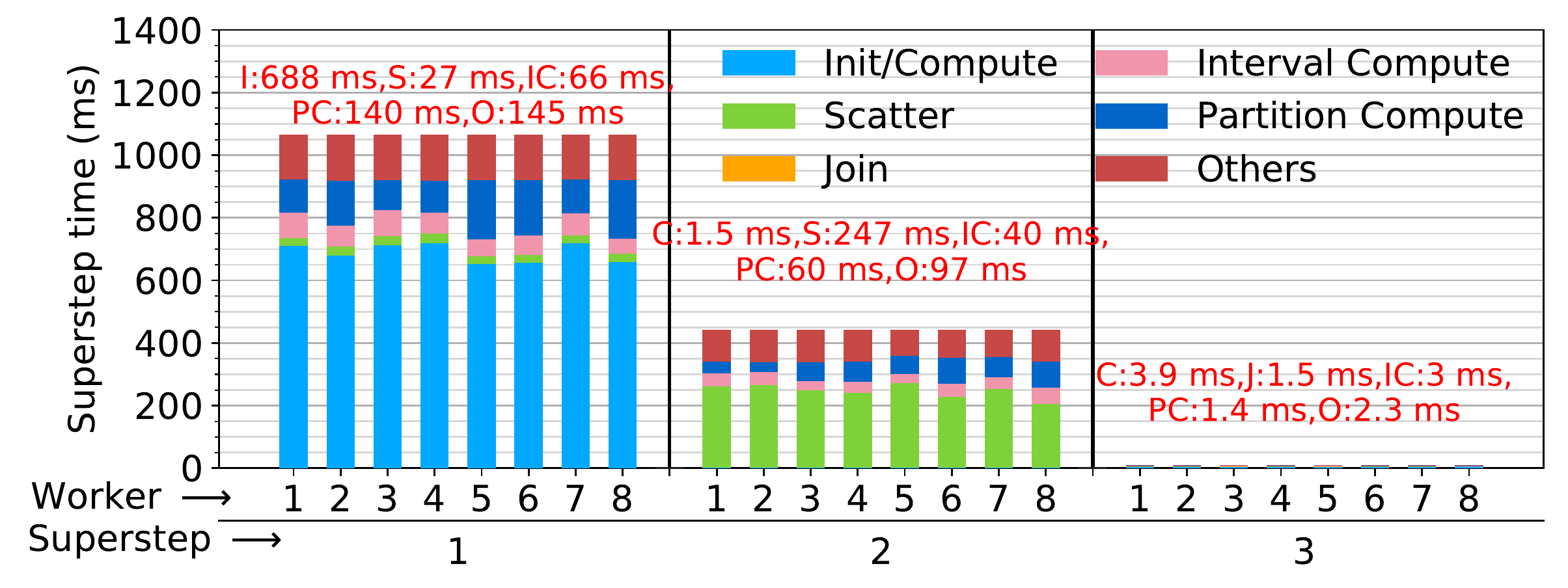}
		\caption{\addc{Stacked bar plot of component execution times in each superstep, averaged over all queries of query type $Q7$ on 100k:A-S graph. Header labels indicate average component time across Workers in a superstep.}%
		}
		\label{fig:stacked_7}

	\end{minipage}%
	\qquad \\
	\begin{minipage}{.65\textwidth}
		\centering%
		\includegraphics[width=0.95\textwidth]{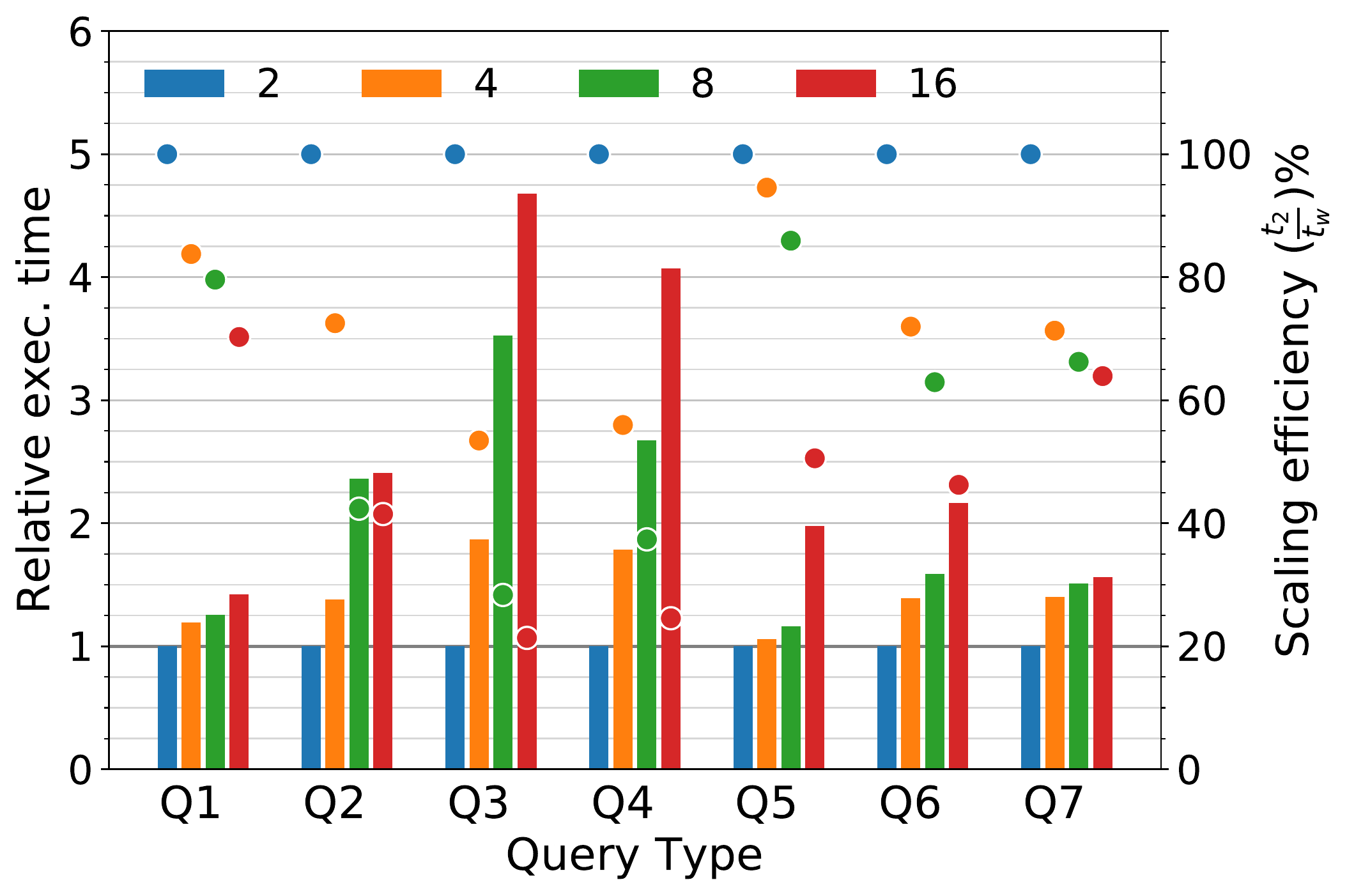}
		\caption{\addc{Relative execution time (left axis, bar) and Scaling efficiency\%=$\frac{t_2}{t_w}\%$ (right axis, circle) for Worker counts $w=\{4,8,16\}$, relative to $w=2$ for \emph{Weak Scaling} runs with $(w \times 6.25k)$:F-S graphs}}
		\label{fig:scaling_rel}
	\end{minipage}%
	\addcEnd
\end{figure}

\addc{
	We evaluate the weak scaling capabilities of \graphdb using the static Facebook-distribution graphs. We use 4 different system resource sizes -- $2$, $4$, $8$ and $16$ Workers, with 1 compute node per Worker, and the graph sizes increase proportional to the Worker count -- 12.5k:F-S, 25k:F-S, 50k:F-S and 100k:F-S. %
	This attempts to keep the workload per Worker constant across the scaling configurations, with the per-Worker vertex and edge counts remaining within $\pm 18\%$ and $\pm 23\%$ of their mean, respectively. 100k:F-S is partitioned into $512$ partitions ($128$ per vertex type), and other graphs into $256$ partitions ($64$ per vertex type). This ensures that we have enough partitions for the compute threads to process them in parallel across all Workers. We generate and use a $100$ query workload for each query type, for each graph.%
}%

\addc{
	The left Y axis of Figure~\ref{fig:scaling_rel}~(bars) shows for each query type, the average relative execution time when using $w=4,8,16$ Workers, compared to $w=2$ Workers. The right Y axis (circles) shows the $\text{\em scaling efficiency}= \frac{t_2}{t_w}\%$, i.e., time taken on 2 Workers vs. time taken on $w$ Workers. With perfect weak scaling, the relative time should be constant and efficiency $100\%$. 
	The asymmetric nature of graph data structure makes it rare to get ideal weak scaling. However, we do see that query types $Q1$, $Q5$, $Q6$ and $Q7$ offer $\geq 60\%$ scaling efficiency on up to 8 Workers, and all queries but $Q3$ and $Q4$ have $\geq 40\%$ efficiency on up to 16 Workers. %
	$Q3$ and $Q4$ are unable to fully exploit the additional resources due to stragglers among their threads, which are often $10\times$ slower  due to uneven load. These two queries also have the largest result cardinality, which causes more messages to be sent over the network as the number of machines increase. As a result, they have poor scaling efficiency.%
}

\section{Conclusions}%
\label{sec:conclude}
In this article, we have motivated the need for querying over \addc{large} temporal property graphs and the lack of such platforms. We have proposed an intuitive temporal path query model to express a wide variety of requirements over such graphs, and designed the \graphdb distributed engine to implement these at scale over the Graphite ICM platform. Our novel analytical cost model uses concise information about the graph to allow accurate selection \addc{of a} distributed query execution plan \addc{from several choices}. These are validated through rigorous experiments on \addc{8 temporal} graphs with \addc{a $1600$-query workload}, derived from the LDBC benchmark. \graphdb out-performs the baseline graph platforms and gives $<1~sec$ latency for most queries.

As future work, we plan to explore out of core execution models to scale beyond distributed memory, indexing techniques to accelerate performance, more generalized temporal tree and reachability query models, and compare performance with other research prototypes and metrics from literature. \addc{Designing incremental query execution strategies over streaming property-graph updates is also a related and under-explored challenge.} \addc{The \graphdb platform is also finding relevance in analyzing epidemiological networks that form temporal property graphs constructed from, say, digital contact tracing for the COVID-19 pandemic. This may motivate the need for further query operators.} %

\section*{Acknowledgements} The first author of this work was supported by the Maersk CDS M.Tech. Fellowship, and the last author was supported by the Swarna Jayanti Fellowship from DST, India. We thank Ravishankar Joshi from BITS-Pilani, Goa for his assistance with the experiments.

\bibliographystyle{IEEEtran}
\bibliography{paper}

\end{document}